\documentclass[prd,twocolumn,showpacs,preprintnumbers,superscriptaddress,nofootinbib,amsmath,amssymb,nobalancelastpage]{revtex4}
\usepackage{graphicx}
\usepackage{bm}
\usepackage{dcolumn}
\usepackage{amsmath}
\usepackage{amssymb}
\usepackage{color}
\usepackage{url}

\usepackage{aas_macros}

\usepackage{hyperref}

\usepackage{subfigure}

\newcommand{\cm}{{\mathrm{cm}}}
\newcommand{\second}{{\mathrm{s}}}
\newcommand{\GeV}{\mathrm{GeV}}

\newcommand{\yr}{\mathrm{yr}}

\newcommand{\sig}{\sigma}
\newcommand{\sigv}{\langle \sig v \rangle}
\newcommand{\mass}{M}
\newcommand{\Pw}{\mathbf{P}_\mathrm{w}}

\newcommand{\degr}{^\circ}
\newcommand{\nhat}{\boldsymbol{\hat{\textbf{n}}}}
\newcommand{\PSF}{{\rm PSF}}
\newcommand{\Aeff}{A_\mathrm{eff}}
\newcommand{\prob}{\mathrm{P}}
\newcommand{\HO}{H_0}
\newcommand{\HA}{H_1}
\newcommand{\Hb}{H_b}
\newcommand{\Hsb}{H_{s+b}}
\newcommand{\crit}{\mathcal{C}}
\newcommand{\Ex}{\mathrm{E}}

\newcommand{\SNR}{\mathrm{SNR}}
\newcommand{\Tobs}{T_{\rm obs}}
\newcommand{\tr}{\mathrm{tr}}

\begin{document}

% Title & Authors
\title{A Comprehensive Search for Dark Matter Annihilation in Dwarf Galaxies}

\author{Alex Geringer-Sameth}
\email{alexgs@cmu.edu}
\affiliation{McWilliams Center for Cosmology, Department of Physics, Carnegie Mellon University, Pittsburgh, PA 15213}
\affiliation{Department of Physics, Brown University,  Providence, RI 02912}

\author{Savvas M. Koushiappas}
\email{koushiappas@brown.edu}
\affiliation{Department of Physics, Brown University,  Providence, RI 02912}

\author{Matthew G. Walker}
\email{mgwalker@andrew.cmu.edu}
\affiliation{McWilliams Center for Cosmology, Department of Physics, Carnegie Mellon University, Pittsburgh, PA 15213}

\date{\today}

\begin{abstract}

We present a new formalism designed to discover dark matter annihilation occurring in the Milky Way's dwarf galaxies. The statistical framework extracts all available information in the data by simultaneously combining observations of all the dwarf galaxies and incorporating the impact of particle physics properties, the distribution of dark matter in the dwarfs, and the detector response. The method performs maximally powerful frequentist searches and produces confidence limits on particle physics parameters. Probability distributions of test statistics under various hypotheses are constructed exactly, without relying on large sample approximations. The derived limits have proper coverage by construction and claims of detection are not biased by imperfect background modeling. We implement this formalism using data from the Fermi Gamma-ray Space Telescope to search for an annihilation signal in the complete sample of Milky Way dwarfs whose dark matter distributions can be reliably determined. We find that the observed data is consistent with background for each of the dwarf galaxies individually as well as in a joint analysis. 
The strongest constraints are at small dark matter particle masses. Taking the median of the systematic uncertainty in dwarf density profiles, the cross section upper limits are below the pure $s$-wave weak scale relic abundance value ($2.2 \times 10^{-26} \, \cm^3\second^{-1}$) for dark matter masses below 26~GeV (for annihilation into $b\bar{b}$), 29~GeV ($\tau^+\tau^-$), 35~GeV ($u \bar{u}, d \bar{d}, s \bar{s}, c \bar{c}$, and $gg$), 6~GeV ($e^+e^-$), and 114 GeV ($\gamma\gamma$). For dark matter particle masses less than 1~TeV, these represent the strongest limits obtained to date using dwarf galaxies.

\end{abstract}


\maketitle

%%%%%%%%%%%%%% Introduction
%%%%%%%%%%%%%%
\section{Introduction}

By the mid-1970s the quest to extend particle physics beyond the Standard Model had begun to receive valuable contributions from cosmology. Astronomical observations revealed dark matter to be the dominant form of matter in the universe and it appeared to consist of some non-baryonic substance \cite{1978ApJ...223.1015G,1978AJ.....83.1050S,1979ARNPS..29..313S,1982PhRvL..48..223P,2000ApJ...542..281A}. This immediately provided a fundamental question for particle physics: what theoretical extension to the Standard Model can explain the nature of this component? It was realized that a new particle, if it were massive ($\sim \GeV$ scale) and interacted with the Standard Model via weak interactions, would automatically exist in the universe with an abundance roughly equal to that observed for dark matter (e.g.~\cite{zel1965advances,1965PhL....17..164Z,1966PhRvL..17..712C,1977PhRvL..39..165L,1977PhLA...69...85H,1979PhLB...82...65W,1979ARNPS..29..313S,1985PhRvD..32.3261B,1986PhRvD..33.1585S,1988NuPhB.310..693S,1991NuPhB.360..145G}). In parallel, particle physicists had been exploring fine-tuning issues with the Standard Model which seemed to point toward the existence of new physics (and new particles) at the weak scale $\sim \mathcal{O}(100\, \GeV)$ (e.g., \cite{1974PhRvL..33..451G,1981PhRvD..24.1681D}). The discovery of these particles would simultaneously revolutionize both particle physics and cosmology.

Equally important to the theory, weakly interactive massive particles (WIMPs) are within reach of discovery through a variety of experiments. In astrophysics, one of the most promising approaches has been the search for dark matter annihilation into Standard Model particles. The process of annihilation is, in fact, fundamental to the relic abundance argument motivating WIMP dark matter. The measured abundance of dark matter in the universe sets a lower bound on the annihilation cross section: if the (s-wave, velocity-averaged) annihilation cross section $\sigv$ is approximately $2.2 \times 10^{-26} \, \cm^3 \second^{-1}$~\cite{2012PhRvD..86b3506S} the relic abundance of a WIMP will equal that observed for dark matter, with a smaller cross section {\em overproducing} the dark matter abundance, and a higher cross section (up to limits derived from quantum mechanical or astrophysical arguments \cite{1990PhRvL..64..615G,2001PhRvL..86.3467H,2000PhRvL..85.3335K,2007PhRvL..99w1301B}) underproducing the amount of dark matter in the universe (and implying an additional dark matter component). 

This relic abundance (or thermal) cross section provides a very natural target for experimental sensitivity. In the same way that new particles are hoped to be seen at the LHC at the electroweak scale, we may reasonably expect to discover evidence for dark matter annihilation if we can probe annihilation cross sections of order $10^{-26} \, \cm^3 \second^{-1}$ (indeed there is a great mutual interest between these two searches). Conversely, a study that excludes the possibility of dark matter with a thermal cross section provides far-reaching constraints on particle physics beyond the Standard Model.

The highest flux of gamma-rays from dark matter annihilation likely comes from the Galactic Center (GC) as it is close-by and because the dark matter distribution at the GC may exhibit a high density cusp \cite{1999PhRvL..83.1719G,2001PhRvD..64d3504U,2002PhRvL..88s1301M,2004PhRvL..92t1304M,2014arXiv1406.4856F,2014PhRvD..89f3534L}.  In recent years, several groups using data from the  Large Area Telescope (LAT) onboard the Fermi Gamma-ray Space Telescope have presented evidence of a gamma-ray excess towards the GC that can be interpreted as coming from dark matter annihilation (with a cross section near the thermal one) \cite{2011PhLB..697..412H,2011PhRvD..84l3005H,2012PhRvD..86h3511A,2013PhRvD..87l9902A,2013PhRvD..88h3521G,2014arXiv1402.4090A,2014arXiv1402.6703D,2014arXiv1406.6948Z,2014arXiv1409.0042C}. However, the confounding difficulty of using the GC to detect a dark matter signal is the presence of bright astrophysical sources (point-like and diffuse) at the GC and near the plane of the Milky Way. A physical understanding of {\em all} of these backgrounds is required in order to rule them out as an explanation for the gamma-ray excess.

This essential problem of backgrounds calls for a search in quieter, cleaner environments where the detection of a gamma-ray excess would be much more compelling as a dark matter signal. This motivates the use of Milky Way dwarf galaxies as annihilation targets, an idea first suggested by  \cite{1990Natur.346...39L} and further explored in~\cite{2004PhRvD..70b3512B,2006PhRvD..73f3510B,  2007PhRvD..75b3513C,2006JCAP...03..003P,2007PhRvD..75h3526S,2010JCAP...01..031S, 2008ApJ...678..614S,2008ApJ...678..594W,2011APh....34..608H,2011PhRvL.107x1303G,2011PhRvL.107x1302A,2012APh....37...26M,2012PhRvD..86f3521B,2013arXiv1309.4780H,2013PhRvD..88h2002A,2013PhRvD..88h3535N,2013JCAP...03..018S,2013ApJ...773...61S,2014arXiv1409.1572C,2014JCAP...02..008A} (among others). These sources have very high mass-to-light ratios and are free of any known astrophysical sources of gamma-rays. They are also relatively nearby (at tens to hundreds of kiloparsecs).

Still, the annihilation signal from a typical dwarf galaxy is likely an order of magnitude or so smaller than from the GC (e.g.~\cite{2011MNRAS.418.1526C,0004-637X-801-2-74}). Therefore, it is necessary to perform a maximally-sensitive analysis of the gamma-ray data, taking into account everything known about the expected annihilation signal, backgrounds, and detector response.

This includes using data from all the dwarf galaxies simultaneously. At present, we have a quantitative determination of the dark matter distribution in twenty dwarf galaxies~\citep{0004-637X-801-2-74}. Analyzing two identical dwarfs together is equivalent to observing one for twice the time. Therefore, it might seem that Fermi's six year all-sky survey is equivalent to 120 years of high-quality observation of a dwarf galaxy! Such an observation would overcome the lower luminosity of the dwarfs relative to the Galactic Center. There is, unfortunately, a caveat: the Milky Way dwarfs are not all at the same distance from Earth and they do not have the same distribution of dark matter within them. A joint analysis should take this into account by, in some sense, weighting the different dwarfs according to their expected gamma-ray fluxes.

Alternatively, it would seem that Atmospheric Cherenkov Telescopes (ACTs) and other pointed instruments might as well spend their time on the single best target. However, in this case too there are reasons to allocate time to multiple sources: accessibility of a source depends on season and sky position; diversifying observations provides some control over systematics and hedges against the uncertainty in the dark matter distribution of the various dwarfs. Therefore, these experiments too will benefit from a joint analysis \cite{VERITASjointdwarfs}.

In this paper we present a weighting scheme that is optimal: it gives rise to a search that is more likely to discover an annihilation signal than any other (using the same data). The statistical framework extracts all the available information in the data by incorporating the impact of particle physics properties on the expected signal, the distribution of dark matter in Milky Way dwarf galaxies, the astrophysical and instrumental backgrounds, and the detector response. The method uses the framework of frequentist statistics and the probability distributions of test statistics under various hypotheses are constructed exactly, without making large sample size approximations or relying on background models. Therefore, obtained confidence intervals have proper coverage by construction and claims of detection are not biased by imperfect background modeling.

The framework is quite general and does not have anything to do with dark matter searches per se (though it was inspired by this problem) or even with gamma-ray analysis. However, the presentation will develop the statistical method in parallel with an application to the joint analysis of Milky Way dwarfs with the Fermi LAT. The dark matter analysis can be taken as a template for other studies using this method.

We begin in Sec.~\ref{sec:DMsignal} by reviewing the form of the expected dark matter signal in dwarf galaxies and then describe the preparation of the Fermi LAT gamma-ray data. In Sec.~\ref{sec:dwarfdata} we introduce the dwarf galaxies used in the analysis and summarize how their density profiles are determined.

Our statistical framework is developed in Sec.~\ref{sec:statisticalframework}. This section is more formal and pedagogical, defines the problem as one of frequentist hypothesis testing, and derives the ``most powerful'' form of the test statistic we use and how to compute its probability distribution.

We then return to the gamma-ray data and describe how the background is dealt with in Sec.~\ref{sec:background} (though this empirical technique can apply in other studies generally). In Sec.~\ref{sec:optimizing} we begin to apply the statistical technique to the dark matter search, showing how it allows us to optimize the data cuts and decide which dark matter hypotheses we need to test.

Before obtaining results from the observed data, Sec.~\ref{sec:summaryofprocedure} gives a step-by-step summary of the entire procedure for applying the statistical framework to the dark matter search. This section is written more ``algorithmically'' and can be read, in lieu of Sec.~\ref{sec:statisticalframework}, as instructions for practically applying the method.

Section~\ref{sec:results} presents the results of the search for annihilation in the twenty Milky Way dwarf galaxies and the resulting limits obtained on the annihilation cross section for various channels. Section~\ref{sec:discussion} contains a discussion of many aspects of the study including a comment on a potential positive detection, the Galactic Center signal, systematic uncertainties, and an exploration of the future sensitivity of Fermi to dark matter annihilation in dwarfs (e.g. over the mission lifetime and using the Pass 8 data reduction).

Appendices contain a derivation of the compound Poisson distribution that governs the test statistic, the FFT procedure used to numerically obtain it, and details of the numerical convolution of the dark matter halo $J$-profiles with the Fermi LAT point spread function.

%%%%%%%%%% DM flux from dwarfs
%%%%%%%%%%
\section{Expected dark matter signal}
\label{sec:DMsignal}

\subsection{Dark matter flux}
The observable signal of dark matter annihilation in dwarf galaxies is governed by the distribution of dark matter within the system as well as the particle physics of the interactions. The annihilation rate per volume is
\begin{equation}
r= \frac{1}{2} \sigv n^2,
\label{eqn:ratepervol}
\end{equation}
where $\sigv$ is the total annihilation cross section multiplied by the relative velocity of two particles and averaged over the dark matter velocity distribution, and $n$ is the number density of dark matter particles\footnote{Equation~\eqref{eqn:ratepervol} assumes that dark matter is its own antiparticle. If this is not the case, and the abundance of particle and antiparticle is the same, the factor of $\frac{1}{2}$ disappears but each factor of $n$ must be replaced by $\frac{1}{2}n$ (particle and antiparticle each constitute half the total dark matter density). The result is that the annihilation rate is half what it is when dark matter is its own antiparticle. Throughout this work, we will assume Eq.~\eqref{eqn:ratepervol} --- for the case of distinct particle and antiparticle one may replace $\sigv$ by $\frac{1}{2}\sigv$ everywhere (e.g. cross section limits increase by a factor of 2, but so does the thermal cross section).}  (see e.g. \cite{1990eaun.book.....K,1991NuPhB.360..145G,1988NuPhB.310..693S}).

Dark matter annihilation can proceed through multiple channels (e.g. annihilation into $b$ quark pairs or into $\tau$ leptons). Except for neutrinos, all Standard Model particles produced generically give rise to gamma-rays. The total number of gamma-rays produced per annihilation per energy interval is given by
\begin{equation}
\frac{dN_\gamma (E)}{dE} = \sum\limits_i B_i \frac{dN_{\gamma,i}(E)}{dE}. 
\label{eqn:dNdE}
\end{equation}
The branching ratio $B_i$ is the probability that an annihilation proceeds through channel $i$ and $dN_{\gamma,i}/dE$ is the number of gamma-rays produced per annihilation per gamma-ray energy by the products of channel $i$. This energy spectrum depends on the kinetic energy of the standard model particles produced in an annihilation. The kinetic energy in turn depends on the mass of the dark matter particle: for dark matter annihilating at rest, the available energy for the products is twice the mass of the particle. Conservation of momentum requires that for annihilation into a pair of particles of the same mass (e.g. particle-antiparticle pairs) each member of the pair has an energy equal to the dark matter particle mass.  In this work we adopt the annihilation spectra of~\citet{2011JCAP...03..051C} that were generated from fits to a suite of {\tt Pythia} \cite{2008CoPhC.178..852S} simulations, and include potentially important electroweak corrections (see discussion in \cite{2011JCAP...03..051C} as well as 
\cite{2007PhRvD..76f3516K,2008PhRvD..78h3540B,2009PhRvD..80l3533K,2010PhRvD..82d3512C,2011JCAP...03..019C}). 

When looking in a particular direction $\nhat$, the gamma-rays we detect at Earth are due to annihilation at all points $\ell \nhat$ along the line of sight ($\ell$ being distance). The gamma-ray flux (number of photons from the direction $\nhat$ per energy, solid angle, area, and time) is given by
\begin{equation}
\begin{aligned}
\frac{dF(E,\nhat)}{dE d\Omega}&= \int\limits d\ell \, \ell^2 \, r(\ell\nhat) \frac{dN_\gamma(E)}{dE}\frac{1}{4\pi \ell^2} \\
                                                     &= \frac{\sigv}{8\pi \mass^2} \frac{dN_\gamma(E)}{dE}  \int\limits d\ell \, \rho^2(\ell\nhat),
\label{eqn:DMflux}
\end{aligned}
\end{equation}
where we have used the relation $\rho = \mass n$ connecting the dark matter mass density $\rho$, number density $n$, and particle mass $\mass$. Equation~\eqref{eqn:DMflux} is valid for the flux from nearby, zero-redshift objects like Milky Way dwarf spheroidals, where there is no attenuation of the gamma-rays on their way to Earth. We write the flux using the mass density rather than the number density because observations of the motions of stars in dwarf galaxies are used to estimate $\rho$ (see Sec.~\ref{sec:dwarfdata}).

The expression for the expected flux separates into two simple factors. The first depends only on particle physics: the dark matter mass, annihilation cross section, branching ratios, and the Standard Model physics of gamma-ray production. The second term quantifies the distribution of dark matter in the astrophysical system. We refer to this term as the $J$-profile
\begin{equation}
\frac{dJ(\nhat)}{d\Omega} =  \int\limits d\ell \, \rho^2(\ell\nhat),
\label{eqn:Jdef}
\end{equation}
with units of $[\GeV^2 \cm^{-5} \mathrm{sr}^{-1}]$.

\subsection{Instrument response}

To find the number of events that will actually be detected we must convolve the dark matter flux Eq.~\eqref{eqn:DMflux} with the instrument response. The expected number of events detected by Fermi is given by
\begin{equation}
\frac{dN(E_r, \nhat_r)}{dE_r d\Omega_r} = \int\limits_E \int\limits_\Omega dE d\Omega \frac{dF(E,\nhat)}{dEd\Omega} R(E_r, \nhat_r | E, \nhat),
\end{equation}
where the subscript $r$ stands for reconstructed. The quantity $R(E_r, \nhat_r | E, \nhat) dE_r d\Omega_r$ is the instrument response: the probability that a photon with true energy $E$ and direction $\nhat$ will be reconstructed with an energy in the interval $dE_r$ around $E_r$ and in the solid angle $d\Omega_r$ around direction $\nhat_r$.

The instrument response is a complicated quantity that depends on the properties of the detector and the observation strategy. For a Fermi LAT observation of a fairly localized, steady source the instrument response can be broken up into three factors \cite{2012ApJS..203....4A}: exposure ($\epsilon$), point spread function ($\PSF$), and energy dispersion ($D$):
\begin{equation}
R(E_r, \nhat_r | E, \nhat) = \epsilon(E) \PSF(\nhat_r | E,\nhat) D(E_r | E),
\end{equation}
where the exposure $\epsilon$ has units of area $\times$ time and, for a localized source, depends only on energy.

Except for dark matter annihilation directly into two photons, the spectrum $dN_\gamma / dE$ is much broader than Fermi's energy dispersion. The exposure and point spread function, as well, have a slowly varying $E$-dependence over the width of the energy dispersion. Therefore, we neglect the energy dispersion (equivalent to assuming perfect energy reconstruction) unless considering annihilation into a photon final state.

Using Eqs.~\eqref{eqn:DMflux} and~\eqref{eqn:Jdef}, the expected number of detected dark matter events is (dropping the subscript $r$'s)
\begin{equation}
\frac{dN(E, \theta)}{dE d\Omega} = \frac{\sigv}{8\pi \mass^2} \frac{dN_\gamma(E)}{dE} [(J \ast \PSF)(E,\theta)] \epsilon(E),
\label{eqn:expectedcounts}
\end{equation}
where $J \ast \PSF$ is the 2-dimensional convolution of the $J$-profile with the $\PSF$, and $\theta$ is the angular separation between the center of the dwarf and the reconstructed direction of the event.

We consider spherically symmetric dark matter halos so that $dJ(\nhat)/d\Omega$ is a function only of the angular separation between $\nhat$ and the direction towards the center of the dwarf. The LAT's point spread function is also assumed to be circularly symmetric. Therefore, the 2-d convolution can be performed using one-dimensional Hankel transforms that we compute numerically (at each energy) using our implementation of Hamilton's efficient FFTLog algorithm~\cite{2000MNRAS.312..257H}. The details of this convolution are discussed in Appendix~\ref{app:JconvPSF}. The upshot is that the relevant ``observation''-space of the expected signal is two dimensional: a photon has an energy and an angular separation from the direction towards the dwarf.

When we consider annihilation into a photon final state Eq.~\eqref{eqn:expectedcounts} is modified by replacing $dN_\gamma/dE$ with the convolution of $dN_\gamma/dE$ with the energy dispersion $D$. We estimate the energy dispersion as a Gaussian with standard deviation 10\% of the true energy. This seems to be a reasonable approximation over the relevant energy range ($E > 10 \, \GeV$)~\cite[Fig.~69]{2012ApJS..203....4A}. The annihilation spectrum in this case is a delta function $dN_\gamma(E)/dE = 2\delta(E-\mass)$ centered on the dark matter mass. Convolution with the energy dispersion is simply a Gaussian, normalized to 2, with mean $M$ and standard deviation $0.1M$.

\subsection{Pass 7 Fermi-LAT data}
We use the publicly available data from the Fermi Science Support Center \cite{FSSC}, and select {\tt ULTRACLEAN } photons of {\tt evclass=4} with energies between [0.2-1000] GeV in the mission elapsed time interval of [239557417 - 423617437] seconds (August 4, 2008 to June 4, 2014). The photons are selected using the provided {\tt gtselect} tool within a Region of Interest of radius of 15 degrees centered on each dwarf and with a zenith angle cut set to {\tt zmax=100}. This data is processed following all standard recommendations and caveats \cite{Cicerone} regarding good time intervals using the  {\tt gtmktime} tool with the recommended filter of {\tt  DATA\_QUAL==1 \&\& LAT\_CONFIG==1 \&\& ABS(ROCK\_ANGLE)<52}. We generate a livetime cube using {\tt gtltcube} and compute exposure and point spread functions (PSFs) with {\tt gtpsf} by using the {\tt P7REP\_ULTRACLEAN\_V15} instrument response function.

%%%%%%%%% DWARFS 
\section{Dwarf galaxies}
\label{sec:dwarfdata}
The dominant systematic we face is the uncertainty in the dwarf $J$-profiles. The estimation of a dwarf's density profile is based on the positions and (spectroscopically obtained) line-of-sight velocities of its member stars~\citep{simon07,walker09a}. Statistically, these observational quantities respond to the gravitational potential of the system as described by the Jeans equation~\citep{2008gady.book.....B,2007PhRvD..75h3526S,2008ApJ...678..614S,2013pss5.book.1039W,2013NewAR..57...52B,2013PhR...531....1S}. Because the dwarfs are dark matter dominated, the gravitational potential is determined by the dark matter density profile.

In \citet{0004-637X-801-2-74} we presented a uniform analysis of the stellar kinematic data (projected positions and line-of-sight velocities) from the 20 Milky Way dwarfs for which such data are available.  Briefly, a likelihood function employs the Jeans equation to relate empirical distributions of position and velocity to a parametric, spherically-symmetric density profile of the form
\begin{equation}
\rho(r)=\rho_s [r/r_s]^{-\gamma}[1+(r/r_s)^{\alpha}]^{(\gamma-\beta)/\alpha}.
\label{eq:rhoprofile}
\end{equation}
We used the software package MultiNest~\citep{2008MNRAS.384..449F,2009MNRAS.398.1601F} to generate samples from the posterior PDFs of the five free parameters in Eq.~\eqref{eq:rhoprofile} as well as a sixth ``nuisance'' parameter that specifies the ratio of the velocity dispersions in radial and tangential directions.

The results of this analysis can be thought of as exploring the parameter space to find regions which give a reasonable fit to the available kinematic data. Due to the degeneracy between mass and velocity anisotropy, as well as the limited number of observed stars, the likelihood function is agnostic to very different types of halos so long as they fit a basic relationship between $\rho_s$ and $r_s$. For example, the analysis allows halos with very large values of $\rho_s$ coupled with small $r_s$. These halos correspond to density spikes at the centers of the dwarf galaxies. We are able to rule these out using a cosmological plausibility argument --- essentially requiring that the perturbation that formed the halo was not too rare. Additionally, the likelihood is unable to distinguish between different values of $r_s$ once $r_s$ is beyond the distance of the measured stars. This makes sense as the stars do not feel the potential far outside their current orbits. We adopt the most conservative choice (in terms of expected annihilation signal) by truncating the halos at the distance to the outermost member star (corresponding to an angle $\theta_{\rm max}$). This prevents the halos with unreasonably large $r_s$ values from inflating the integral in Eq.~\eqref{eqn:Jdef}. Finally, the likelihood function does not distinguish between cusped ($\gamma >0)$ and cored ($\gamma =0$) profiles. We do not apply any external judgement to this question and our sample halos reflect the large allowed range of inner slopes $0 <\gamma < 1.2$. See~\citep{0004-637X-801-2-74} for detailed explanations of the procedure.

The Jeans analysis should be thought of as generating realizations of halos which reasonably fit the stellar kinematic data. For the dark matter search this induces a systematic uncertainty. When we present the results of the search and limits on the annihilation cross section we will separate this systematic from the statistical uncertainty induced by our finite photon statistics. This is done by redoing the analysis separately for different realizations of halo parameters. The systematic ``band'' that results from this repetition should be thought of as reflecting our imperfect knowledge of the dwarf density profiles. As our knowledge of the dwarfs' internal structure improves, these bands will shrink.

Assumptions of the Jeans analysis include adopting dynamic equilibrium and spherical symmetry, a Plummer profile for the stellar distribution, and a constant velocity anisotropy parameter. It is important to note that these are approximations, which may not hold in the Milky Way dwarfs. We refer the reader to \citet{2015MNRAS.446.3002B}, who find that relaxing some of these assumptions has an effect for classical dwarfs with large kinematic samples. However, in the case of ultrafaint dwarfs, uncertainties in density profiles are dominated by errors due to limited numbers of observed stars, not by modeling assumptions. In our results, it is the uncertainty in the ultrafaint dwarf profiles that dominates the systematic uncertainty in the cross section limits.

\begin{figure*}
\includegraphics[scale=0.7]{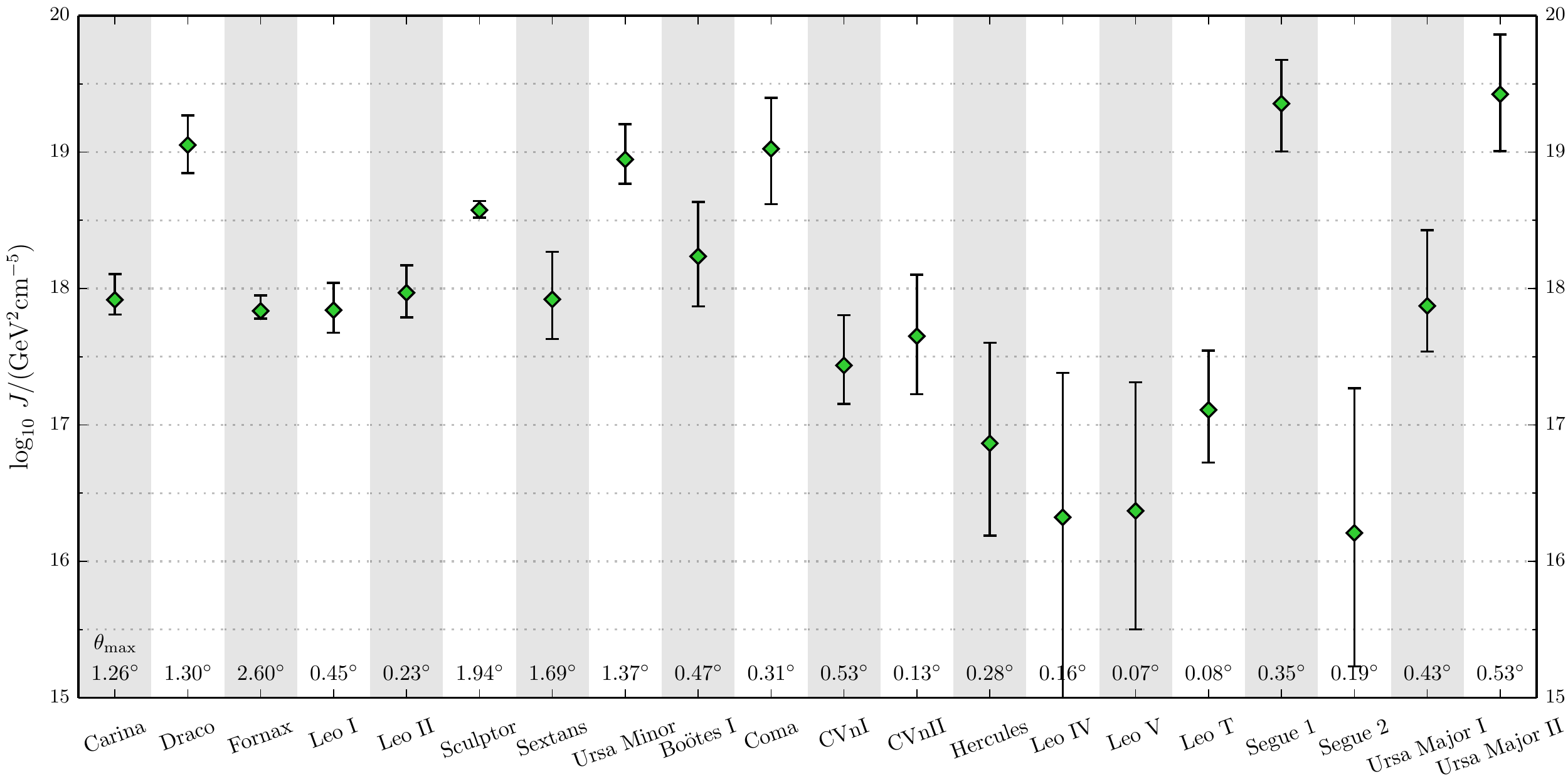}
\caption{Annihilation $J$-profiles  for all dwarf galaxies used in this analysis. The error bars show the $1\sigma$ allowed range in the value of $J(\theta_{\rm max})$ based on the analysis of stellar velocities described in \citet{0004-637X-801-2-74}.
\label{fig:Jtotall}}
\end{figure*}

Integrating the $J$-profile over a solid angle of radius $\theta$ yields the ``$J$ value'' $J(\theta)$. This is a measure of the amplitude of the dark matter annihilation flux from a dwarf. Figure~\ref{fig:Jtotall} is reproduced from~\cite{0004-637X-801-2-74} and shows the $J(\theta_{\rm max})$ values of the 20 dwarfs used in this analysis.

In this work we perform a joint analysis of the 20 dwarfs whose $J$-profiles were determined in~\citep{0004-637X-801-2-74}: Bo\"otes I, Canes Venatici I, Canes Venatici II, Carina, Coma Berenices, Draco, Fornax, Hercules, Leo I, Leo II, Leo IV, Leo V, Leo T, Sculptor, Segue 1, Segue 2, Sextans, Ursa Major I, Ursa Major II, and Ursa Minor. When we perform the search for annihilation in individual dwarfs we also include the nearby satellite Willman 1. However, we cannot reliably determine the density profile of this object as it shows strong evidence for tidal disruption and/or non-equilibrium kinematics~\citep{2011AJ....142..128W}, violating an assumption underlying the Jeans equation. In our framework, combining data from different dwarfs requires knowledge of the $J$-profiles and so we do not include Willman 1 in the joint analysis.

\section{Statistical Framework}
\label{sec:statisticalframework}

In the frequentist paradigm we interrogate the data through the framework of hypothesis testing. For example, to find out whether observations of a dwarf show evidence for dark matter annihilation we may start by testing the hypothesis $\Hb$: the observed data $D$ were generated from background processes only. We find a way to calculate the probability $\prob(D|\Hb)$ of observing $D$ if $\Hb$ were true. If this probability is small, say $\prob(D|\Hb) = 0.01$, the hypothesis is ``rejected at 99\% significance'' --- i.e. it is very unlikely to have measured the data we did if there were no dark matter annihilation.

Confidence intervals on dark matter model parameters can be generated by performing an ensemble of hypothesis tests. For simplicity, imagine that dark matter annihilation is governed by two parameters, the particle mass $\mass$ and the velocity-averaged annihilation cross section $\sigv$. For every possible pair of values of these parameters we perform the hypothesis test ``dark matter has mass $\mass$ and annihilation cross section $\sigv$.'' We classify a point in parameter space by whether its associated hypothesis is rejected at a given level $\alpha$ (e.g. $\alpha= 0.05$ for a 95\% confidence region). That is, we divide the parameter space into allowed regions where $\prob(D|\mass, \sigv) > \alpha$ and excluded regions where $\prob(D|\mass, \sigv) < \alpha$. The allowed region constitutes an $\alpha$-level confidence region for mass and cross section. The interpretation of the two regions is straightforward (e.g. for $\alpha=0.05$): whatever the true values of $\mass$ and $\sigv$ are, there is only a 5\% chance that the hypothesis associated with those true values will be rejected. Equivalently, there is a 95\% chance that the constructed confidence region contains the true parameters.

The implementation of this scheme is made possible by the construction of a test statistic $T$, a single number that is a function of the data we measure. The test statistic is a random variable and when we make a measurement we sample this variable. For a given hypothesis, a probability distribution function (PDF) governs the measurement of $T$, with the observed value denoted $\Tobs$. Before making the measurement, we decide on a critical region $\crit$ of $T$-space such that $\prob(T\in\crit|H) = \alpha$. Should $\Tobs$ be measured to lie in the critical region we reject the hypothesis $H$ at level $\alpha$.

The use of a test statistic allows us to make precise the ``probability of observing the data given a hypothesis''. For this purpose it is useful to choose a test statistic that reflects how ``signal-like'' or ``background-like'' the data are, with larger values of $T$ indicating the presence of a signal (e.g. dark matter annihilation). For instance, when testing the hypothesis $\Hb$ that there is no dark matter annihilation we may identify a special value $T^*$ and define the critical region as $\crit: T > T^*$, where $\prob(T>T^*|\Hb)=0.01$ (i.e. $\alpha=$ 0.01). The interpretation of $\crit$ is that there is only a 1\% chance of the data being so ``signal-like'' if there were no dark matter annihilation. If the measured $\Tobs$ is larger than $T^*$ the hypothesis $\Hb$ is rejected at 99\% significance.

Constraints on the particle physics parameters take the form of upper limits on the annihilation cross section. Upper limits on $\sigv$ are generated by choosing the critical region to be $\crit: T < T^*$, where $\prob(T < T^* | \mass, \sigv) = \alpha$. We will reject the hypothesis that dark matter has a particular mass $\mass$ and cross section $\sigv$ if $\Tobs$ is found to be smaller than $T^*$ (i.e. the measurement is too background-like). This choice of critical region for $T$ (i.e. $T<T^*$ as opposed to $T>T^*$) generates upper limits on the cross section: for large cross sections $T^*$ will increase since the data is likely to be more ``signal-like''. For sufficiently large cross sections the associated hypothesis will be always be rejected, leading to upper limits on $\sigv$.

\subsection{General form of the test statistic}
\label{sec:generalformofT}

In principle, $T$ can be an arbitrary function of the data. However, some functions are better, in a well-defined sense, than others. Here we detail the construction of an optimal test statistic.

The gamma-ray data are in the form of a list of discrete detector events. We wish to jointly analyze the gamma-ray signal from multiple targets simultaneously and to take full advantage of the information contained in the data. Each event is assigned a numerical weight $w(Q)$ based on its properties $Q$ and the hypothesis we are testing. We use a test statistic that is simply the sum of the weights of all the events in the entire data set
\begin{equation}
T = \sum\limits_{i=1}^N w(Q_i),
\label{eqn:Tdef}
\end{equation}
where $i$ runs over all detected events. The total number of events $N$ and the collection $\{Q_i\}$ are random variables.

For the data set we are working with, the dark matter physics is encoded in three properties of each detected event: which dwarf field the event came from $\nu$, the reconstructed energy of the photon $E$, and the reconstructed direction of the photon $\theta$ (i.e. the angular separation between the event and the direction toward the dwarf galaxy). Therefore, in our study $Q=(\nu, E, \theta)$ is the set of these three variables, the first being discrete and the latter two continuous. 

This general form for the test statistic is capable of reproducing many other analyses by making particular choices for the weight function. For example, a standard event counting analysis can be performed by setting $w(Q)=1$ for events in some energy range and within some angular separation of one of the dwarfs, and setting $w(Q)=0$ for all other events. In this case the test statistic $T$ just counts the number of events detected. As a second example, the analysis performed in \cite{2011PhRvL.107x1303G} is recovered by having $w(Q)$ be a function only of which dwarf field the event came from (and not of the energy or angular separation of the event). The test statistic then becomes a simple weighted sum of counts observed from each dwarf.

\subsection{Designing the weight function}
\label{sec:designingweight}
Given this general form of test statistic the important work lies in designing the weight function. Here we show that there is a statistically most powerful choice of weight function.

Recall that $\alpha$ denotes the probability of rejecting the hypothesis when the hypothesis is true. The power of a statistical test is the probability of rejecting the hypothesis when the hypothesis is false (i.e. when it {\em ought} to be rejected). Therefore, we seek a test statistic that maximizes the power for a given $\alpha$. The power of a test is an ambiguous concept because it depends on what the truth actually is. That is, a test that is powerful at rejecting $\HO$ when $\HA$ is true may not be powerful at rejecting $\HO$ when $H_2$ is true \cite[\S 21.16-18]{Kendall5th}. We therefore restrict our task to finding the test statistic that maximizes the power of rejecting $\HO$ for a single suitable alternative hypothesis $\HA$.

As discussed above, for constructing limits we test hypotheses of the form ``the dark matter particle has mass $\mass$ and annihilation cross section $\sigv$''. For these cases we take the alternative hypothesis to be $\Hb$, the background-only hypothesis of no dark matter annihilation. This gives the most constraining upper limits on $\sigv$ if dark matter has an annihilation cross section too low for the instrument to detect. When performing a search for annihilation we ask whether we can reject the hypothesis $\Hb$. In this case the test statistic is chosen to maximize the power versus an alternative hypothesis that dark matter has a particular mass and a small annihilation cross section. That is, the test is designed to be sensitive to weak signals. The choice of particle parameters besides the cross section will be dealt with using a ``trials factor'': we may test $\Hb$ against hypotheses corresponding to several different masses and branching ratios.

Below we present two approaches for constructing a most-powerful weight function $w(Q)$. The first is heuristic and more intuitive, the second more rigorous. Both yield similar conclusions.

It will be useful to write the test statistic Eq.~\eqref{eqn:Tdef} in an alternate form by introducing a new set of random variables that are easier to work with. The random variable $Z_Q$ is the number of events that are detected having properties in an infinitesimal bin centered on $Q$. Using the set $Q=(\nu,E,\theta)$ described above, $Z_Q$ is the number of events from dwarf $\nu$ that have energy between $E$ and $E+dE$ and were detected between $\theta$ and $\theta+d\theta$ from the location of the dwarf. The sizes of these bins are infinitesimal so that $Z_Q$ is almost always 0 and is occasionally 1. Making a measurement is equivalent to measuring the infinite collection of $Z_Q$ for all possible $Q$: $D = \{Z_Q\}$ (for a finite set of $Q$, $Z_Q$ will be 1; for the rest $Z_Q$ will be 0). The weight of an event with properties $Q$ is denoted $w_Q$. The test statistic can be written in terms of the variables $Z_Q$ as
\begin{equation}
T = \sum\limits_Q w_Q Z_Q,
\label{eqn:Tdiscrete}
\end{equation}
where the sum is over all possible properties of a detected event. In our case the notation $\sum_Q$ is shorthand for $\sum_\nu \int_E \int_\theta$. The test statistic is determined by the infinite collection of random variables $\{Z_Q\}$ and the infinite collection of numerical weights $\{w_Q\}$. Defining a weight function $w(Q)$ is equivalent to fixing values for each of the $w_Q$.

In our situation it is useful to write each $Z_Q$ as the sum
\begin{equation}
Z_Q = X_Q + Y_Q,
\label{eqn:ZequalsXplusY}
\end{equation}
where $X_Q$ is the number of events detected with properties $Q$ that originated from dark matter annihilations in a dwarf galaxy (signal events) and $Y_Q$ is the number of detected events originating from all other sources (background events). The collection $\{X_Q\}$ are independent random variables and are also independent of all of the $\{Y_Q\}$. The probability distribution for $X_Q$ is
\begin{equation}
\prob(X_Q) = \begin{cases}
                        1-s_Q & \text{for $X_Q= 0$},\\
                        s_Q &  \text{for $X_Q=1$},                        
                        \end{cases}
\label{eqn:Xprob}
\end{equation}
where $s_Q$ is the (infinitesimal) expected number of detected dark matter events having properties $Q$. For the dark matter search $s_Q$ will be given by Eq.~\eqref{eqn:expectedcounts} multiplied by the infinitesimal element $dEd\Omega$, i.e. it is the expected number of dark matter annihilation events detected from dwarf $\nu$ with energy between $E$ and $E+dE$ and angular separation between $\theta$ and $\theta +d\theta$.

The probability distribution describing $Y_Q$ may not be as simple because different $Y_Q$ may be correlated (e.g. if the background has a contribution from unresolved sources). In deriving an optimal choice of weights we will make the assumption that the $Y_Q$ are independent and each is described by
\begin{equation}
\prob(Y_Q) = \begin{cases}
                        1-b_Q & \text{for $Y_Q= 0$},\\
                        b_Q &  \text{for $Y_Q=1$},                        
                        \end{cases}
\label{eqn:Yprob}
\end{equation}
with $b_Q$ the expected number of background events having properties $Q$. Because of this assumption of independence, the choice of weights may not be strictly the most powerful but we expect the deviations from optimality to be minimal. However, it is important to note that the calculation of the PDF of $T$ will not use this simplifying assumption and will correctly incorporate any correlations present in the background.

\subsubsection{Signal-to-noise method}
To construct confidence regions we test the hypothesis that dark matter is present and has a particular set of particle physics parameters. This test is to be most powerful against the alternative that the data are generated by background processes only. The two hypotheses are referred to as $\Hsb$ and $\Hb$.

The problem of maximizing the power of $\Hsb$ versus $\Hb$ can be visualized as trying to maximally separate the PDFs of $T$ for the two hypotheses (see Fig.~\ref{fig:sigvforsearch} as an illustration). The specific shapes of the PDFs are controlled by the weight function $w(Q)$. An approximate way of describing the PDFs is by their means and standard deviations: $\mu_{s+b}$, $\mu_b$, $\sigma_{s+b}$, and $\sigma_b$. The ``separation'' of the two PDFs can be quantified by a signal-to-noise ratio:
\begin{equation}
{\rm SNR} = \frac{\mu_{s+b} - \mu_b}{\sigma_b}
\label{eqn:SNRdef}
\end{equation}

We will write the quantities in the above equation in terms of the weights $w_Q$ and find the collection of $w_Q$ that maximizes the signal-to-noise ratio. Using Eqs.~\eqref{eqn:Tdiscrete}, \eqref{eqn:ZequalsXplusY}, \eqref{eqn:Xprob}, \eqref{eqn:Yprob}, the independence of the $\{X_Q\}$ and $\{Y_Q\}$, and the fact that  all $s_Q$=0 if $\Hb$ is true, it is straightforward to show that
\begin{equation*}
	\begin{aligned}
	\mu_{b} \equiv \mathrm{E}\left[ T| \Hb \right] &=  \sum\limits_Q w_Q b_Q,\\
	\mu_{s+b} \equiv \mathrm{E}\left[ T| \Hsb \right] &= \sum\limits_Q w_Q (s_Q+b_Q),\\
	\sigma_{b}^2 \equiv  \mathrm{Var}\left[ T| \Hb \right] &=  \sum\limits_Q w_Q^2 b_Q, % \\
	%\sigma_{s+b}^2 \equiv \mathrm{Var}\left[ T| \Hsb \right] &=& \sum\limits_Q w_Q^2 (s_Q+b_Q) .
	\end{aligned}
\end{equation*}
where $\mathrm{E}[T|H]$ and $\mathrm{Var}[T|H]$ are the mean and variance of $T$ under the hypothesis $H$. Inserting these results into Eq.~\eqref{eqn:SNRdef} yields
\begin{equation}
{\rm SNR} = \frac{ \sum\limits w_Q s_Q}{ \sqrt{ \sum\limits w_Q^2 b_Q }}.
\label{eqn:SNR}
\end{equation}
We find the weights that maximize this quantity by differentiating it with respect to an arbitrary weight $w_R$ and setting the derivative to zero. This leads to the following condition that holds for each set of properties $R$:
\begin{equation*}
w_R \frac{b_R}{s_R} = \frac{\sum\limits w_Q^2 b_Q}{ \sum\limits w_Q s_Q}.
\end{equation*}
The solution to this set of equations is
\begin{equation}
w_Q = \frac{s_Q}{b_Q}.
\label{eqn:wsnr1}
\end{equation}
Had we used $\sigma_{s+b}$ instead of $\sigma_b$ in the definition of ${\rm SNR}$ (Eq.~\eqref{eqn:SNRdef}) the resulting optimal weights would be
\begin{equation}
w_Q = \frac{s_Q}{s_Q+b_Q}.
\label{eqn:wsnr2}
\end{equation}
Note that even though $s_Q$ and $b_Q$ are each infinitesimal their ratio is finite.

This argument tells us that each event should be given a weight determined by the ratio of the expected signal to the expected background for events of that type. This makes intuitive sense: events which are more likely to be signal are given a larger weight than those likely to be due to background processes. One consequence of this weighting, applied to the dark matter search, is that events which have an energy larger than the mass of the dark matter particle we are considering in $\Hsb$ will be ignored (given a weight of 0) because they must be due to background.

Inserting the weights in Eq.~\eqref{eqn:wsnr1} back into Eq.~\eqref{eqn:SNR} gives a useful interpretation for the test statistic's numerical value:
\begin{equation}
{\rm SNR}^2 = \sum\limits w_Q s_Q = \mathrm{E}[T|H_s].
\label{eqn:SNR2}
\end{equation}
The expected value of the test statistic due to signal events (e.g. dark matter annihilation) is simply the square of the expected signal to noise ratio of detection. This relation provides a computationally trivial method for approximating the expected results of experiments. For a given set of parameters describing the source of interest (e.g. dark matter mass and annihilation cross section), one can quickly see at what level an optimally sensitive search can distinguish signal from background. The sum $\sum w_Q s_Q$ in Eq.~\eqref{eqn:SNR2} incorporates information about the signal, background, and detector properties.

\subsubsection{Likelihood ratio method}
\label{sec:likelihoodratiomethod}
An alternative derivation of the optimal weights is based on a theorem from classical statistics. The Neyman-Pearson lemma \cite[\S 21.10]{NeymanPearson1933,Kendall5th} states that the most powerful test between two simple hypotheses, such as $\Hsb$ and $\Hb$, can be performed by using the likelihood ratio as the test static. The likelihood $\prob(D|H)$ is the probability of observing the data $D$ if the hypothesis $H$ were true. In our case, to test the hypothesis $\Hsb$ against the alternative $\Hb$ we use the likelihood ratio
\begin{equation}
\Lambda = \frac{\prob(D|\Hsb)}{\prob(D|\Hb)}
\label{eqn:LR}
\end{equation}
and reject the hypothesis $\Hsb$ if $\Lambda$ is found to be smaller than a critical value $\Lambda^*$. This critical value is determined by $\alpha$, the desired level of the test: $\prob(\Lambda < \Lambda^*|\Hsb) = \alpha$.

In the case under consideration (independent $\{X_Q\}$ and $\{Y_Q\}$) it is easy to write down the likelihoods under the two hypotheses. Let $\{Q_i \mid i=1 \dots N\}$ denote the properties $Q$ of the $N$ observed events. That is, $Z_Q$ was found to be 0 for all but the finite set $\{Q_i\}$ for which $Z_Q=1$. The probability of measuring this collection of $Z_Q$ under the two hypotheses is
\begin{equation}
\begin{aligned}
\prob(\{Z_Q\}| \Hb) &= \prod\limits_Q (1 - b_Q) \prod\limits_{i} b_{Q_i}, \\
\prob(\{Z_Q\}| \Hsb) &= \prod\limits_Q (1 - s_Q - b_Q) \prod\limits_{i} (s_{Q_i} + b_{Q_i}). 
\label{eqn:likelihoods}
\end{aligned}
\end{equation}
In these equations, the first product contains the infinite set of all $Q$ except for the finite set $\{Q_i\}$ while the second product only contains $N$ factors corresponding to the $\{Q_i\}$. In the limit that the binning of event space becomes infinitesimal, $b_Q$ and $s_Q$ approach zero and it makes no difference whether the first product omits a finite collection of $Q$. One can also show that in this limit the infinite products converge exactly to exponentials:
\begin{equation}
	\begin{aligned}
	\prod\limits_Q (1 - b_Q) &\rightarrow \exp \left( - \sum\limits_Q b_Q \right), \\
	\prod\limits_Q (1 - s_Q - b_Q) &\rightarrow \exp \left( - \sum\limits_Q (s_Q + b_Q) \right).
	\label{eqn:infprod}
	\end{aligned}
\end{equation}

Using Eqs.~\eqref{eqn:likelihoods} and \eqref{eqn:infprod}, the likelihood ratio Eq.~\eqref{eqn:LR} is given by
\begin{equation*}
\Lambda = \exp \left( - \sum\limits_Q s_Q \right) \prod\limits_{i} \left(1 + \frac{s_{Q_i}}{ b_{Q_i}} \right).
\end{equation*}
It makes no difference if we use $\log \Lambda$ as the test statistic since the logarithm is a monotonic function:
\begin{equation*}
\log \Lambda = - \sum\limits_Q s_Q + \sum\limits_{i=1}^N \log \left(1 + \frac{s_{Q_i}}{ b_{Q_i}} \right) .
\end{equation*}
The first term is a constant that does not depend on the data and can be ignored. This leaves us with a test statistic that is most powerful at distinguishing $\Hsb$ from $\Hb$:
\begin{equation}
T = \sum\limits_{i=1}^N \log \left(1 + \frac{s_{Q_i}}{ b_{Q_i}} \right),
\label{eqn:Tloglike}
\end{equation}
where $\Hsb$ should be rejected if $T$ is below $T^*$, specified by the condition $\prob(T<T^*|\Hsb) = \alpha$. Comparing Eq.~\eqref{eqn:Tloglike} with Eq.~\eqref{eqn:Tdef} we see that if we set the weight function to be
\begin{equation}
w(Q) = \log \left(1 + \frac{s_{Q}}{ b_{Q}} \right)
\label{eqn:wloglike}
\end{equation}
the test statistic Eq.~\eqref{eqn:Tdef} is equivalent to a likelihood ratio test statistic. Therefore, using the weight function Eq.~\eqref{eqn:wloglike} gives rise to the most powerful test statistic. This will be the test statistic we use throughout this work. Note that if we are testing the hypothesis $\Hb$ and want the test to be optimally sensitive to $\Hsb$ we can use precisely the same weight function as in Eq.~\eqref{eqn:wloglike}. The only difference is that $\Hb$ will be rejected when $T$ is {\em larger} than $T^*$, where $T^*$ is determined by $\prob(T>T^*|\Hb) = \alpha$.

It is interesting to observe that the log-weighting in Eq.~\eqref{eqn:wloglike} is, in some sense, a compromise between the two weighting schemes presented in Eqs.~\eqref{eqn:wsnr1} and~\eqref{eqn:wsnr2}. Considered as functions of $x\equiv s_Q/b_Q$, we see that $x/(1+x) \le \log(1+x) \le x$ for all physical values of $x$ (non-negative $s_Q$ and $b_Q$). When we are considering a very weak signal ($s_Q \ll b_Q$) all three become equivalent to Eq.~\eqref{eqn:wsnr1}. In this case the test statistic is actually independent of the annihilation cross section since $\sigv$ enters as a multiplicative factor in $s_Q$ (Eq.~\eqref{eqn:expectedcounts}) and two test statistics are equivalent if they differ by a constant factor. This implies that when searching for the presence of a small signal (i.e. testing the background-only hypothesis) the test statistic is optimal against alternatives $\Hsb$ with any (small) cross section (keeping other model parameters fixed). The issue of sensitivity to weak signals is discussed again at the end of Sec.~\ref{sec:defsigv90}.

\subsection{Probability distribution of the test statistic}
\label{sec:pdfofT}

Here we derive the PDF of the test statistic defined by Eq.~\eqref{eqn:Tdef} for any choice of weight function $w(Q)$. First note that $T$ is the sum of two terms
\begin{equation}
T = T_s + T_b,
\label{eqn:Tdefsb}
\end{equation}
where $T_s$ is the total weight of all detected photons originating from dark matter annihilation in dwarfs (signal) and $T_b$ is the total weight of all other detected events (background). The signal events and background events are statistically independent of one another. Therefore, the PDF of $T$ is the convolution of the PDFs of $T_s$ and $T_b$. In this study, determination of the PDF of $T_b$ will be done by empirically sampling a background region as described in Sec.~\ref{sec:background}. The background may also be treated in the same way as $T_s$, which we now proceed to describe.

To find the PDF of $T_s$ note that the number of detected signal events $N_s$ is a random variable distributed according to a Poisson distribution. The weights of the detected signal events $\{ w(Q_i) \mid i=1\dots N_s\}$ are independent and identically distributed random variables. Therefore, the random variable $T_s$ is the sum of independent variables where the number of terms in the sum is itself a Poisson random variable. Such a quantity is distributed according to a compound Poisson distribution (e.g. \cite{adelson1966,Embrechts:2009ly}).

This compound Poisson distribution is determined by two quantities. The first is the mean $\mu$ of the Poisson distribution determining the total number of signal events observed. In terms of the definitions given in Eq.~\eqref{eqn:Xprob} we have $\mu = \sum_Q s_Q$. The second input is the single-event weight distribution $f(w)$. Specifically, $f(w)dw$ is the probability that a detected signal event has properties $Q$ that cause it to be given a weight $w(Q)$ between $w$ and $w+dw$. It is completely determined from the collection $s_Q$ once the weight function has been chosen. To compute $f(w)$ we divide the $Q$-space into small tiles (i.e. for each dwarf we divide the $E$--$\theta$ plane into small bins) and find the weight $w_Q$ and the probability $s_Q$ in each tile. The weights $w_Q$ are binned into a histogram where each $w_Q$ adds a probability of $s_Q$ to the histogram. The histogram is then normalized to 1, yielding $f(w)$. This procedure can be made arbitrarily accurate by dividing the $Q$-space into small enough bins. In practice, we choose the bins to be small enough so that our results do not depend on the binning.

The Fourier transform $\phi_T(k)$ of the PDF of a compound Poisson distribution takes a simple form (see Appendix~\ref{app:pdfs}):
\begin{equation}
\phi_T(k) = \exp \left[ \mu \left( \phi_W(k) -1 \right) \right],
\label{eqn:CPft}
\end{equation}
where $\phi_W(k)$ is the Fourier transform of the single-event weight distribution $f(w)$. Numerically, this function can be computed quickly using fast Fourier transforms (FFTs). Working in Fourier-space also makes convolutions efficient --- one simply multiplies the Fourier transforms. Finally, an inverse FFT is used to obtain the PDF. Details of the numerical computation of the PDF of $T$ as well as the derivation of Eq.~\eqref{eqn:CPft} are given in Appendix~\ref{app:pdfs}.

%%%%%%%%%%% BACKGROUND
\section{Diffuse background}
\label{sec:background}

Knowledge of the diffuse background enters the analysis in two places. First, all of the weighting schemes we consider (e.g. Eq.~\ref{eqn:wloglike}) require $b_Q$, the expected number of background events with properties $Q$. Second, no matter the weighting scheme, the PDF of the test statistic due to background $T_b$ must be found in order to construct the PDF for the observed test statistic (see Sec.~\ref{sec:pdfofT}).

\subsection{Background model for weighting}
\label{sec:bgmodelforweighting}

To construct $b_Q$ used in the weighting function we make the assumption that the background process is isotropic over the small region surrounding each dwarf ($\theta <0.5^\circ$). Therefore, the background is simply defined by an energy spectrum (different for each dwarf). We estimate the spectrum by using the detected events in a region of radius $10\degr$ centered on the dwarf. Such a region is small enough that changes in the instrument exposure across the field of view will not affect the inferred spectrum. Events within $0.8\degr$ of sources from the Fermi LAT second source catalog~\citep{2012ApJS..199...31N} are masked. The size of the mask is based on the LAT point spread function at the energies we consider ($E > 1\, \GeV$), though for some very bright sources these masks are enlarged to prevent contamination (as a check we doubled all mask sizes and found no effect on any of our results).

The energy spectrum of these background events is modelled as a piecewise function $f(E)$. For energies below $10\, \GeV$ we replace each event with a Gaussian of width 20\% of the measured energy, giving a kernel density estimate. Above $10\, \GeV$ we splice on a power law with exponential cutoff. The form is $f(E) = f_0 (E/E_0)^\gamma \exp[(E-E_0)/E_c]$, where $E_0=10\,\GeV$ and $f_0$ is the kernel density estimate of the spectrum at $10\,\GeV$. We choose this smooth fitting function to avoid noise in the kernel density estimator due to the relatively low number of observed events with high energies. It is nonetheless flexible enough to model the shape of the observed background spectrum. The energy spectrum is divided by the solid angle of the background region to give an expected number of background events per energy interval per solid angle.

We note that we try only to model the shape of the {\em observed} background and do not seek to understand the sources of background or even estimate its intrinsic flux. Thus, the background spectrum we derive already includes the effects of the energy-dependent effective area\footnote{It would be straightforward to include knowledge of the exposure across the field of view to estimate the background spectrum in the central $\theta <0.5^\circ$ region. This would be necessary if the exposure varied strongly across the field of view.}.

The function $b_Q$, the expected number of background events with energy between $E$ and $E+dE$ and angular separation between $\theta$ and $\theta+d\theta$, is simply
\begin{equation}
b_Q = f(E) \, dE \, 2\pi \sin(\theta) d\theta,
\label{eqn:bQestimate}
\end{equation}
where $f(E)$ is the estimate of the background energy spectrum per solid angle.

\subsection{Test statistic due to background}
\label{sec:bgsampling}

There are two ways to derive the PDF of the test statistic due to background events (Sec.~\ref{sec:pdfofT}). The simplest is to assume that Eq.~\eqref{eqn:bQestimate} fairly describes the background processes. This requires that individual background events are independent of one another, are described by the energy spectrum $f(E)$, and that the background is isotropic within the central Region of Interest (ROI) centered on the dwarf. The probability distribution of $T_b$ of Eq.~\eqref{eqn:Tdefsb} is then a compound Poisson distribution, completely analogous to $T_s$.

However, there is good reason to believe that the processes which give rise to the actual background violate all of these assumptions. For example, the presence of unresolved sources gives rise to a non-Poisson counts distribution. It will also induce correlations between number of counts and the energy distribution of the events. The function $f(E)$, though a good fit to the average energy distribution, cannot describe the variation in energy spectrum caused by unresolved sources.

Therefore, following \citet{2011PhRvL.107x1303G,2012PhRvD..86b1302G}, we perform a sampling of the background region to {\em empirically construct} the distribution of $T_b$. This procedure works no matter what weighting function is chosen. We place trial regions of interest (ROIs) throughout the background region. Specifically, the trial ROIs are overlapping circles of radius $0.5\degr$ with centers spaced by approximately $0.25\degr$ covering the field of view out to $10\degr$. Trial ROIs that overlap with the central ROI, any of the source catalog masks, or the boundary of the $10\degr$ region are discarded.

The events in each trial ROI are weighted accorded to the chosen weighting scheme and then summed, giving a total background weight for each trial ROI. The probability distribution of these weights over all the trial ROIs is an empirical determination of the PDF of $T_b$  due to background {\em in the central ROI}. The fundamental assumption made is that whatever processes give rise to the background nearby each dwarf galaxy are at work in the direction towards the dwarf.

This procedure is robust to the choice of $b_Q$ (e.g. Eq.~\ref{eqn:bQestimate}). No matter what weighting function is chosen, the sampling will give the correct distribution of $T_b$ for that weighting. If the form of $b_Q$ does not reflect the true background the test statistic will not be optimal (see Sec.~\ref{sec:designingweight}) but confidence limits will still have proper coverage and detection significances will have the correct probabilistic interpretation.

It is important to note that this procedure is not based on any physical understanding of what generates the background --- it seeks only to describe the distribution of a single observed quantity $T_b$. It therefore includes any correlations caused by unresolved sources and detector effects such as misreconstructed cosmic rays. The procedure does not directly require any knowledge of the instrument response of the detector (effective area, point spread function), though these weakly enter in the choice of the size of the field of view and the size of the masks on known point sources. Incidentally, the empirical background distribution can be used in a ``$P(D)$ analysis'' to understand the physical sources of background including dark matter annihilation (see e.g.~\cite{2009JCAP...07..007L,2009PhRvD..80h3504D,2010PhRvD..82l3511B}). We discuss the effects of assuming a compound Poisson origin of $T_b$ in Sec.~\ref{sec:commentonexcess}.

%%%%%%%%%%%
\section{Designing searches and limits}
\label{sec:optimizing}

\subsection{Hypothesis tests}

The search for a dark matter signal and placing limits on the mass and cross section are done with separate hypothesis tests. First, we perform the search by asking whether the observed data is consistent with the hypothesis $\Hb$ that there is no annihilation in the dwarf galaxies (the alternative is $\Hsb$, that dark matter has a particular mass, cross section, and branching ratios).

If $\Hb$ cannot be rejected we construct limits by testing the ensemble of hypotheses $\Hsb$ to find which dark matter properties are ruled out (i.e. what collection of masses and cross sections are rejected at, say, 95\% confidence). In both cases the optimal test statistic is determined by two hypotheses under consideration, $\Hsb$ and $\Hb$, one being the null hypothesis and the other the alternative.

\subsection{Expected results}

A powerful benefit of being able to find the PDF of $T$ for any hypothesis is the straightforward computation of expected results. Computing the PDF of the test statistic is equivalent to simulating the results of the observations under a particular hypothesis. Instead of simulating large numbers of realizations of the raw data (e.g. collections of signal and background events) we can compute, exactly, the probability distribution of the test statistic that would have been derived from the raw data. Therefore, without ``uncovering'' the actual photon data we can easily predict how our methods are likely (in a well-defined sense) to perform. 

For example, suppose we wish to predict how strong our upper limits will be if there was no dark matter annihilation in the dwarf galaxies. We can perform the usual hypothesis test of $\Hsb$. However, instead of using the actual observed data to compute $\Tobs$ we can {\em assume} that the observed test statistic will just be sampled from the background-only PDF of $T$. That is, we compute $\prob(T \mid \Hb)$ and take $\Tobs$ to be some quantile of this distribution. A central estimate of the expected limit can be found by taking $\Tobs$ to be the median of the background-only distribution. To find the statistical uncertainty in the limit we can compute limits when $\Tobs$ is at the 16th and 84th percentiles of the background distribution. This gives a range where the upper limit is likely to be found (under the background-only hypothesis).

Likewise, we can simulate the results of a search for dark matter annihilation by sampling $\Tobs$ from the PDF of $T$ including the component $T_s$ due to dark matter annihilation. That is, we test the hypothesis $\Hb$ but take the observed test statistic to be a quantile of the distribution $\prob(T \mid \Hsb)$. We can therefore ask ``How likely are we to make a detection if the cross section has a particular value?'', i.e. the power of the test.

When searching for a signal we will need to test the background-only hypothesis $\Hb$ against multiple signal hypotheses $\Hsb$, with different dark matter properties. It is important to determine how ``finely-grained'' the $\Hsb$ are. For example, for how many trial dark matter masses should be searched for? Should the search be performed for different annihilation channels? Or will a dark matter signal be detected regardless of the specific alternative hypothesis we are testing against?

There are several additional benefits to being able to compute the distribution of expected results. In frequentist statistical analysis it is vital that the choice of test statistics and critical regions not be influenced by the observed data. One issue that has not been addressed is the decision on which events to include in the analysis. In our case this entails selecting which dwarf galaxies to consider. We also need to decide on the energy range of the events we consider and the maximum angular separation from a dwarf an event can have. These choices define our Regions of Interest (ROIs) for each dwarf galaxy. We can use the expected limits formalism to find out how different choices of ROIs will affect the particle physics limits.

Finally, we can use this formalism to make predictions for future experiments: with more observation time, different detector properties, and different targets, how strong are the dark matter limits expected to be? Examples of such projections are carried out in Sec.~\ref{sec:future}.

\subsection{Smallest detectable signal}
\label{sec:defsigv90}
As discussed at the end of Sec.~\ref{sec:likelihoodratiomethod}, when searching for weak signals with $s_Q \ll b_Q$ (e.g. for sufficiently small $\sigv$) the weighting becomes independent of $\sigv$ (since $\sigv$ just acts to scale the test statistic by a constant factor). However, to squeeze as much power out of the search as possible we will always use the most powerful set of weights, Eq.~\eqref{eqn:wloglike}, which requires choosing a scaling for the signal (equivalently, choosing $\sigv$) used in the weighting.

We wish to make the search sensitive to the weakest possible signal we can hope to detect. We therefore find the smallest cross section $\sigv_{90}$ such that there is a 90\% chance of making a $3\sigma$ detection if the true cross section were $\sigv_{90}$ (and $\sigv_{90}$ is used in the weighting Eq.~\ref{eqn:wloglike}). In other words, we cannot reasonably hope to make a detection if the dark matter particle has a smaller cross section. The quantity $\sigv_{90}$ is a useful measure of the sensitivity of a dark matter search (i.e. instrument and observation + analysis framework).

The specific value of $\sigv_{90}$ will depend on the mass and branching ratios of the particle we are searching for. The definition of $\sigv_{90}$ is illustrated below in Fig.~\ref{fig:sigvforsearch}.
\begin{figure}[h]
\centering
\includegraphics{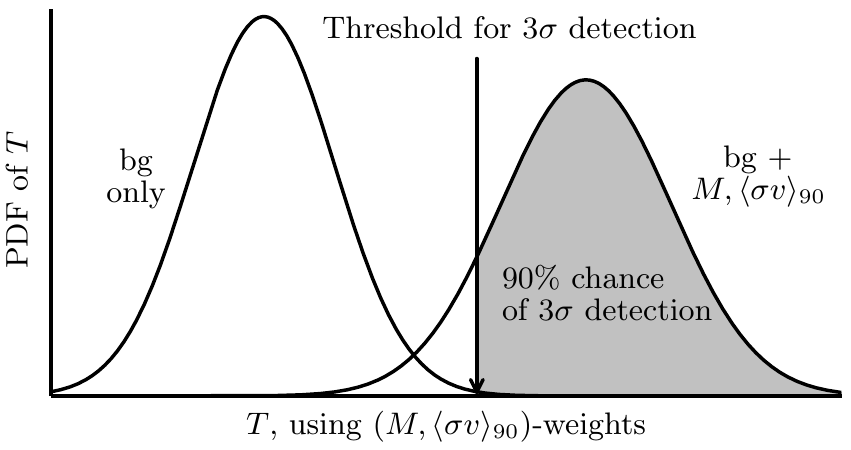}
\caption{Definition of $\sigv_{90}$, the cross section for which there is a 90\% chance of making a $3\sigma$ detection. The PDF on the left is the distribution of the test statistic if there is no dark matter annihilation (background-only). On the right is the PDF for $T$ if dark matter has mass $M$ and cross section $\sigv_{90}$. If the actual cross section is $\sigv_{90}$ there is a 90\% chance of observing the test statistic in the shaded region.
\label{fig:sigvforsearch}}
\end{figure}

If the true value of the cross section is different from $\sigv_{90}$ then the search will not be the most powerful.  For a range of masses, cross sections, and annihilation channels we have compared the power of a search using the true value of $\sigv$ in the weighting to that when using $\sigv_{90}$. The increase in power is negligible except for cross sections much smaller than $\sigv_{90}$ where we would have no hope of making a detection even if using the true value of $\sigv$ in the weighting.

\subsection{ROI size and energy range to consider
\label{sec:roiparams}}

We must first decide, for each dwarf, what energy range of events to consider and over what angular range. For example, the Fermi LAT detects photons down to a few tens of MeV. But is it really helpful to include all these low energy events in the analysis?

In principle, the event weighting method can only benefit by including more data. This can be seen heuristically by plugging the weights Eq.~\eqref{eqn:wsnr1} back into Eq.~\eqref{eqn:SNR} to find that the expected signal-to-noise ratio when using optimal weights is $\left(\sum s_Q^2 / b_Q\right)^{1/2}$. Including a larger parameter space of events corresponds to including more $Q$-terms in the sum, increasing the signal-to-noise ratio (since each term is positive).

We face diminishing returns, however, by including terms that contribute little to the sum. And, in practice, including low energy events as well as those at large angular separations from the dwarf can be detrimental. The reason is contamination by nearby gamma-ray sources. For instruments such as Fermi, the point spread function becomes quite broad toward lower energies. It is important that events from a nearby source do not leak into the ROI containing a dwarf galaxy. Such excess events are not accounted for by the background component and would be mistakenly attributed to dark matter annihilation. Making the radius of the ROI as small as possible also helps with this issue. Furthermore, when masking the known gamma-ray sources in the background region, we would prefer to use as small masks as possible in order to have many independent samples of the background. Finally, the power law-like spectra of the events causes the data set to be dominated by huge numbers of low energy events. If low energy events are not helpful in searching for or constraining dark matter annihilation it is computationally efficient to ignore them (which also avoids any systematic errors they may introduce).

To quantify the tradeoff between search sensitivity and ROI size and energy range we explore how $\sigv_{90}$ depends on the ROI parameters. We expect that lowering the low energy threshold will decrease $\sigv_{90}$ up to a point. Beyond this point the inclusion of lower energy events will not make the search much more sensitive. A similar result should hold with the radius of the ROI: including events at large separations from the dwarf will not measurably improve sensitivity.

\begin{figure*}
\centering
\subfigure{\includegraphics{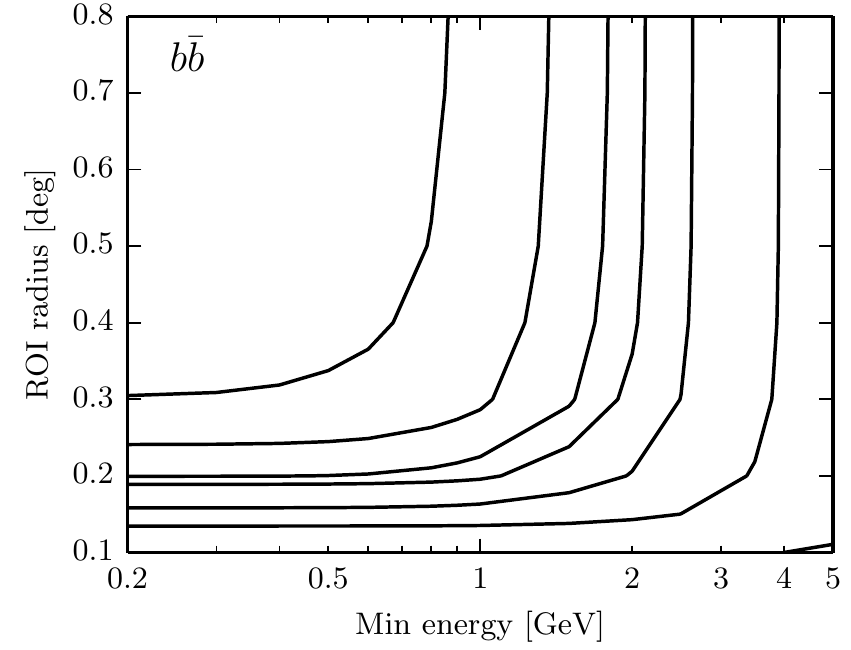}}
\subfigure{\includegraphics{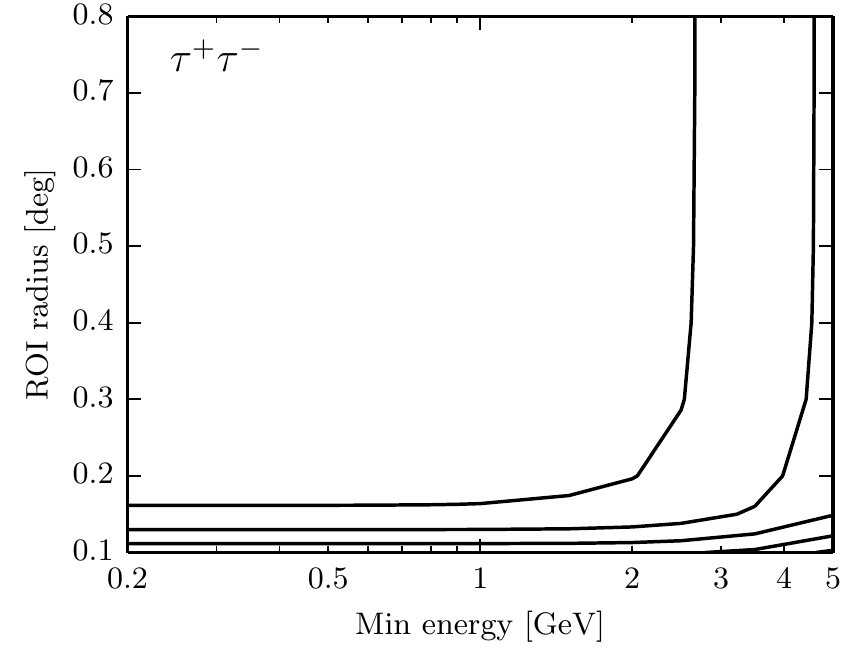}}
\caption{Dependence of search sensitivity on ROI parameters. Each contour corresponds to a different mass of the dark matter particle with $M= 10, 20, 30, 40, 50, 100$, and $500$ GeV going from top left to bottom right. The contour shows the ROI parameters which increase $\sigv_{90}$ by 50\% from the minimum possible $\sigv_{90}$ found using minimum energy $= 0.4$ GeV and ROI radius $=2\degr$. The left and right panels are for annihilation into $b\bar{b}$ and $\tau^+\tau^-$. (For the  $\tau^+\tau^-$ channel the 100 and 500 GeV contours are outside the range of the figure).
\label{fig:EminvsROIsize}}
\end{figure*}

Figure~\ref{fig:EminvsROIsize} illustrates the dependence of $\sigv_{90}$ on the choice of ROI parameters. Each curve in the figure corresponds to a dark matter particle mass, with $M= 10, 20, 30, 40, 50, 100$ and $500$ GeV going from the top left to the bottom right. The two panels correspond to the annihilation channels $b\bar{b}$ and $\tau^+\tau^-$. The curve for a given mass passes through the ROI parameters which result in $\sigv_{90}$ being 1.5 times greater than the minimum possible $\sigv_{90}$. This minimum is found by including events with energy above 0.4 GeV and an ROI radius of $2\degr$ (at 0.4 GeV the PSF is approximately $2\degr$, larger than the distance from many dwarfs to their nearest gamma-ray sources).

The figure shows that the two properties of the ROI (its minimum energy and angular radius) are essentially independent: in order to achieve a given $\sigv_{90}$ one needs to meet requirements for both minimum energy and ROI radius. There is very little tradeoff between the two, i.e. we cannot substantially increase sensitivity by reducing ROI radius while including lower energy events, for example.

\begin{figure*}
\centering
\includegraphics{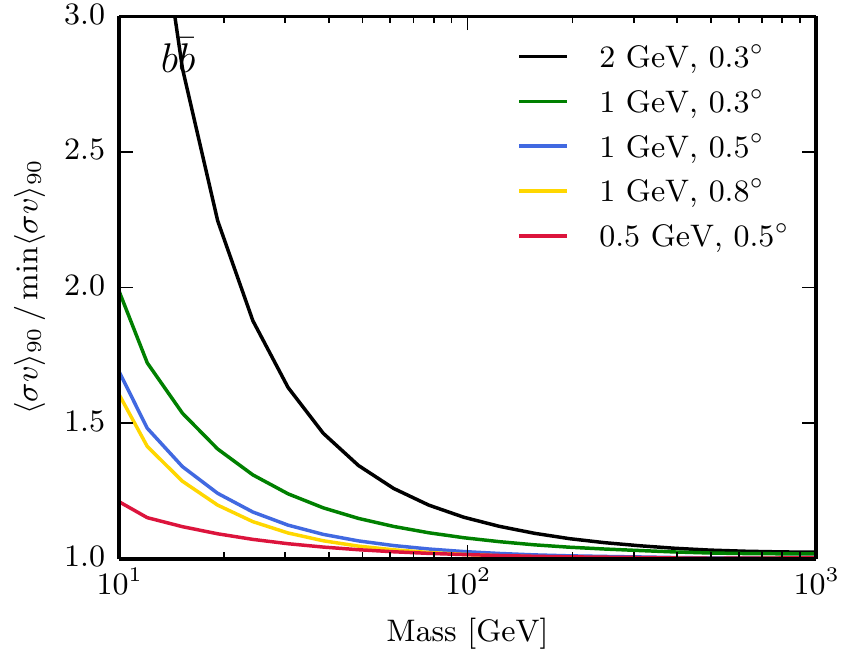}
\includegraphics{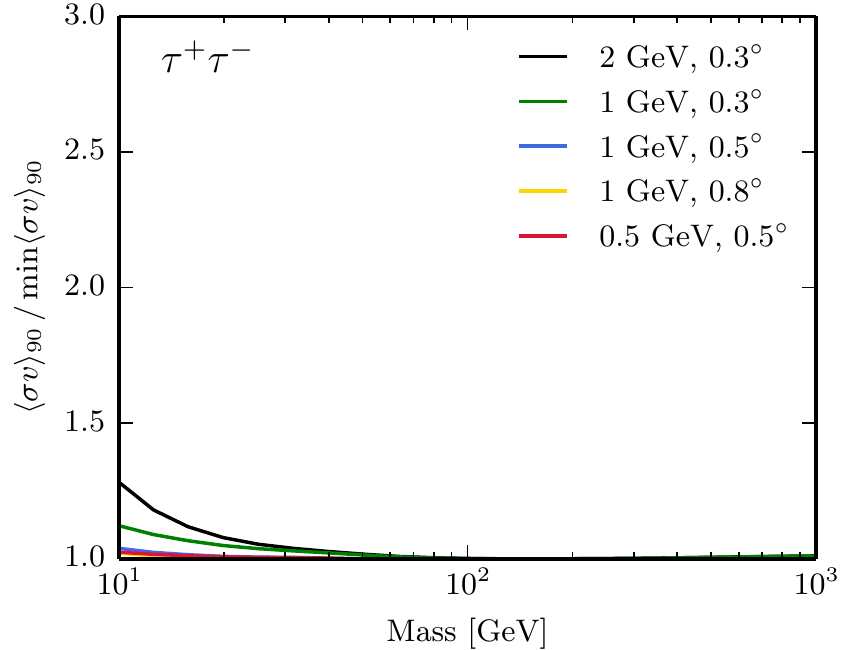}
\caption{Dependence of search sensitivity on ROI parameters. Each curve corresponds to a choice of minimum energy and ROI radius. The $y$-axis measures the relative increase in $\sigv_{90}$ from its minimum value (found using minimum energy $= 0.4$ GeV and ROI radius $=2\degr$) as a function of dark matter particle mass. The two panels are for annihilation into $b\bar{b}$ and $\tau^+\tau^-$.
\label{fig:massvsminsigv90}}
\end{figure*}

Figure~\ref{fig:massvsminsigv90} presents a different way of looking at the issue. Here we plot the fractional increase in $\sigv_{90}$ from its minimum value as a function of mass ($x$-axis) and annihilation channel (solid vs. dashed lines).

For annihilation into tau leptons, we are assured full sensitivity if we use a minimum energy of 2 GeV and an ROI radius of $0.3\degr$. This holds for any dark matter particle mass. For annihilation into quarks the situation is different for low mass WIMPs because the photon spectrum for $b\bar{b}$ is softer than for $\tau$'s. For the lowest mass WIMPs it may be necessary to use events down to 0.6 GeV and ROI sizes of $0.5\degr$ in order to achieve the best sensitivity. However, as discussed above using such low energy events may be detrimental because of source contamination.

\begin{figure*}
\centering
\subfigure{\includegraphics{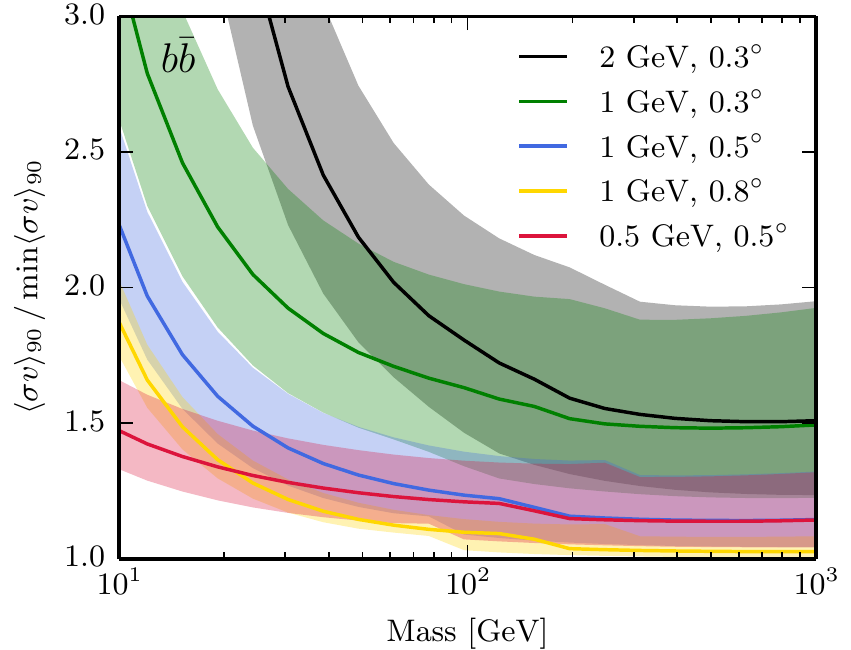}}
\subfigure{\includegraphics{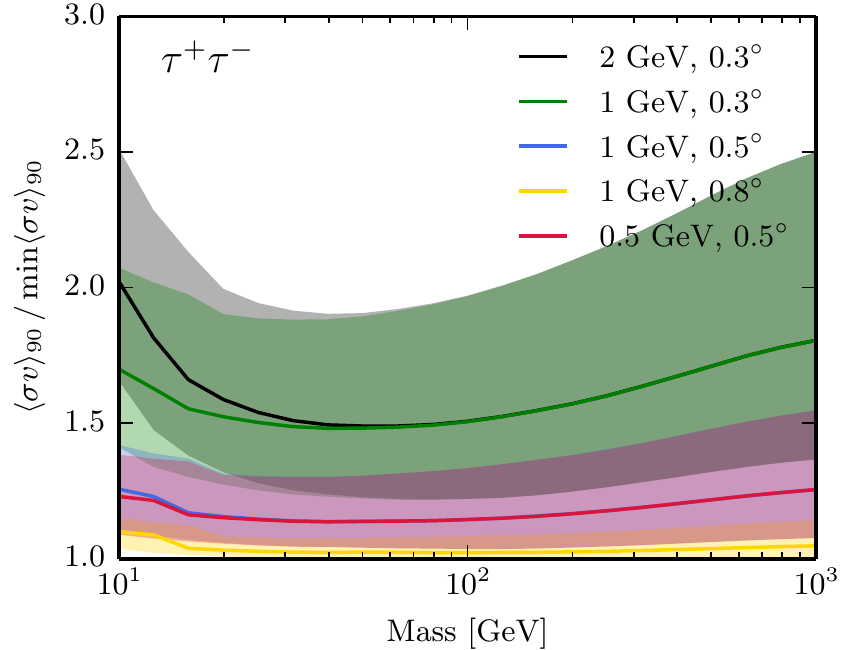}}
\caption{Dependence of sensitivity on ROI parameters for an extended source. The figure is constructed the same as for Fig.~\ref{fig:massvsminsigv90} but using the extended emission expected from Draco. The shaded bands are the $\pm 1\sigma$ quantiles for the ratio of sensitivities when considering all the allowed halos for Draco. The solid lines are the median ratios. The left panel is for annihilation into $b\bar{b}$ while the right is for $\tau^+\tau^-$.
\label{fig:massvsminsigv90Draco}}
\end{figure*}

Figures~\ref{fig:EminvsROIsize} and~\ref{fig:massvsminsigv90} were produced using the instrument response functions corresponding to the sky location of the dwarf Segue 1 and by assuming that the dwarf under consideration is a point source of gamma-rays. The situation is complicated by the extended emission expected from some of the dwarf galaxies. In Fig.~\ref{fig:massvsminsigv90Draco} we show how the sensitivity depends on the ROI parameters for the dwarf Draco, one of the most spatially extended dwarfs (only Sextans is more extended, but it's flux is likely an order of magnitude fainter, see~\cite[Table 2]{0004-637X-801-2-74}). The figure is constructed as in Fig.~\ref{fig:massvsminsigv90} but now the shaded regions show the $\pm 1 \sigma$ uncertainty when considering all the allowed halos for Draco. The solid lines are the median ratio, over all density profiles, of $\sigv_{90}$ to the minimum possible $\sigv_{90}$. The left panel is for annihilation into $b\bar{b}$ and the right is for $\tau^+\tau^-$.

As expected, for extended dwarfs like Draco, the reduction in sensitivity when using a small ROI radius is more severe. For annihilation into $\tau^+\tau^-$, using an ROI radius of $0.5\degr$ and a minimum energy of 1 GeV will be enough for near-optimal search sensitivity, regardless of the level of spatial extension. For annihilation into $b\bar{b}$ the same ROI parameters are suitable for dark matter masses above around 30 GeV. For lower masses, however, a minimum energy of 0.5 GeV is required for  near-optimal sensitivity (in the absence of nearby sources).

When setting cross section limits we use a similar procedure to find suitable ROI parameters. For any choice of minimum energy and ROI radius we can compute an expected limit. We have found that when $\sigv_{90}$ is replaced by the expected cross section limit, the analogs of Figs.~\ref{fig:EminvsROIsize},~\ref{fig:massvsminsigv90}, and~\ref{fig:massvsminsigv90Draco} are very similar. Therefore, ROI parameters chosen for the search are suitable for computing cross section limits as well.

\subsection{Event classes}
\label{sec:eventclasses}

We have also considered the more discrete choice of which Fermi LAT event class to use. The LAT event classes are involved sets of cuts, applied to the raw data, that produce the processed event files used in high-level analyses. The Fermi collaboration recommends the SOURCE class for analyzing localized sources (such as dwarf galaxies). More stringent filtering to remove cosmic ray backgrounds is used in the CLEAN and ULTRACLEAN event classes, intended for diffuse analysis of large sky regions and for studies where it is vital to push the rate of misidentified cosmic rays below the level of the isotropic gamma-ray background (the high purity coming at the expense of reduced detector effective area). In our framework it is straightforward to compute expected limits for data reduced with each of these event classes. We have found that the effects are negligible: using ULTRACLEAN instead of SOURCE weakens the expected limits by a few percent. The ULTRACLEAN class, however, has a greatly reduced event rate which makes our unbinned analysis more computationally efficient. All results in this paper are derived using Pass 7 Reprocessed ULTRACLEAN data.

\subsection{Searching for a point source versus an extended source}
\label{sec:pointvsextended}

Here we characterize the reduction in sensitivity if we perform the search using a point source alternative hypothesis when the truth is, in fact, that the dwarf is extended. It is desirable to perform searches against as few alternative hypotheses as possible. If we do not lose sensitivity when searching for point source emission, even for an extended halo, we are justified in performing a single search for point-like emission.

We calculate $\sigv_{90}$ using two different weighting schemes but always assuming the halo is an extended object. The minimum possible $\sigv_{90}$ occurs when the signal model used for the weighting coincides with the true signal, i.e. for the weighting scheme using an extended source model. We then compute $\sigv_{90}$ using a point source signal model in the weighting (i.e. replace the $J$-profile in Eq.~\eqref{eqn:expectedcounts} with a two-dimensional delta function). That is, if we mistakenly search for point-like emission when the actual emission is extended, what is the minimum $\sigv$ we are sensitive to?

For a minimum energy of 1 GeV and an ROI radius of $0.5\degr$ the reduction in sensitivity is around 30\% for Draco's most extended profiles and significantly smaller for the rest. For dwarfs besides Draco the effect will be much less. Therefore, for searches in individual dwarfs we use point source models.

\subsection{How many trial masses?}

In the event weighting framework, a search for annihilation is performed by testing the background-only hypothesis $\Hb$ against an alternative hypothesis $\Hsb$ that dark matter has a specific mass, set of branching ratios, and annihilation cross section $\sigv_{90}$ (which depends on the branching ratios and mass). A search using a particular alternative $\Hsb$ may not be very sensitive if the truth is, in fact, different from $\Hsb$.

This is dealt with by performing, one at a time, tests of $\Hb$ against an array of alternative hypotheses with different masses and branching ratios. We can quantify how many alternatives are necessary by computing the power of the tests for a given true mass and set of branching ratios.

For every pair of masses ($M, M_\tr$) we assume that dark matter has a true mass $M_\tr$ and cross section $\sigv_{90}$ (associated with $M_\tr$). We then find the power (probability of making a $3\sigma$ detection) when testing the background-only hypothesis against an alternative that dark matter has mass $M$ and cross section $\sigv_{90}$ (associated with $M$). When $M = M_\tr$ the power of the test will be 90\% since this condition defines $\sigv_{90}$. We wish to find out how different $M$ can be from the true $M_\tr$ before the power of the test suffers.

\begin{figure}
\centering
\subfigure{\includegraphics[scale=0.87]{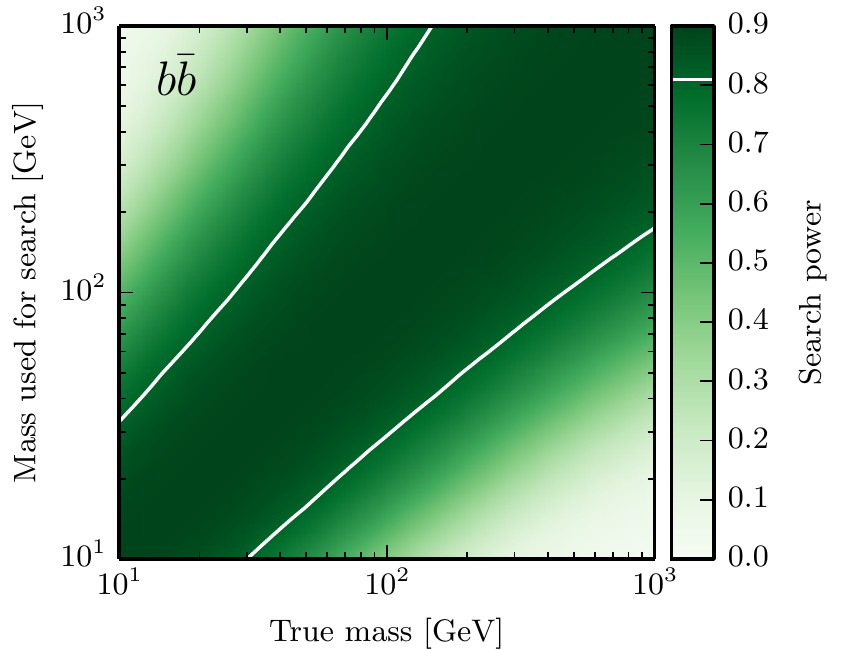}}\\
\subfigure{\includegraphics[scale=0.87]{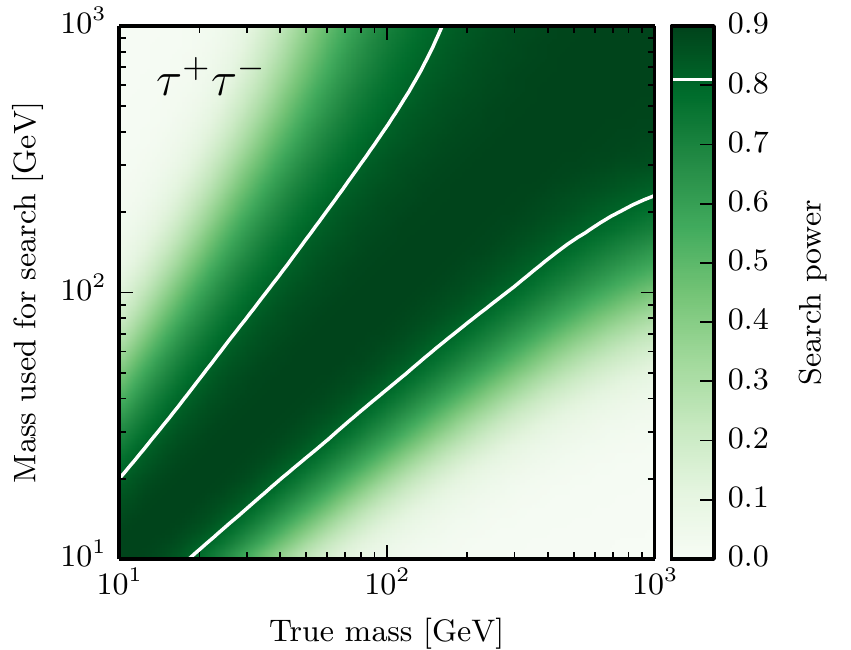}}\\
\subfigure{\includegraphics[scale=0.87]{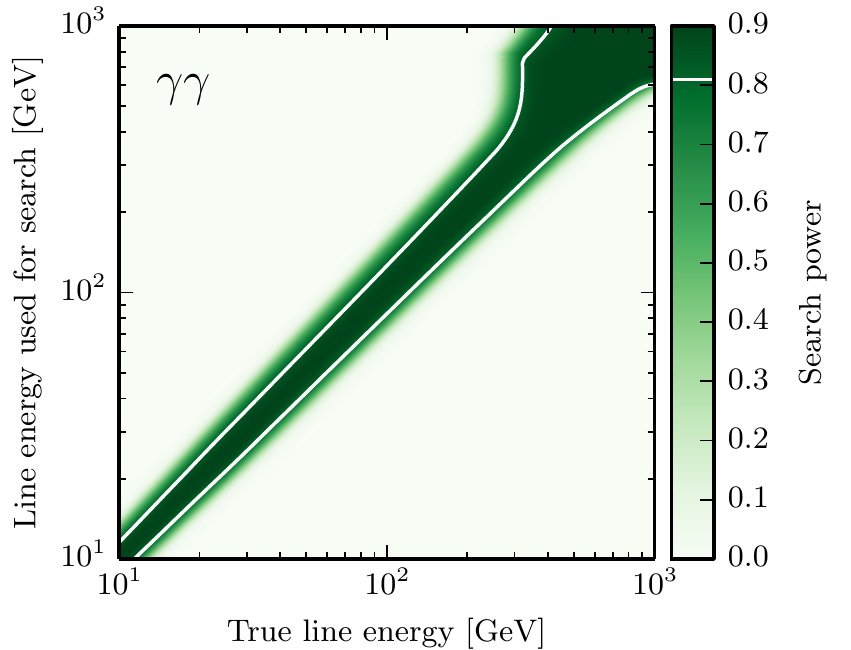}}
\caption{Statistical power of the search when the true mass differs from that used in the alternative hypothesis. The color shows the probability of making a $3\sigma$ detection if dark matter has true mass $M_\tr$ ($x$-axis) but the search is performed using mass $M$ ($y$-axis) as the alternative hypothesis in the weighting. The white contours show where the power is reduced by 10\% from its maximum (occurring along the line $y=x$). The top panel is for annihilation into $b$ quarks, middle for $\tau$ leptons, and bottom for gamma-ray line emission. Very few alternative hypotheses need to be tested in order to have near-maximum power for any dark matter mass in the range 10 GeV to 1 TeV for $b\bar{b}$ and $\tau^+\tau^-$. For line emission it is important to search with a finer spacing in dark matter mass (equivalently, photon line energy).
\label{fig:trialmasses}}
\end{figure}

Figure~\ref{fig:trialmasses} shows the power when searching for dark matter with mass $M$ ($y$-axis) when the true mass is $M_\tr$ ($x$-axis). The color represents the probability of making a $3\sigma$ detection. The three panels are for dark matter annihilating into $b$ quarks, $\tau$ leptons, and directly into two gamma-rays. The white contours indicate where the power has dropped by 10\% from its maximum of 90\% (occurring along the line $y=x$).

It is apparent that searching for a very few trial masses is sufficient to explore the full range of dark matter masses. For example, in a search for annihilation into $b\bar{b}$, we need only perform three searches (at, say, 10 GeV, 100 GeV, and 1 TeV) to be sure we are sensitive to any mass in the range from 10 GeV to 1 TeV. For annihilation into $\tau^+\tau^-$ searches for masses of 10 GeV, 40 GeV, and 300 GeV are sufficient.

For a line search we need more trials, about 15 to cover the line energies between 10 GeV to 1 TeV. The shape of the band reflects the ``window function'' nature of the weighting. I.e. the optimal weighting scheme gives zero weight to events outside a narrow energy region where the line photons are expected. The width of the window is determined by the energy dispersion of the detector, estimated as 10\%.

The strange behavior at the top right is due to two effects. The broadening in the vertical direction at around 400 GeV occurs because above this energy we expect zero background. Therefore, any weighting scheme that gives a nonzero weight to signal photons above 400 GeV will yield a detection (as long as at least one event above this energy is observed). This is confirmed by noticing that the expected number of signal events, for a cross section of $\sigv_{90}$, tends towards 2.3 as the line energy increases. In a background-free experiment a $3\sigma$ detection is made by observing a single event and so the probability of a detection is $1-\exp(-2.3) = 0.9$ as expected.

The broadening of the band only occurs in the direction of larger trial masses because the weighting for lower masses will include energies where some background is likely. The broadening ends in a ``kink'' where the trial mass weighting ``window'' no longer includes the true line energy.

%%%%%%%% Summary of procedure
\section{Summary of procedure}
\label{sec:summaryofprocedure}

The previous sections have been presented in a general way to develop the framework and indicate that it can apply to many types of studies. Here we will detail the procedure used to conduct searches and produce limits on dark matter annihilation in dwarf galaxies. This section is presented in a step-by-step manner and can be used as a template for other analyses.

The first step is to decide on a set of Region of Interest (ROI) parameters. In this study we include ULTRACLEAN events with energies between 1 GeV and 300 GeV that are reconstructed within $0.5\degr$ of each of twenty dwarf galaxies. This choice of energy and ROI radius is a compromise between sensitivity and the need to avoid contamination by nearby point sources (most of the dwarfs have a gamma-ray point source between $1\degr$ and $2\degr$ away; see Secs.~\ref{sec:dwarfdata},~\ref{sec:roiparams}, and~\ref{sec:eventclasses} regarding this choice).

%Weight function
\subsection{The weighting function}
Every test requires a weighting function (see Secs.~\ref{sec:generalformofT} and~\ref{sec:designingweight}) and we adopt the weighting function defined by Eq.~\eqref{eqn:wloglike}. This function assigns a numerical weight to an event given three properties: the dwarf field it came from $\nu$, its reconstructed energy\footnote{In our implementation we use $\log_{10}E$ as the energy variable for convenience.} $E$, and the angular separation of the event from the direction toward the dwarf~$\theta$. Explicitly,
\begin{equation}
w(Q) = w(\nu, E, \theta) = \log\left(1 + \frac{s(\nu,E,\theta)}{b(\nu,E,\theta)}\right),
\label{eqn:summary_wdef}
\end{equation}
where $s$ and $b$ are the expected number of detected signal and background events from dwarf $\nu$, with energy between $E$ and $E+dE$ and angular separation between $\theta$ and $\theta + d\theta$. (These functions are the $s_Q$ and $b_Q$ of Sec.~\ref{sec:designingweight}, just with the event properties $Q$ written explicitly.) In computing the weight, the differentials cancel and one can just use the expected number of events per energy per solid angle for $s$ and $b$ in Eq.~\eqref{eqn:summary_wdef}.

The weighting depends on the dark matter mass $\mass$, cross section $\sigv$, and branching ratios $B_i$ through the function $s(\nu,E,\theta)$,
\begin{equation}
s(\nu,E,\theta) = \frac{dN(\nu,E,\theta)}{dEd\Omega} dE\, 2\pi \sin(\theta)d\theta,
\label{eqn:summary_sdef}
\end{equation}
with $dN/dEd\Omega$ defined in Eqs.~\eqref{eqn:expectedcounts} and~\eqref{eqn:dNdE}. Note that in Eq.~\eqref{eqn:expectedcounts} that the symbols $J$, PSF, and $\epsilon$ should all be considered to contain a subscript $\nu$ since they differ for each dwarf.

The energy spectrum for background events $b(\nu,E,\theta)$ is defined in Eq.~\eqref{eqn:bQestimate}. See Sec.~\ref{sec:bgmodelforweighting} for details.

We denote the set of particle parameters $(\mass, \sigv, B_i)$ used in the weighting function by $\Pw$. To be explicit about the dependence of the weight on all parameters we can write the weighting function as $w(\nu, E, \theta \mid \Pw)$.

\begin{figure}
\centering
\subfigure{\includegraphics{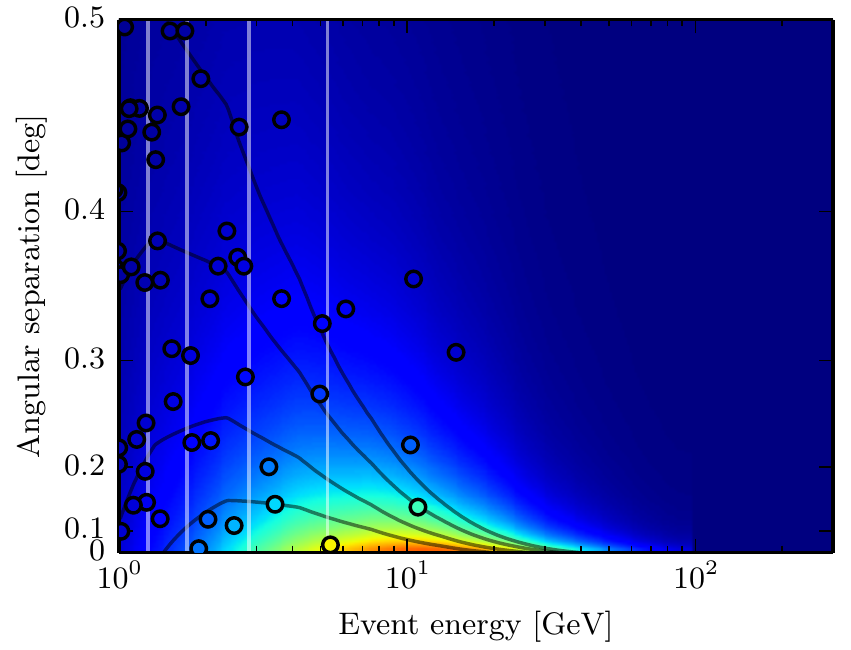}}
\caption{Illustration of the weighting scheme for a 100 GeV particle annihilating into $b\bar{b}$. Color represents the weight given to an event with a particular energy and angular separation from the dwarf (going from zero in blue to a maximum value in red). The black (light gray) contours show where the signal (background) events are expected to be detected, with successive contours enclosing 25\%, 50\%, 75\%, and 90\% of the expected events (bottom to top for signal contours, left to right for background). The filled circles show the observed events from the dwarf Segue 1.
\label{fig:weightingscheme}}
\end{figure}

Figure~\ref{fig:weightingscheme} illustrates the weighting function for the dwarf Segue 1 using a dark matter mass of $100\,\GeV$ and annihilation into $b$ quarks. The color represents the weight that would be assigned to an event with a particular energy and angular separation from Segue 1. The contours show the expected number of signal events (black contours) and background events (light gray), with the successive contours enclosing 25\%, 50\%, 75\%, and 90\% of the events. The small circles are the observed events, with color indicating the weight they are assigned. We see that there is a ``sweet spot'' at an energy around a tenth of the dark matter mass, where events are given the highest weight: events at this energy have the highest odds of being due to dark matter annihilation rather than background.

%Computing PDF
\subsection{Computing the PDF of $T$}
The weighting function defines the test statistic $T$ through Eq.~\eqref{eqn:Tdef}. The probability distribution describing $T$ must be determined under the hypothesis $\Hsb$: dark matter has mass $M$, cross section $\sigv$, and branching ratios $B_i$. It is important to note that this set of $M$, $\sigv$, and $B_i$ can differ from that used to define the weighting function. In particular, $\Hb$ (the background-only hypothesis) is just the special case of $\Hsb$ where $\sigv=0$.

The procedure for computing the PDF of $T$ under the hypothesis $\Hsb$ as described in Sec.~\ref{sec:pdfofT}. It is easiest to do this with FFTs: use Eq.~\eqref{eqn:CPft} to find the FFT of $T_s$. For each dwarf, compute the PDF of $T_b$ using the sampling procedure described in Sec.~\ref{sec:bgsampling} and take the FFT of each dwarf's $T_b$ distribution. Multiply all the FFTs together and perform the inverse FFT to convolve $T_s$ with $T_b$ and obtain the PDF of the test statistic $T$.

\subsection{Search for annihilation}
\label{sec:summary_search}
The search for a signal is the testing of the null hypothesis $\Hb$: background-only, against the alternative $\Hsb$: dark matter has mass $M$, cross section $\sigv_{90}$, and branching ratios $B_i$.

Given a choice of mass and branching ratios, find $\sigv_{90}$. This is done by constructing a function that returns the power of the search as a function of $\sigv$. Find the value $\sigv_{90}$ such that the power is 90\%. To compute the power given a value of $x$ for the cross section:
\begin{enumerate}
\item Adopt a weighting function which uses $\sigv=x$ in $\Pw$.

\item Determine the PDF of $T$ under the background-only hypothesis $\Hb$ (i.e. that $\sigv=0$).

\item Find the critical value of the test statistic $T^*$ corresponding to a $3\sigma$ detection. I.e. the value of $T^*$ such that $\prob(T < T^* \mid \Hb) = 0.9987$.

\item Determine the PDF of $T$ under the hypothesis $\Hsb$ that $\sigv=x$.

\item Compute the probability of making a $3\sigma$ detection by finding the probability that $T$ is greater than $T^*$. I.e. find $\prob(T > T^* \mid \Hsb)$. This is the power of the search for dark matter with cross section $\sigv=x$.

\end{enumerate}
See Fig.~\ref{fig:sigvforsearch} for an illustration of the definition of $\sigv_{90}$.

Once $\sigv_{90}$ has been found, performing the actual search is simple:
\begin{enumerate}
\item Adopt a weighting function which uses a cross section $\sigv_{90}$ in $\Pw$.

\item Find the {\em observed} value of the test statistic: apply the weighting function to the observed events in the ROI's for the dwarfs. Add up the weights of all the observed events to obtain the observed test statistic $\Tobs$. Explicitly,
\begin{equation}
\Tobs = \sum\limits_i w(\nu_i, E_i, \theta_i \mid \Pw),
\label{eqn:summary_Tobs}
\end{equation}
where $i$ runs over all the events in the ROIs centered on the dwarfs (see Eq.~\eqref{eqn:Tdef}).

\item Determine the PDF of $T$ under the background-only hypothesis $\Hb$ (i.e. that $\sigv=0$).

\item Find the probability of measuring $T$ to be less than $\Tobs$ under the background-only hypothesis (i.e. $\prob(T < \Tobs \mid \Hb)$). This is the ``significance'' of the detection of an annihilation signal (see Sec.~\ref{sec:statisticalframework}).
\end{enumerate}

%%%%% limits
\subsection{Constructing limits}
\label{sec:summary_limits}
To find upper limits on the annihilation cross section the procedure is similar to that used for the search. For a given mass and set of branching ratios, construct a cumulative distribution function (CDF) for the test statistic. This is a function of $\sigv$ that returns the probability that $T$ is smaller than observed if $\sigv$ were the true value of the cross section. The 95\% upper limit on the cross section is the value of $\sigv$ for which this function is 0.05. To find the value of the CDF for a given cross section $x$: 
\begin{enumerate}
\item Adopt a weighting function which uses $\sigv=x$ in $\Pw$.

\item Find the {\em observed} value of the test statistic: apply the weighting function to the observed events in the ROI's for the dwarfs. Add up the weights of all the observed events to obtain the observed test statistic $\Tobs$ (see Eq.~\ref{eqn:summary_Tobs}).

\item Determine the PDF of $T$ under the hypothesis $\Hsb$ that $\sigv=x$.

\item Find the probability of measuring $T$ to be smaller than $\Tobs$ under the $\Hsb$ hypothesis. I.e. find $\prob(T < \Tobs \mid \Hsb)$. If this probability is less than 5\% then $\sigv=x$ is rejected at 95\% confidence.
\end{enumerate}

\subsection{Expected limits}
\label{sec:summary_expectedlimits}
To find the range of $\sigv$ where the limit is likely to be found construct a CDF slightly modified from the one used to find the observed limits. Replace Step 2 above by:
\begin{enumerate}
\item Determine the PDF of $T$ under the background-only hypothesis $\Hb$ (i.e. that $\sigv=0$).
\item Set $\Tobs$ to be an appropriate quantile from this distribution. E.g. the 16th, 50th, or 84th percentile.
\end{enumerate}
For example, the expected limit $\sigv_E$ found using the 16th percentile has the following interpretation: if there were no dark matter annihilation (background-only) there is a 16\% chance that the observed limit will be stronger than $\sigv_E$. So if the expected limits for both the 16th and 84th percentile are found (call them $\sigv_{E1}$ and $\sigv_{E2}$) we can say that there is a 68\% that the observed limit will lie in the range between $\sigv_{E1}$ and $\sigv_{E2}$.

%%%%%%%%% RESULTS

\section{Results}
\label{sec:results}

\subsection{Search for annihilation in individual dwarfs}
\label{sec:results_indiv}
As discussed in Sec.~\ref{sec:pointvsextended} when performing a search we lose minimal sensitivity by assuming the dwarf emission profiles are point sources. Under this assumption, the spatial emission profile is described simply by an amplitude $J=J(\theta_{\rm max})$. The annihilation cross section is completely degenerate with this $J$ value. Therefore, when searching for emission from an individual dwarf we can remain completely agnostic about its $J$ value. The assumed $J$ value does not influence the results of the search (it just changes $\sigv_{90}$ used in the alternative hypothesis).

\begin{figure*}
\includegraphics[scale=0.68]{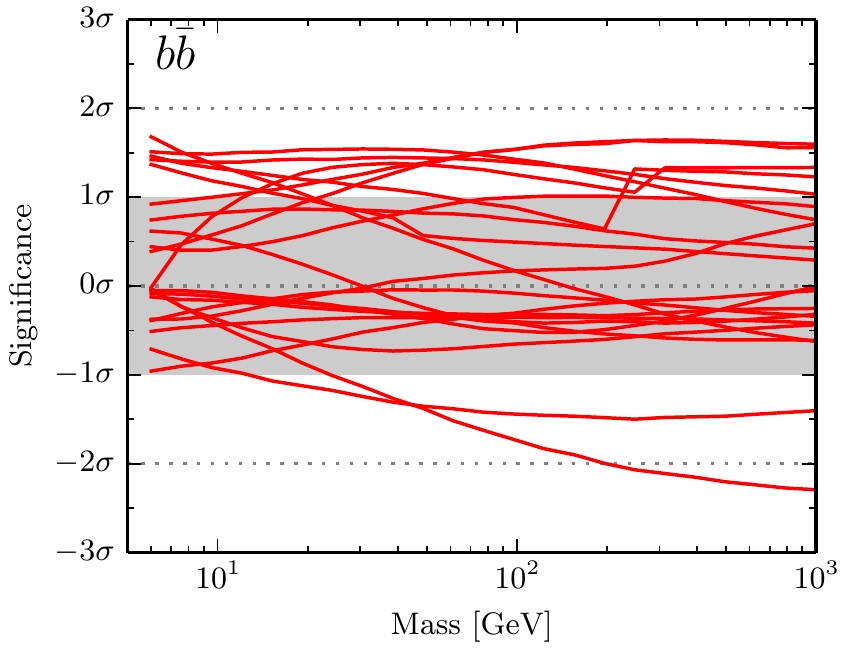}
\includegraphics[scale=0.68]{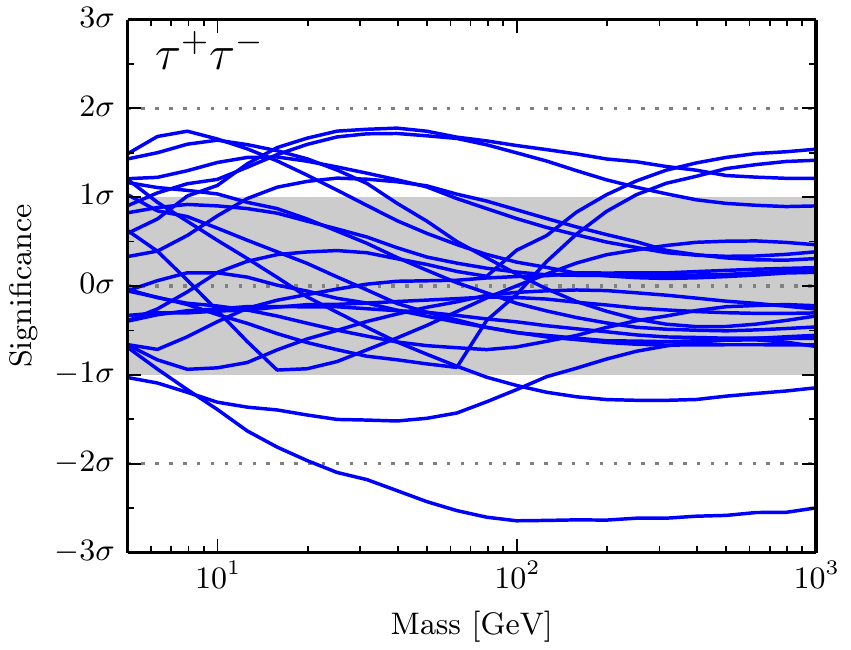}
\includegraphics[scale=0.68]{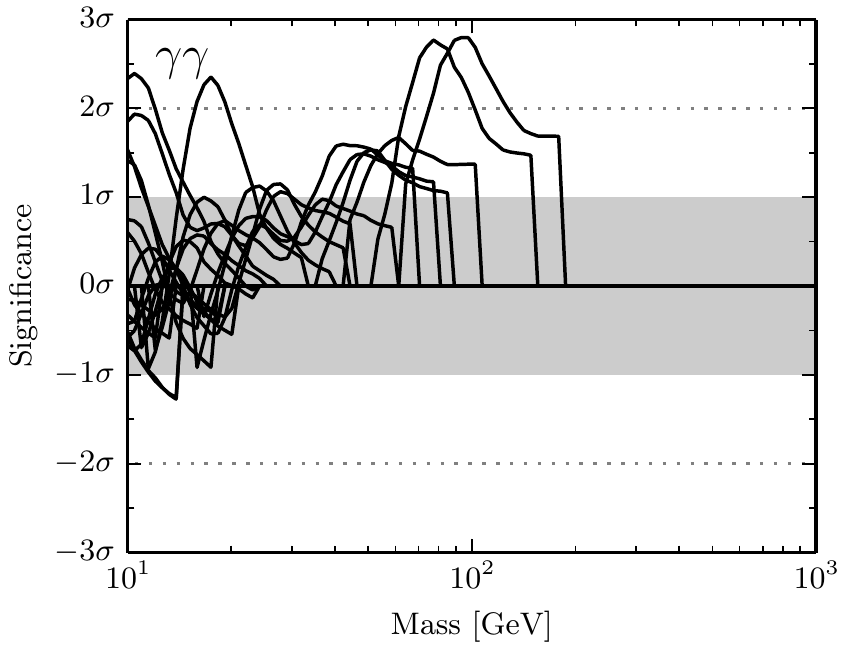}
\caption{Results of the search for annihilation in each dwarf separately. For each mass, annihilation channel, and dwarf galaxy the significance of the events in the central ROI is computed according to the background-only probability distribution. The $y$-axis shows the corresponding ``sigma value'' obtained from a standard normal distribution. Each line represents one dwarf galaxy. There is no strong detection of annihilation in any of the dwarfs individually. 
\label{fig:searchindiv_oneplot}}
\end{figure*}

\begin{figure*}
\includegraphics[scale=0.9]{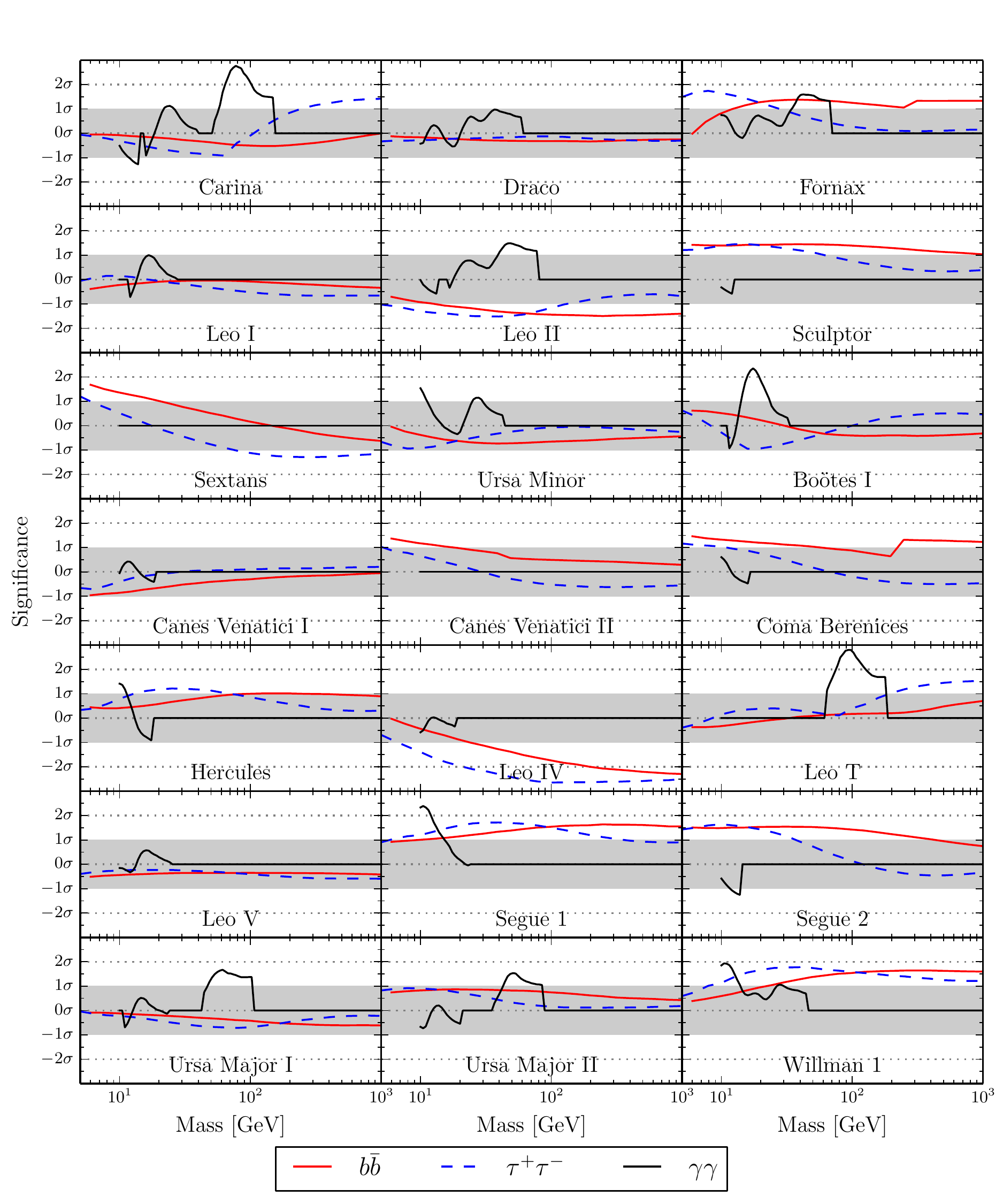}
\caption{Results of the search for annihilation in each dwarf separately. For each mass, annihilation channel, and dwarf galaxy the significance of the events in the central ROI is computed according to the background-only probability distribution. The $y$-axis shows the corresponding ``sigma value'' obtained from a standard normal distribution. Each line represents one dwarf galaxy. There is no strong detection of annihilation in any of the dwarfs individually.
\label{fig:searchindiv_separate}}
\end{figure*}

Figure~\ref{fig:searchindiv_oneplot} show the results of the search for annihilation in individual dwarfs. A search is performed separately for each mass and separately for annihilation into $b$ quarks (left panel), into $\tau$ leptons (middle panel), and into $\gamma\gamma$ (right panel).  Figure~\ref{fig:searchindiv_separate} separates the results for each dwarf galaxy, showing all three annihilation channels in the same plot (solid red lines for $b\bar{b}$, dashed blue for $\tau^+\tau^-$, and solid black for $\gamma\gamma$). We include the object Willman 1 when performing the individual searches. This object is nearby but due to its irregular kinematics we cannot reliably estimate its $J$-profile (see discussion in Sec.~\ref{sec:dwarfdata}). Therefore, it cannot be included in the joint search described in the next section.

The significance of the detection is found using the method in Sec.~\ref{sec:summary_search}. The probability of measuring $T$ to be less than $\Tobs$ is converted into ``sigma units'' using the inverse CDF of a standard normal distribution and is plotted as the $y$-coordinate in Figure~\ref{fig:searchindiv_oneplot}. If the events in the ROI are drawn from the background distribution there is a 68\% chance that the significance lies between $-1\sigma$ and $1\sigma$, a 95\% chance the significance is between $-2\sigma$ and $2\sigma$, etc.

The distribution of significances in Figure~\ref{fig:searchindiv_oneplot} is consistent with the background-only hypothesis. None of the dwarfs individually show a significant $\Tobs$ that could be due to dark matter annihilation. It is important to note that the significances shown are ``local'': they represent the results of a single hypothesis test. The more appropriate statistical question is to find the probability that {\em any} of the searches give an unusually large $\Tobs$. The correction to the significance can be made using a ``trials factor'', which would take into account the number of independent masses, annihilation channels, and dwarfs that were tested. Because none of the dwarfs show any evidence for a signal, it is unnecessary to precisely quantify the trials factor --- incorporating a trials factor can only decrease the significance.

The kinks in the curves occur due to the presence of individual high energy events. For high mass dark matter (especially for the hard $\tau$ spectrum) the weighting scheme gives large weights to the high energy photons. The kinks occur where the mass of the particle rises above a high energy event, causing the observed weight to suddenly jump. That is, a single high energy event ``outweighs'' a large number of low energy events --- the search is signal dominated for these dark matter masses. Notice that the kinks are always followed by an increase in significance, never a decrease, as expected.

\subsection{Joint search}

One of the main novelties of the technique presented here is that it enables an optimally sensitive search of the entire data set. That is, we can include all the events from all the dwarfs in the weighting to do a simultaneous search among all the dwarfs. For this procedure knowledge of the dark matter distributions in the dwarfs becomes important since the weighting requires the relative strength of the signal amongst the dwarfs (see Eqs.~\ref{eqn:DMflux} and~\ref{eqn:wloglike}).

As discussed in Sec.~\ref{sec:dwarfdata}, there is a systematic uncertainty in the determination of $J$-profiles. To deal with this uncertainty we perform multiple hypothesis tests where we explore the range of allowed density profiles. Specifically, we draw sample halos for the dwarfs according to their posterior probability distributions (see Sec.~\ref{sec:dwarfdata}) and then perform the joint search using this realization of $J$-profiles. We repeat this procedure many times, generating an ensemble of significance curves which explore the possible dark matter distributions in the dwarfs. The $J$-profiles are just used in the weighting function --- the hypothesis being tested is that the data come from background processes only.

\begin{figure*}
\centering
\includegraphics[scale=0.68]{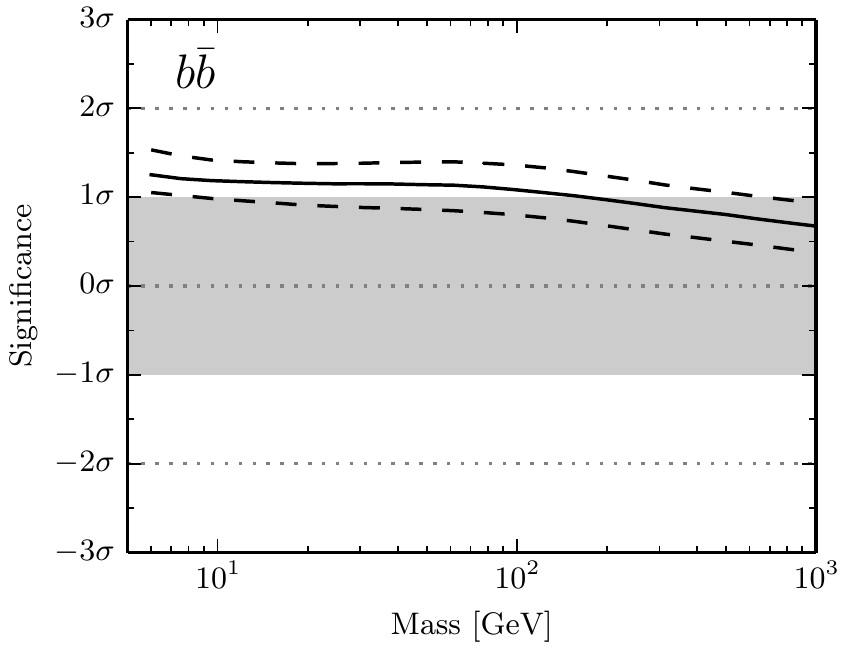}
\includegraphics[scale=0.68]{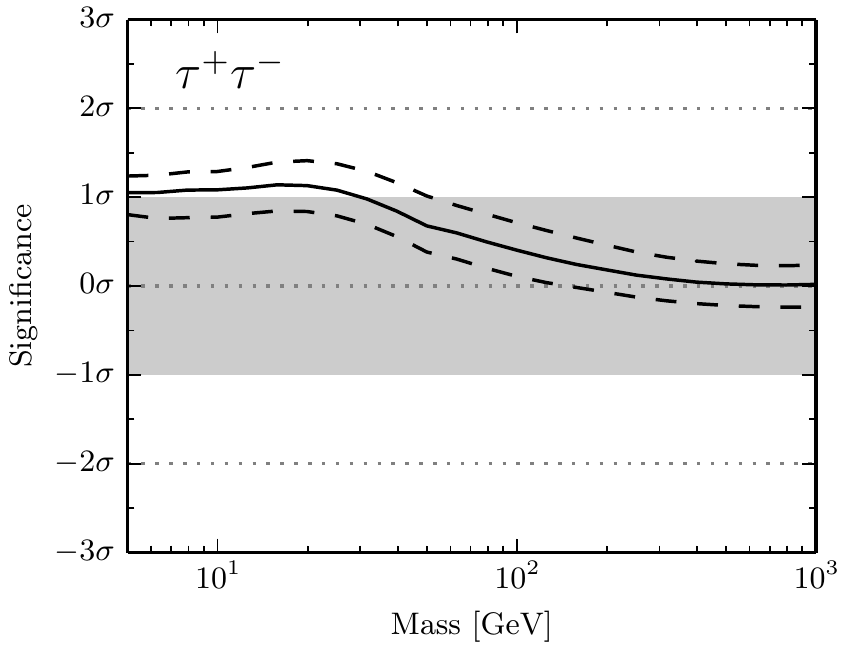}
\includegraphics[scale=0.68]{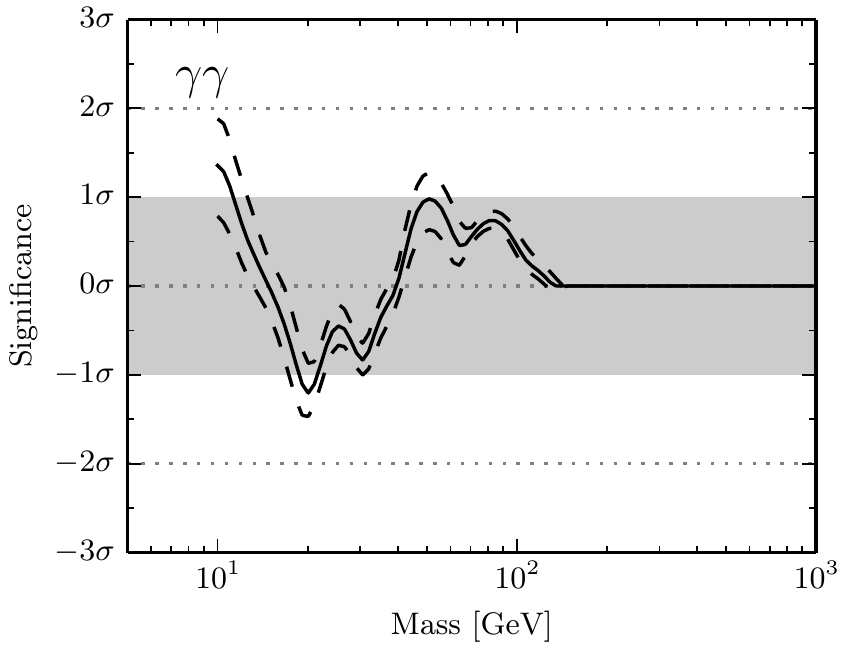}
\caption{Results of the joint search for annihilation in 20 dwarf galaxies. Dwarfs are assumed to be point sources with $J$ values drawn from the posterior chains (1-bin chains for the ultrafaints). At each mass, 250 searches are performed by drawing $J$ values from their distributions, and the bands represent 16th to 84th percentiles of the significance of detection. The significance on the $y$-axis does not take into account any trials factor resulting from performing the search at multiple masses and for multiple realizations of the $J$ values.
\label{fig:searchjoint}}
\end{figure*}

The results of the joint search are plotted in Figure~\ref{fig:searchjoint} where we consider annihilation separately into $b\bar{b}$ (left panel), into $\tau^+\tau^-$ (middle panel), and into $\gamma\gamma$ (right panel). For each value of mass, among the ensemble of significance curves, we find the median (black line) and 16th and 84th percentiles (dashed lines) of the significances.

Combining all observations in this way results in the most sensitive search for dark matter annihilation in dwarf galaxies. There does not appear to be an excess signal over background in the combined data set. We therefore proceed to set limits on the dark matter annihilation cross section.

%%%%%%%%%% LIMITS
\subsection{Cross section limits}
\label{sec:results_limits}

\begin{figure*}
\centering
\includegraphics[scale=0.68]{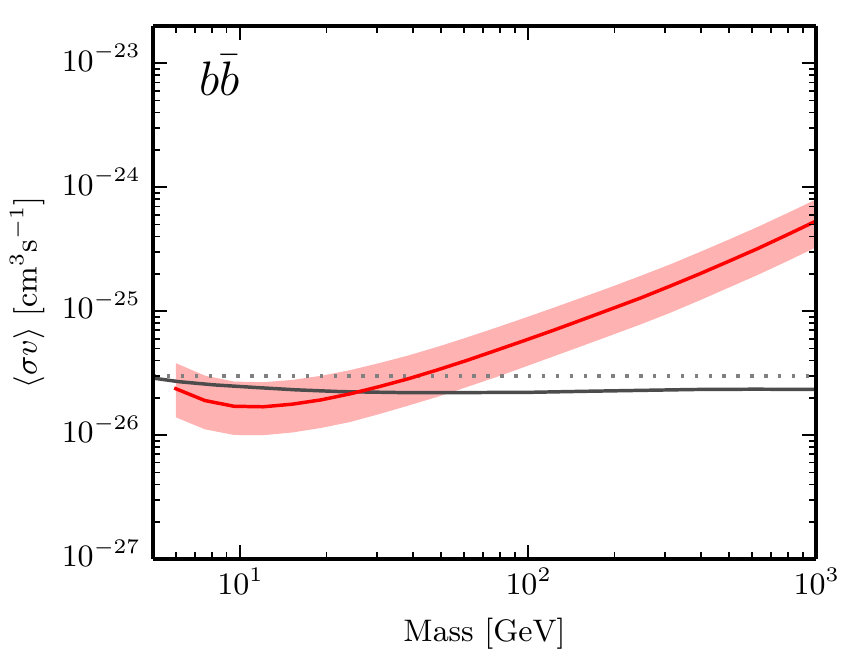}
\includegraphics[scale=0.68]{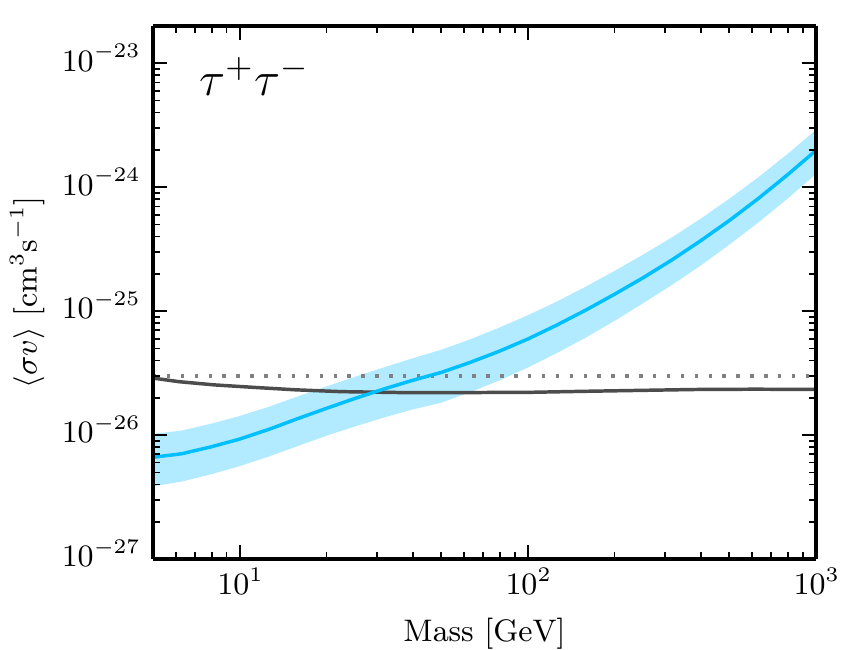}
\includegraphics[scale=0.68]{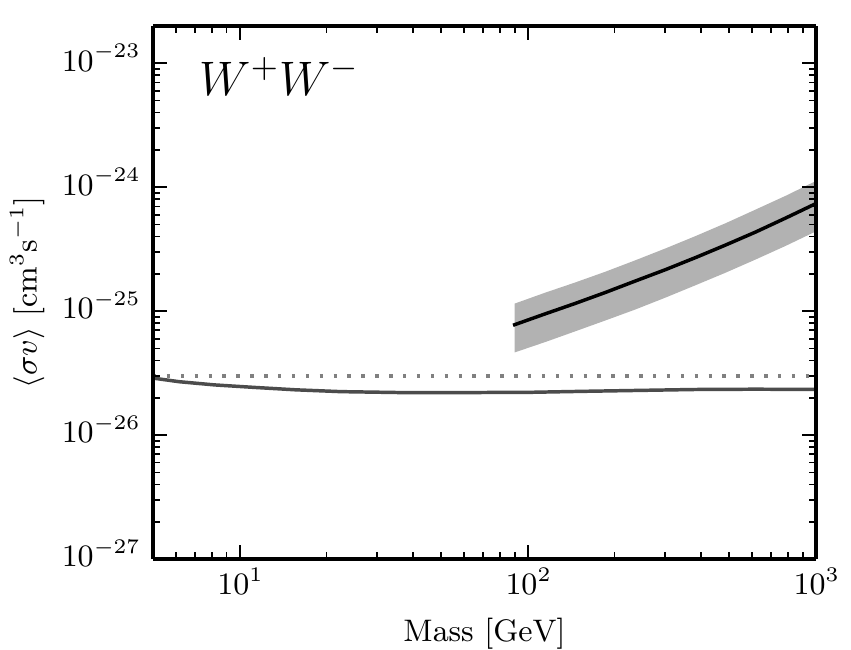} \\
\includegraphics[scale=0.68]{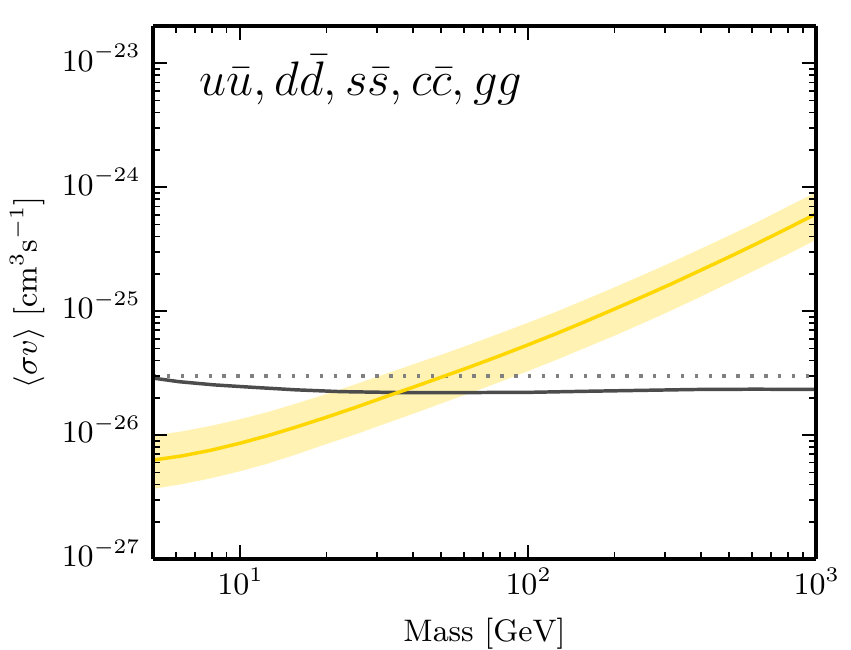}
\includegraphics[scale=0.68]{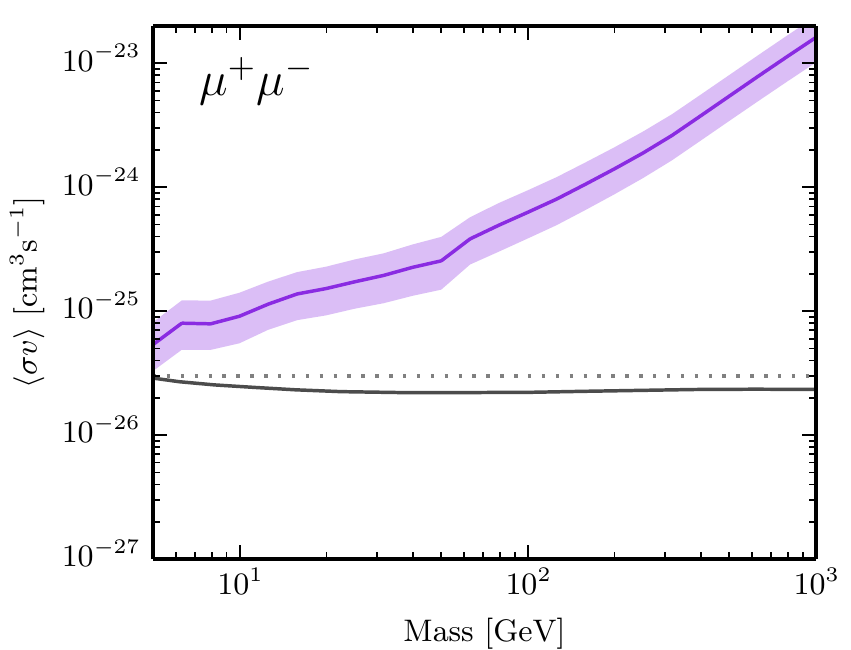}
\includegraphics[scale=0.68]{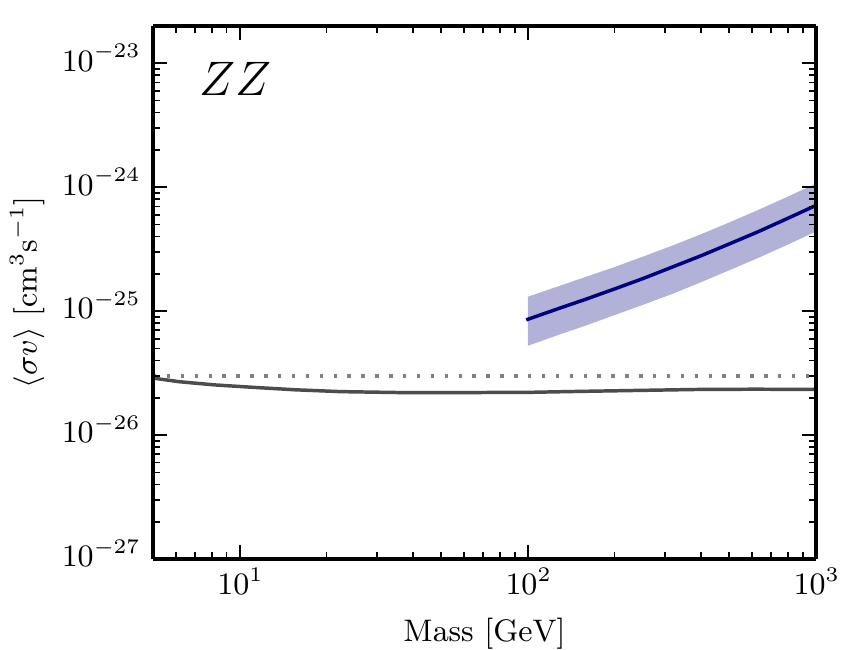} \\
\includegraphics[scale=0.68]{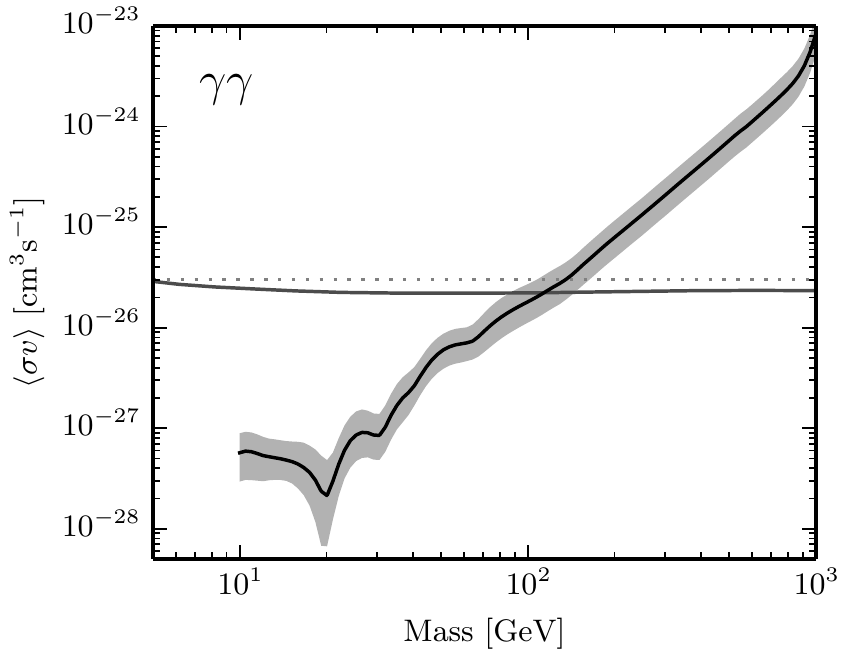}
\includegraphics[scale=0.68]{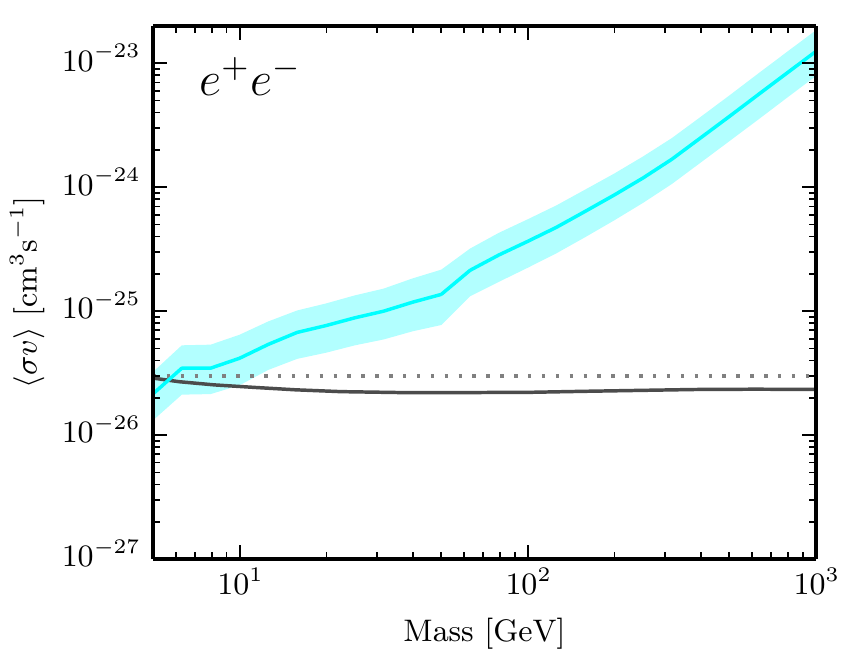}
\includegraphics[scale=0.68]{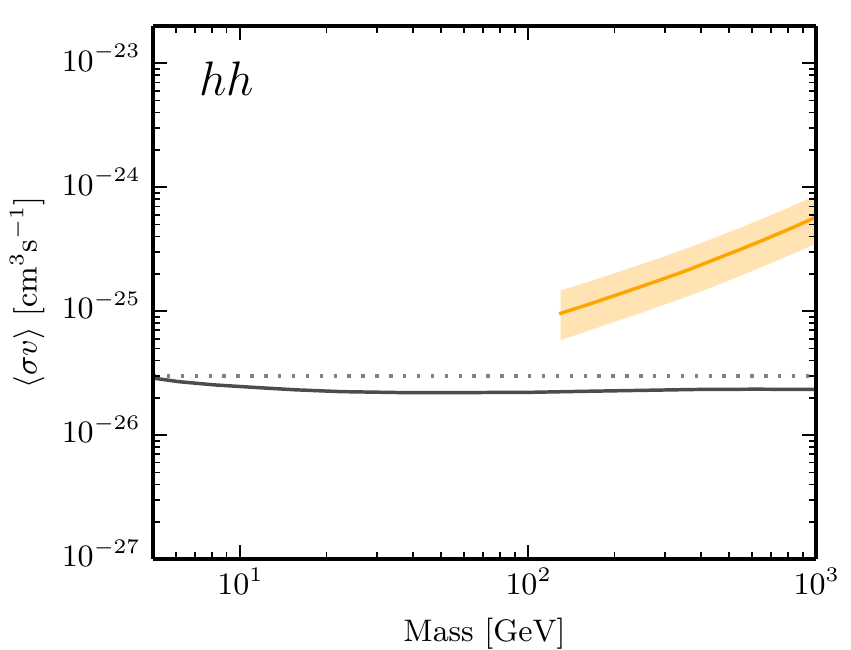} \\

\caption{Annihilation cross section limits from the joint analysis of 20 dwarf galaxies. The shaded band is the systematic $1 \sigma$ uncertainty in the limit derived from many realizations of halo $J$-profiles of the dwarfs consistent with kinematic data. The solid line depicts the median of this distribution of limits over the halo realizations. The thin dashed line corresponds to the benchmark value of the required relic abundance cross section ($3 \times 10^{-26} {\mathrm{cm}}^3/{\mathrm{s}}$), while the solid horizontal line corresponds to the detailed calculation of this quantity derived by \citet{2012PhRvD..86b3506S}. The observed limits are below this latter curve for masses less than $[0, 26, 54]~\GeV$ (for annihilation into $b\bar{b}$), $[18, 29, 62]~\GeV$ ($\tau^+\tau^-$), $[21, 35, 64]~\GeV$ ($u \bar{u}, d \bar{d}, s \bar{s}, c \bar{c}$, and $gg$), $[87,114, 146]~\GeV$ ($\gamma\gamma$), and $[5, 6, 10]~\GeV$ ($e^+ e^-$), where the quantities in brackets are for the $-1\sigma$, median, and $+1\sigma$ levels of the systematic uncertainty band. A machine-readable file tabulating these limits is available as Supplemental Material.
\label{fig:combinedlimits_syst}}
\end{figure*}

Limits on the annihilation cross section are computed using the procedure of Sec.~\ref{sec:summary_limits}. We combine the data from the twenty dwarf galaxies analyzed in~\citep{0004-637X-801-2-74} to produce a single set of limits. The cross section limits are computed for each annihilation channel assuming a 100\% branching ratio. Figure~\ref{fig:combinedlimits_syst} shows the cross section upper limits.

The shaded band in each figure quantifies the systematic uncertainty in the limit due to our uncertain knowledge of the $J$-profiles of the individual dwarfs. If we had perfect knowledge of the density profiles we could construct a single upper limit curve. In this work we take the approach of separating out the uncertainty in the dark matter density profiles from the statistical upper limits.

To produce the bands we sample $J$-profiles from their posterior distributions as described in Sec.~\ref{sec:dwarfdata} and~\citep{0004-637X-801-2-74}. For each realization of density profiles we compute upper limits as described in Sec.~\ref{sec:summary_limits}. This gives a collection of limit curves, one per realization, with roughly 1300 to 2500 realizations in total (depending on channel). At each mass we find the 16th, 50th, and 84th percentiles among the realizations. These define the lower and upper edges of the band and the solid line in the middle. The interpretation of these systematic bands will be discussed further in Sec.~\ref{sec:Juncertainty}.

The horizontal solid gray line is the relic abundance (or thermal) cross section computed by \citet{2012PhRvD..86b3506S}. The dotted gray line is $\sigv = 3\times 10^{-26} \,\cm^3 \second^{-1}$, the rough ``canonical'' value used over the past two decades, before experiments attained their current sensitivities. As discussed in the introduction, these curves are lower limits on the annihilation cross section. Therefore, dark matter masses for which our upper limits lie below the thermal cross section are ruled out (assuming 100\% branching fraction into the given channel).

In addition to systematic uncertainties in the limit we quantify the magnitude of statistical fluctuations in the limit. That is, if it were possible to repeat the gamma-ray observations while holding the $J$-profiles fixed we would obtain different observed limits each time. This is due to fluctuations in the finite number of signal and background events detected. The statistical uncertainty is found by computing expected limits as described in Sec.~\ref{sec:summary_expectedlimits}. For each realization of $J$-profiles we find the range of $\sigv$ in which there is a 68\% of finding the limit if the data are due to background processes only. The median expected limit is found the same way. This 68\% statistical uncertainty can be visualized as an error bar centered on the median expected limit. We have found that while the median varies greatly with different realizations of $J$-profiles, the width of the error bar is constant. Therefore, we can quantify the statistical uncertainty in the limit by a single band.

\begin{figure*}
\includegraphics[scale=0.68]{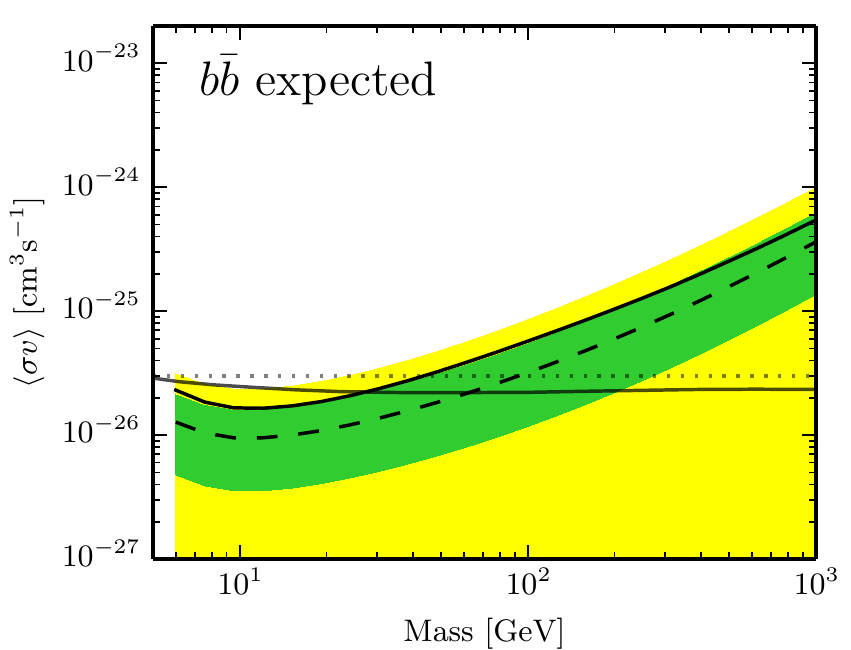}
\includegraphics[scale=0.68]{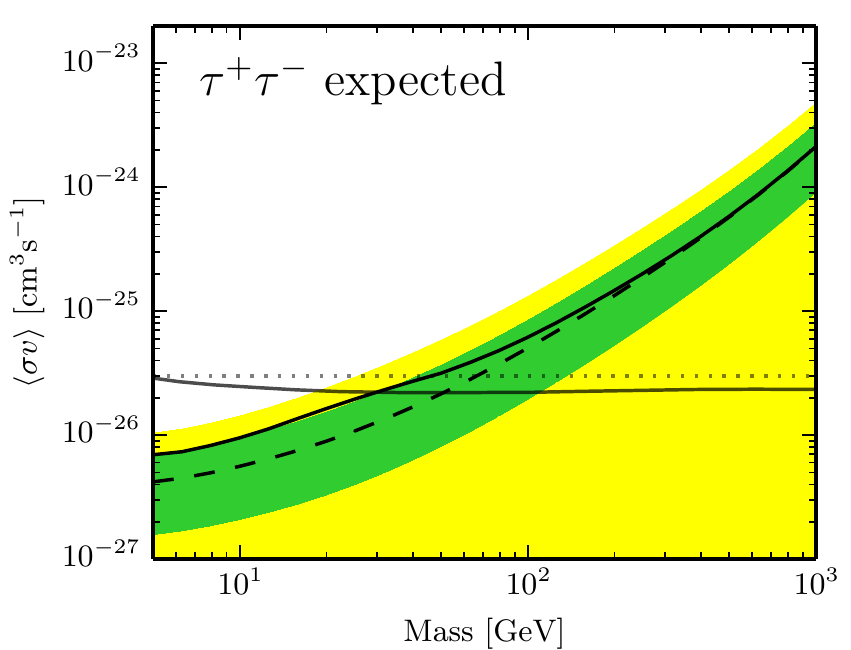}
\includegraphics[scale=0.68]{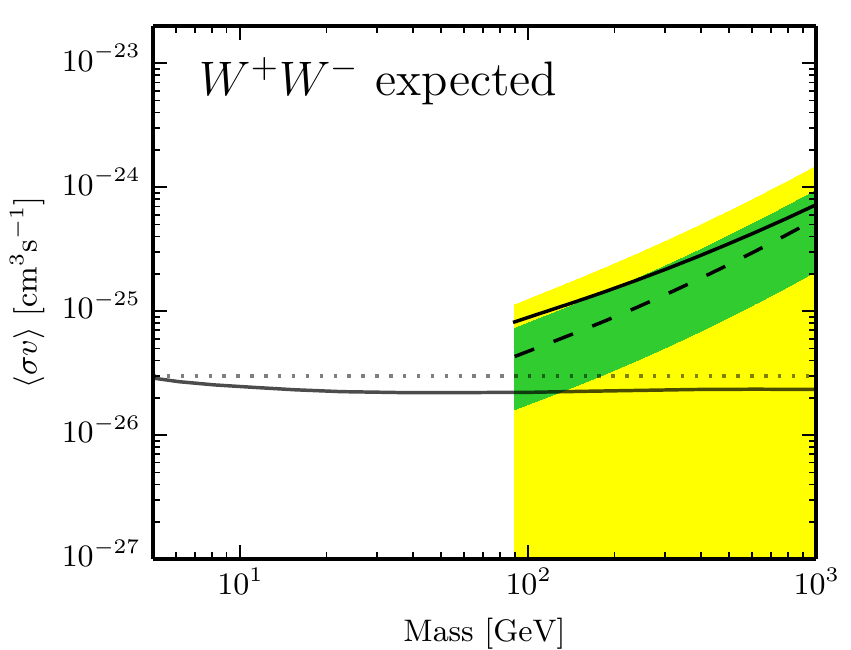}\\
\includegraphics[scale=0.68]{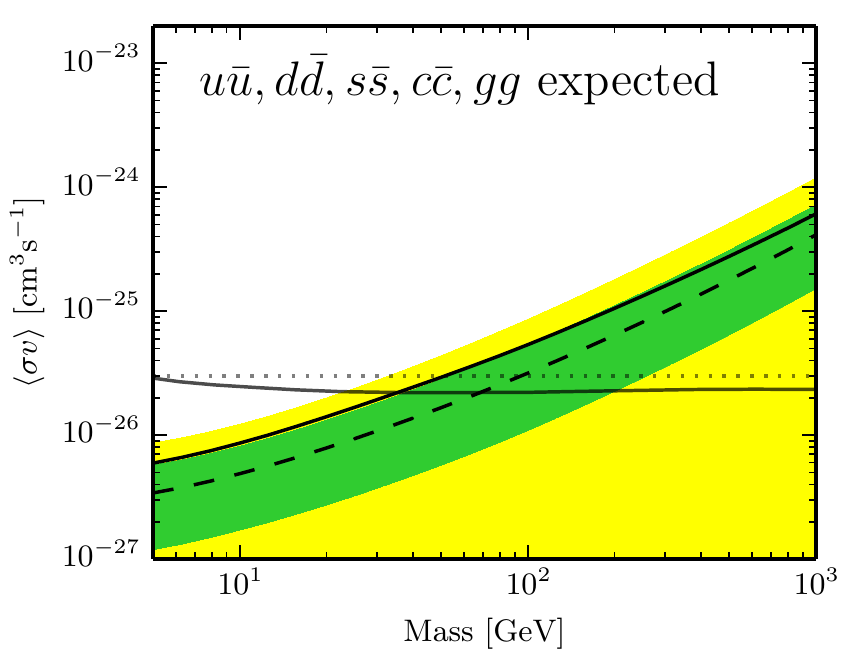}
\includegraphics[scale=0.68]{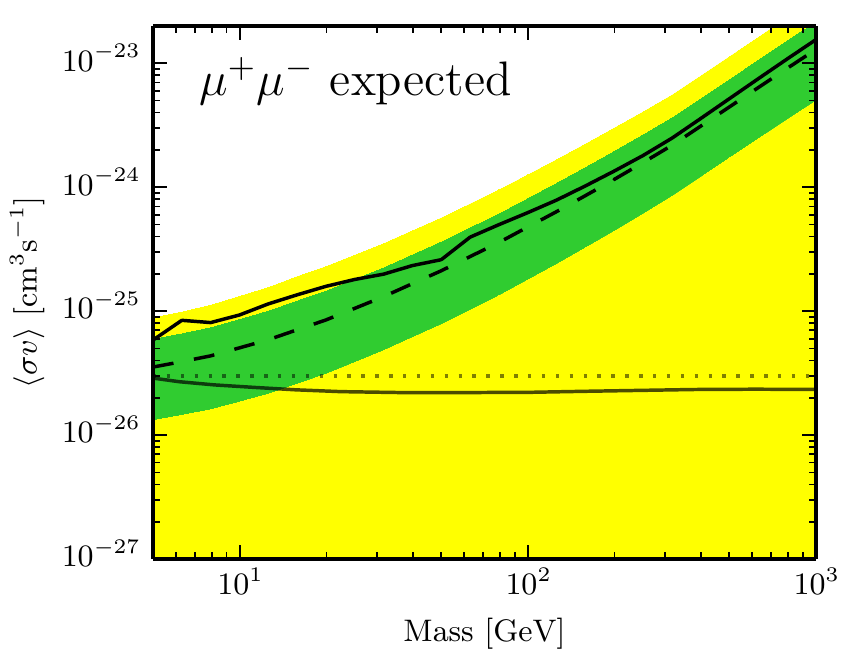}
\includegraphics[scale=0.68]{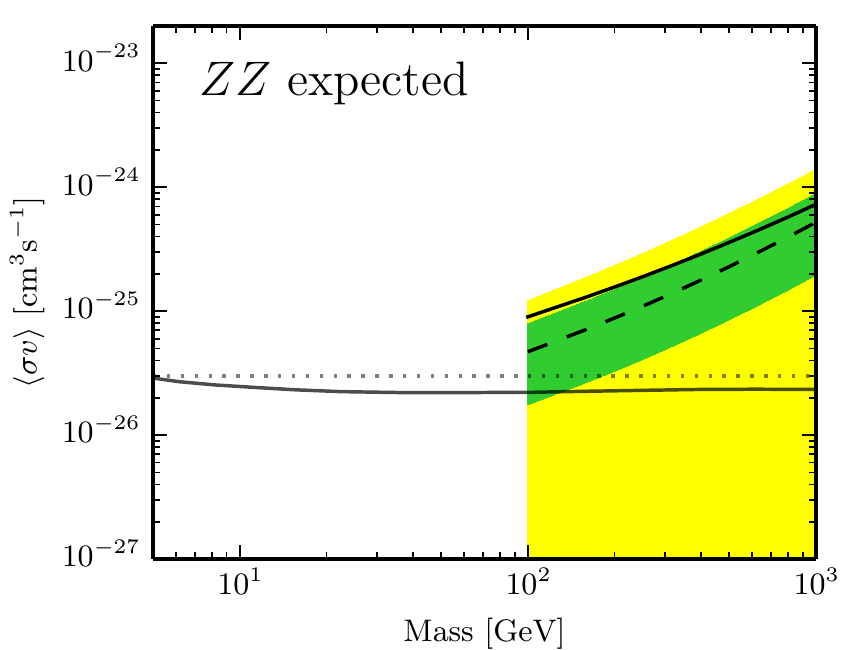} \\
\includegraphics[scale=0.68]{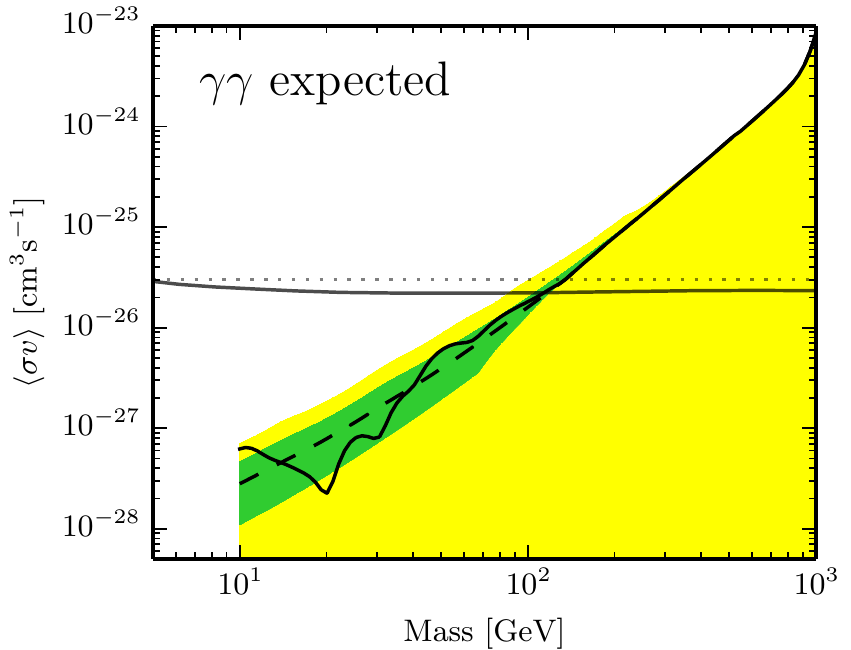} 
\includegraphics[scale=0.68]{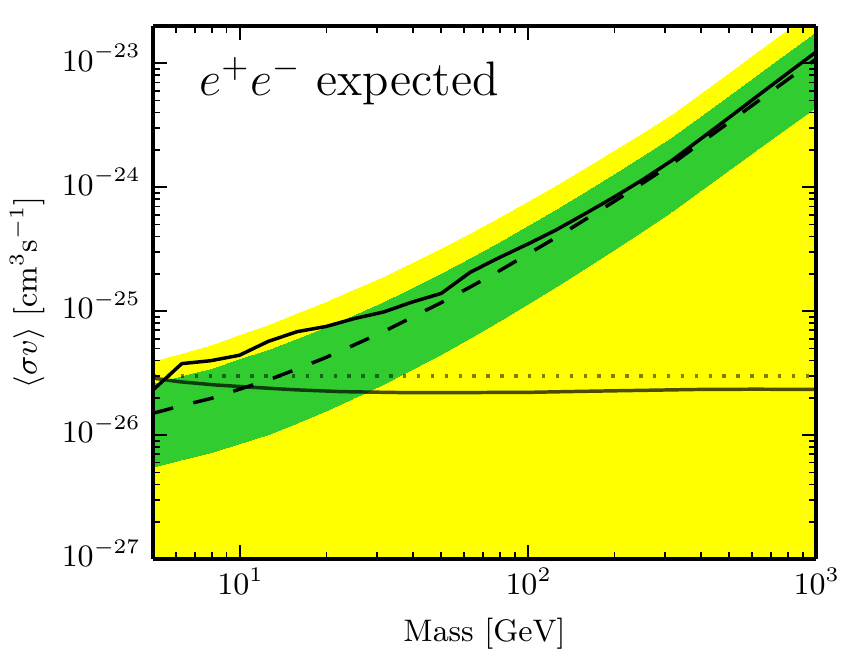}
\includegraphics[scale=0.68]{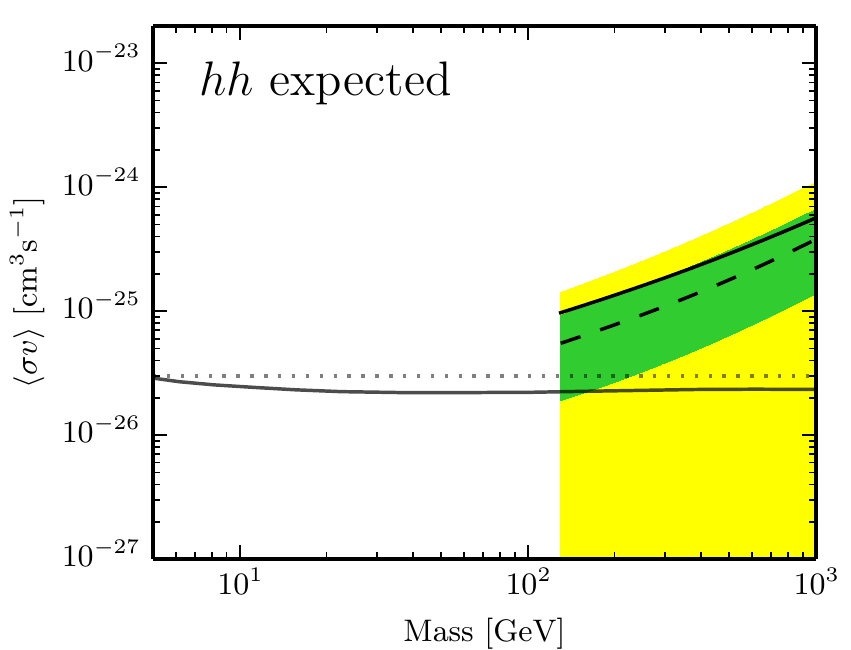} \\
\caption{Expected limits are used to show where the observed limit is likely to be found. There is a 68\% (95\%) of observing the cross section upper limit to lie in the green (yellow) region. The dashed line is the median expected limit (i.e. there is a 50\% chance of observing the limit to be below this line). These ranges are computed for a particular realization of halo density profiles. When sampling different profiles the curves shift up and down but the width of the bands is constant. The solid line is the observed upper limit for this particular realization --- Fig.~\ref{fig:combinedlimits_syst} shows the observed limits including the uncertainty induced by the dwarf density profiles.
\label{fig:combinedlimits_expected}}
\end{figure*}

These statistical fluctuations are shown in Fig.~\ref{fig:combinedlimits_expected}. The figure is produced assuming a single realization of the density profiles for each dwarf (the realization is chosen so that the median expected limit lies close to the median observed limit of Fig.~\ref{fig:combinedlimits_syst}). The dashed line shows the median expected limit. There is an 84\% (16\%) chance of observing an upper limit below the upper (lower) edge of the green band (for the yellow band these probabiilties are 97.7\% and 2.3\%). The green and yellow coloring is inspired by collider searches (see e.g.~\citep[Fig. 8]{2012PhLB..716...30C}) and has the same interpretation: there is 68\% that the observed limit lies in the green band and a 95\% chance the observed limit lies in the yellow band (if the data were generated by backgrounds). We also plot the observed limit (solid line) for this particular realization of density profiles. The yellow band goes to $\sigv=0$ because of our decision to compute 95\% upper limits. There is a strong overlap between the results of the search (Fig.~\ref{fig:searchjoint}) and where the observed limit lies relative to the expected limit.

The lower edge of the yellow band is $\sigv=0$ because of our decision to compute 95\% upper limits. We are assuming the background-only hypothesis $\sigv=0$. When testing this hypothesis there is a 5\% chance that we mistakenly reject it, i.e. draw the upper limit at $\sigv=0$. This is a consequence of the limits having 95\% coverage. It is a manifestation of common occurance in frequentist statistics: negative fluctuations of the background can cause empty confidence intervals.

It is important to note that this ``statistical uncertainty in the limit'' is not an uncertainty in the same sense as that due to the halo profiles. The shaded bands in Fig.~\ref{fig:combinedlimits_expected} quantifies where we would have expected the limits to be found prior to their measurement; or alternatively, how much the observed limits would fluctuate over multiple measurements.

\section{Discussion}
\label{sec:discussion}

\subsection{Comment on an apparent excess
\label{sec:commentonexcess}}
The current work is a generalization of two previous studies~\citep{2011PhRvL.107x1303G,2012PhRvD..86b1302G}. In parallel, the Fermi collaboration has recently published results of a dark matter search in a sample of Milky Way dwarfs~\citep{2014PhRvD..89d2001A}. That work makes use of a likelihood based search, where a many-component background model is fit to a $15\degr$ region surrounding each dwarf. The model consists of known point sources, a diffuse galactic component, and an isotropic flux (modeling extragalactic and unresolved sources as well as instrumental backgrounds). 

In a joint analysis of 15 of the dwarfs an initial interpretation of the likelihood ratio test statistic indicated a signal significance of about $3\sigma$ (corresponding to a TS value of 8.7) for dark matter annihilating into $b\bar{b}$ with a mass between 10 and 25 GeV. The ``excess'' was due primarily to Segue 1, Willman 1, and Ursa Major II, the dwarfs assigned the largest $J$ values in that study. As will be discussed below,~\citep{2014PhRvD..89d2001A} found it necessary to calibrate the detection significance using a post hoc background sampling procedure. 

In this work {\em we find that the observed data is consistent with background for each of the dwarf galaxies individually as well as in a joint analysis}. This difference is perhaps due to the different way that we treat the background compared to the likelihood-based modeling approach. Whereas the Fermi collaboration study assumed that the diffuse background is described by a Poisson process our empirical sampling of the background does not enforce that assumption. To the contrary, we observe that the background is not well described by a Poisson process. In particular, the PDF of the test statistic due to background often has an extended tail to high $T$ values not predicted by a compound Poisson distribution. Thus an observation of a ``large'' $\Tobs$ can be misinterpreted as significant if the background is not understood correctly.

\begin{figure*}
\centering
\subfigure{\includegraphics[scale=.85]{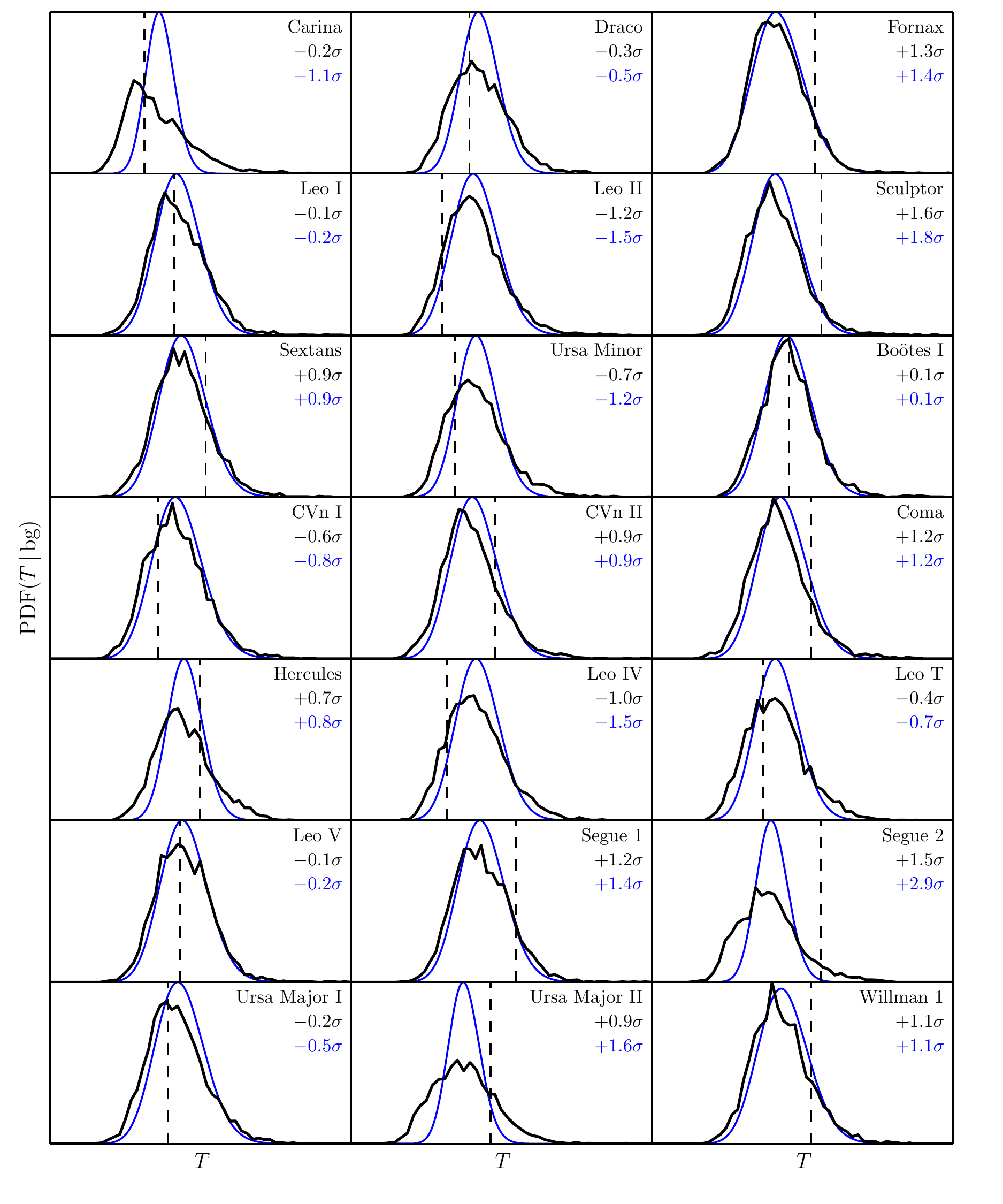}}
\caption{The probability distribution of the test statistic due to background for each dwarf galaxy. The black curve is the empirical background PDF based on field surrounding each dwarf. The blue curve is the PDF assuming that the background is generated by a Poisson process having a spectrum and normalization derived from the surrounding field of view. The vertical dashed line is the observed value of the test statistic in the central ROI surrounding each dwarf. The sigma values shown are the detection significance when the background is modeled empirically (black) or as a Poisson process (blue). The weighting for this search is performed for a 25 GeV particle annihilating into $b$ quarks. This mass and channel correspond to the most significant Fermi detection (10 -- 25 GeV, $b\bar{b}$) in Segue 1, Ursa Major II, and Willman 1.
\label{fig:bgweight_PDFs_25GeV}}
\end{figure*}

This effect can be seen explicitly in Fig.~\ref{fig:bgweight_PDFs_25GeV}. For each dwarf we plot the PDF of the test statistic due to background. The black curve is the empirically derived distribution (Sec.~\ref{sec:bgsampling}) while the blue curve is a compound Poisson distribution based on the observed background energy spectrum (Sec.~\ref{sec:bgmodelforweighting}). In both cases we adopt a weighting function for a search for a 25 GeV particle annihilating into $b$ quarks. The vertical dashed line shows the observed test statistic. The signal significance is shown assuming the two different background PDFs. An assumption of a Poisson background does not describe the actual background in many cases and can lead to a mistakenly large detection significance.

The difficulty in fitting a multi-component Poisson background model is illustrated in Fig.~4 of~\citep{2014PhRvD..89d2001A}. There, ``blank sky locations'' are used to test whether the likelihood ratio test statistic is accurately described by an ``asymptotic'' $\chi^2$ distribution. This sampling of blank sky locations is analogous to the empirical background sampling developed in~\cite{2011PhRvL.107x1303G} and employed in the present work. \citet{2014PhRvD..89d2001A} found that the blank sky PDF of the test statistic deviated from the $\chi^2$ distribution at large values of the test statistic. One of the reasons for the deviation could be that the background model is not flexible enough to describe the true background. \citet{2014arXiv1409.1572C} present evidence that unresolved blazars and radio sources are at least partly responsible for the insufficiency of the background treatment used in~\citep{2014PhRvD..89d2001A}.

The blank sky location sampling of \citet[Fig. 4]{2014PhRvD..89d2001A} reduces the tail probability of a TS = 8.7 observation to a local p-value of $0.13$. This corresponds to a significance of $2.2\sigma$ which can be directly compared to the values shown in our Figs.~\ref{fig:searchindiv_oneplot},~\ref{fig:searchindiv_separate}, and~\ref{fig:searchjoint}. Thus, when calibrating the detection significance using an empirical sampling of the background, the results of \citet{2014PhRvD..89d2001A} are closer in line with what we find. We note that the empirical background sampling is a simple and central component of our framework rather than an additional step used to recalibrate significances derived through a complicated model fitting method.

While the blank sky procedure reduces the maximum local significance in~\citep{2014PhRvD..89d2001A} to 2.2$\sigma$, the largest significance we find in the $b\bar{b}$ channel is still only around 1.5$\sigma$ (see Fig.~\ref{fig:searchjoint}). Note that, in contrast to~\citep{2014PhRvD..89d2001A}, we do not include Willman 1 in the joint search because its $J$-profile cannot be reliably determined. We repeated our search using the same set of 15 dwarfs as in~\citep{2014PhRvD..89d2001A}, treating them as point sources and adopting the $J^{\rm NFW}$ values from Table I of that work. Even in this case, the maximum significance we find is 1.7$\sigma$ (for masses between roughly 10 and 150 GeV). The inclusion of Willman 1 does not seem to be responsible for the discrepancy in detection significance between the two studies.

\subsection{Galactic center}
Over the past few years several groups have used Fermi LAT data to search the Galactic Center region for evidence of dark matter annihilation~\citep{2011PhLB..697..412H,2011PhRvD..84l3005H,2012PhRvD..86h3511A,2013PhRvD..87l9902A,2013PhRvD..88h3521G,2014arXiv1402.4090A,2014arXiv1402.6703D,2014arXiv1406.6948Z,2014arXiv1409.0042C}. These groups report evidence for a spherically symmetric excess of gamma-rays over what is expected from diffuse background processes and point sources. Interpreted as dark matter annihilation, the excess can be best fit with a $\sim 30 \, \GeV$ dark matter particle annihilating into quarks (e.g.~\citep{2014arXiv1402.6703D}, but see also~\citep{2014PhRvD..90d3508L}). Several alternative explanations of the data have been suggested (e.g.~\citep{2011PhLB..705..165B,2012PhRvD..86h3511A,2013PhRvD..87l9902A,2014arXiv1402.4090A,2014PhRvD..90b3015C,2014arXiv1409.0042C,2013PhRvD..88h3521G,2013MNRAS.436.2461M,2014JHEAp...3....1Y}), while~\citet{2014arXiv1406.6027B} conclude that cosmic ray data are already in tension with the dark matter hypothesis. However, since a relative dearth of plausible alternatives have been put forward, it is vital to consider the implications of a dark matter explanation.

It is important to note that while many groups have presented evidence of an excess with concordant dark matter properties {\em all} of the studies use the same data set, namely Fermi LAT observations of the inner galaxy. It is, therefore, of paramount importance to test the dark matter explanation in an independent observation. Gamma-ray observations of dwarf galaxies provide a robust, independent test. If dark matter annihilates in the Galactic Center it also does so in the dwarfs and will generate gamma-rays in a predictable way. The only uncertainty involved is the dark matter distribution within the systems.

\begin{figure}
\centering
\hspace{-0.5cm}
\subfigure{\includegraphics[scale=1]{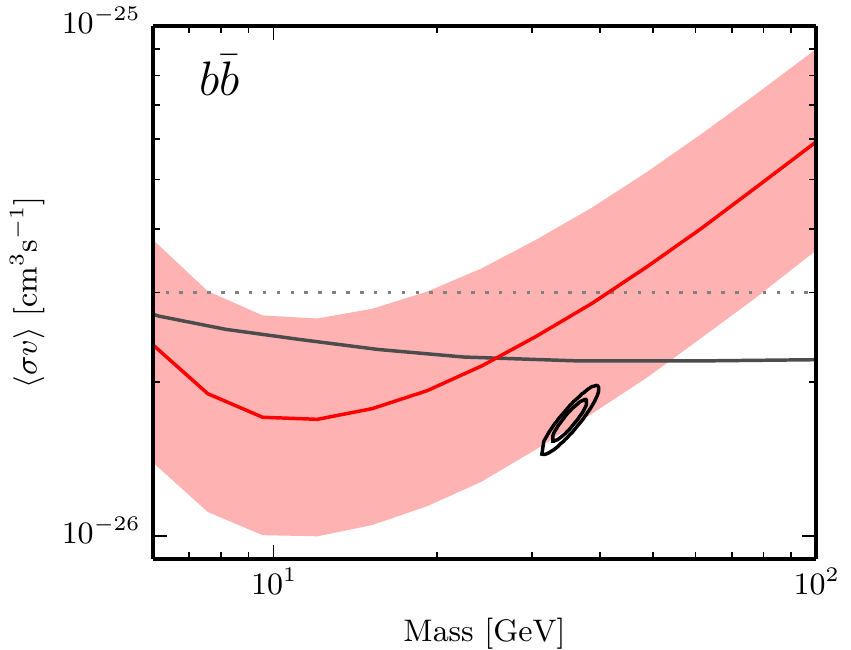}}
\caption{Detail of the 95\% cross section upper limits for annihilation to $b\bar{b}$, comparing with the dark matter interpretation of the Galactic center gamma-ray presented in~\citep{2014arXiv1402.6703D}. As in Fig.~\ref{fig:combinedlimits_syst} the red band represents the range of plausible models for the dark matter distribution in the dwarfs galaxies.
\label{fig:limits_bb_withGCcontours}}
\end{figure}

Figure~\ref{fig:limits_bb_withGCcontours} is a zoom-in of Fig.~\ref{fig:combinedlimits_syst}, showing our observed cross section limits in the mass range suggested by the Galactic Center observations. The black contours are the $1\sigma$ and $2\sigma$ best fitting models from~\citep{2014arXiv1402.6703D, HooperPC} for dark matter annihilating in the Galactic Center (the $b\bar{b}$ channel is shown; the relative positions of the curves for other channels are similar). As before, the red band shows the uncertainty in the limit due to our incomplete knowledge of the dark matter distribution within the dwarfs. 

What this plot indicates is that the dwarfs, despite their larger distance and lower density than the Galactic Center, are able to probe the same dark matter parameter space suggested by the Galactic Center observations. This is achieved by jointly analyzing the data from different dwarfs as well as by constructing a maximally powerful analysis framework which uses all the information available in the data.

While perhaps not satisfying, our current state of knowledge leads us to conclude that observations of the dwarfs {\em may} be in conflict with the dark matter interpretation of the Fermi Galactic Center observations. That is, there are many dark matter density profiles, perfectly consistent with the stellar kinematic data, which exclude the Galactic Center models. At the same time, other allowed density profiles give rise to limits that are too weak to rule out these models.

However, the stellar kinematic data do not allow arbitrarily weak cross sections. As dynamical analyses improve~\citep{2013pss5.book.1039W,2013PhR...531....1S,2013NewAR..57...52B} the systematic band will shrink and our claims about the Galactic Center models will sharpen. Already, they point to the important fact that the dark matter interpretation of the Galactic Center data will be confirmed or ruled out by gamma-ray observations of the Milky Way dwarf population with near-current technology.

\subsection{Systematic uncertainty in dwarf density profiles}
\label{sec:Juncertainty}
In presenting limits on the annihilation cross section we take the approach of separating the statistical uncertainty due to finite photon counts and the systematic uncertainty due to uncertain knowledge of the dwarf halo profiles. This makes clear how our sensitivity to the particle physics depends on what we know about the astrophysics. Our sampling of ``realizations'' of the dwarf $J$-profiles should be thought of as exploring the parameter space of possible dark matter halos that are consistent with the stellar kinematic data. For each realization we compute a statistically rigorous 95\% upper limit on the annihilation cross section. It is difficult, in our opinion, to quote a single upper limit on the cross section which has an unambiguous frequentist statistical interpretation. If, in the future, the range of allowed profiles decreases, the systematic error band will shrink accordingly.

\citet{2011PhRvL.107x1302A,2014PhRvD..89d2001A} incorporate the uncertainty in $J$ as a term in the likelihood function. The idea is essentially to treat the scalar $J$ value as being measured (subject to statistical uncertainty)~\citep{2014arXiv1407.6617C} at the same time as the gamma-rays. In reality, the ``measurement'' of $J$ is actually a Monte Carlo scan of the parameter space describing the halo properties in a Bayesian framework~\citep{0004-637X-801-2-74,2011MNRAS.418.1526C,2013arXiv1309.2641M,2007PhRvD..75h3526S,2008ApJ...678..614S}. The scan gives rise to a posterior probability distribution for $J$ that is often log-normal and is centered on a ``best-fit'' value $\bar{J}$. This posterior distribution is then reinterpreted as the likelihood function for measuring $\bar{J}$ given a true $J$. As a result, a single limit curve is obtained by decreasing the true $J$ values as the cross section is increased until the likelihood decreases by some set amount. 

However, performing this ``inversion'' of the posterior to get the likelihood is not trivial. It does appear to be a simple algebraic manipulation of Bayes' theorem: posterior = likelihood $\times$ prior. If the posterior is log-normal and the prior is flat it seems that likelihood is also log-normal (but with the true $J$ in the denominator of the prefactor, not the ``measured'' $\bar{J}$)~\citep{2014arXiv1407.6617C}. However, the Bayesian analysis that shows the posterior to be log-normal does {\em not} use a flat prior on $J$. In fact it does not explicitly use a prior on $J$ at all. Instead there are priors on the multiple parameters $\alpha, \beta, \dots$ describing the dark matter halo (and in~\citep{2013arXiv1309.2641M,2014PhRvD..89d2001A} additional hyperparameters describing the population of dwarf galaxies). These priors could, in principle, be integrated against a delta function $\delta(J - J(\alpha,\beta,\dots))$ to yield a prior on $J$ but this prior is not necessarily flat. And even if this $J$ prior were found and used to divide the log-normal posterior it would be difficult to interpret the resulting function as the likelihood for $J$ given the ``measurement'' of the number $\bar{J}$. The quantity $\bar{J}$ is an estimator, found by performing the Bayesian analysis of the stellar data, producing a posterior distribution for $\log_{10}J$, approximating it by a normal distribution, and extracting the mean $\overline{\log_{10}J}$ (and standard deviation $\sigma$) from this fitted distribution. The sampling distribution of the estimator $\bar{J}$ (a function of the stellar kinematic data) need not have any relation to the posterior from of the Bayesian analysis.

In the limit of ``large sample sizes'' the central limit theorem shows that the posterior distribution for a parameter approaches a Gaussian centered on the maximum likelihood estimate (MLE) of the parameter with standard deviation $\sigma$~\citep{wasserman2004all}. A second version of the theorem says that the sampling distribution of the MLE is Gaussian, centered on the true value, with the same variance $\sigma$. Such logic may form the basis for using the posterior on $\log_{10}J$ to approximate the sampling distribution of $\overline{\log_{10}J}$. However, the log-normal form for the likelihood is inappropriate and, more importantly, it is far from clear that the stellar kinematic data on the dwarfs constitutes a ``large enough'' sample. The sampling distribution of $\overline{\log_{10}J}$ must be quantified by analysis of simulated sets of stellar observations (e.g.~\citep{gaiachallengewiki,2011ApJ...742...20W,2009MNRAS.395.1079D}).

In this work, we prefer the clean separation of the particle physics uncertainties from the astrophysical uncertainties. This is the most candid way to represent the impact of various uncertainties on the particle physics conclusions. It also separates, as much as possible, the Bayesian scan over dark matter halo parameter space from the rigorous frequentist method used to derive confidence intervals and detection significances. Of course, this separation comes at a cost in the presentation of the results: we cannot quote a single number for a limit or a significance. We show, quantitatively, the impact of the systematic uncertainties on the frequentist statements.

\subsection{Sensitivity scaling}
The framework introduced in this work can be used to derive expected results from future observations or by experiments with different instrumental characteristics.

Equation~\ref{eqn:SNR2} is a good starting point to see how results will scale with, for example, increased observation time. For background-dominated regimes ($s_Q \ll b_Q$) we have $\SNR^2 \sim s_Q^2 / b_Q$, while for small backgrounds we have $\SNR^2 \sim s_Q$.

Increasing the observation time will scale the signal and background by the same factor, while $\sigv$ is a multiplicative factor in $s_Q$. Therefore, cross section limits and sensitivities scale at a rate between $t^{-1}$ and $t^{-1/2}$ depending on whether the observation is signal or background dominated. For Fermi LAT observations, whether we are signal or background-dominated depends on the mass of the dark matter particle and the annihilation channel. For high mass particles and/or hard spectrum channels we approach the signal-dominated regime since the backgrounds fall sharply with increasing energy.

The cosmic ray backgrounds for CLEAN and \mbox{ULTRACLEAN} events are already thought to be subdominant to the extragalactic diffuse gamma-ray emission~\citep[Fig. 28]{2012ApJS..203....4A} so further improvements to cosmic ray discrimination are unlikely to reduce the background substantially. This also indicates that the effective area enters in the same way as observation time and so limits scale with effective area just as with time for instruments which can reject cosmic rays sufficiently well.

The inclusion of additional dwarf galaxies is more unpredictable. In general doubling the number of dwarf galaxies doubles both signal and background. So again, ``number of dwarfs'' scales limits in the same way as observation time. All J values (and backgrounds) being equal, observing one dwarf for eight years or two dwarfs for four yields the same cross section sensitivity.

Of course the $J$-profiles of newly discovered dwarfs are not guaranteed to be similar to those of the known dwarfs. The inclusion of a single nearby object could improve limits substantially. By the same token, the discovery of many distant, low mass dwarfs (e.g. Leo IV and Leo V) will not change the sensitivity to dark matter since they bring in as much background as the promising dwarfs but very little signal.

It is important to note that, in principle, including more dwarfs can only increase the sensitivity. However, this is only true if the proper weighting is used. Adding additional dwarfs in a simple ``stacking'' of the data can actually be detrimental, as can be seen from Eq.~\eqref{eqn:SNR}. If $w_Q = 1$, as in a basic stacking analysis, including dwarfs with lots of background but small expected signal can decrease the signal to noise ratio.

\subsection{The statistical method}

This work (along with~\citep{2011PhRvL.107x1303G}) presents a novel method to treat diffuse and unresolved backgrounds in gamma-ray studies. It can be specifically contrasted with studies where the background takes an assumed form (perhaps with free parameters which are fit). Our procedure has advantages and disadvantages. 

Most importantly, the treatment of the background is model independent. That is, the background is {\em empirically} discovered, not modeled. We do not have to understand the sources of diffuse gamma-rays, or possible misreconstruction of cosmic rays or other instrumental noise. In fact, the empirical sampling makes no use of the instrument response functions at all, save that the exposure is constant across the field of view. We thereby avoid any associated systematics.

The correctness of this method depends on the assumption that processes nearby the dwarf are also at work in the direction of dwarf. However, this assumption can fail if there are actual background gradients across the field of view, if the sources are not masked appropriately, or if effective area changes across field of view.

Specifically, it is difficult to use this sampling procedure in a crowded environment. If there are sources very near to the dwarfs (or even overlapping them) it will be necessary to account for them somehow. There are no tuneable knobs in our background model. Therefore, it is not clear how to include a systematic (e.g. uncertain effective area).

The statistical method, in general, is also designed to be robust. Probability distributions are computed exactly, without relying on asymptotic theorems about the behavior of a test statistic. Therefore, confidence regions have the correct coverage. In contrast it is not clear that the profile likelihood method used in many studies gives confidence regions with proper coverage. It seems that coverage was checked in~\citet{2014PhRvD..89d2001A} using simulated event maps. However, as that study showed, the models used for the simulations do not accurately describe the gamma-ray sky. Do the asymptotic theorems apply if the likelihood model is not flexible enough to include the truth? This is a point that merits further study.

The event weighting method we have introduced is quite general and the ``optimal weighting'' function nicely interpolates between background dominated and signal dominated regimes. That is, if certain parts of ``observation space'' (e.g. certain event energy ranges) are signal dominated they are given a higher weight.

In fact, the weighting can be thought of as a subtle and automatic way of making cuts on the data. For example, consider the ``sliding window technique'' used to search for a gamma-ray line (e.g.~\citep{2007PhRvD..76f3006P,2011arXiv1106.1874B,2012JCAP...07..054B}). In those studies an optimized cut on event energies must be made for each trial line energy. In the line search performed in this work such a window is effectively created by the weight function: events with energies far from the line energy are given a zero weight. A ``cut'' on the data corresponds to a weighting which can only take the values 0 and 1. In this light, the event weighting is seen as a generalized way of optimally cutting the data. We have shown that the ``correct'' (i.e. most powerful) way to do this is by weighting events according the probability that they are due to signal divided by the probability that they are due to background.

The method does not apply simply to dark matter searches or even to gamma-ray observations. It can likely be adapted to many different analyses. The idea is simple: weight different parts of the data according to how likely they are to contain information you care about. Then use the ``total weight'' as a test statistic. It is especially useful for combining observations and can be thought of as a generalization of ``stacking'', which is widely used throughout astronomy. It is currently being applied to an ACT study of dark matter annihilation in dwarfs \cite{VERITASjointdwarfs} but could be simply extended to search for any faint gamma-ray sources. It is an improvement over the commonly used ON/OFF method as it uses the energy and direction information of events and could even be used to weight different data-taking runs according to the varying backgrounds on different nights. At higher energies, experiments like HAWC can likely make use of the event weighting to optimally detect high energy sources of cosmic rays.

\subsection{10 years of ``Pass 8'' data
\label{sec:future}}
We perform a rough estimate of the sensitivity that Fermi can achieve after the completion of the mission. The results presented in this work have used Pass 7 Reprocessed data. Pass 8, currently under development by the Fermi collaboration, is a complete rebuild of the data reduction  (see e.g.~\cite{2013arXiv1303.3514A}). We make use of preliminary estimates of the Pass 8 instrument response functions and assume a mission lifetime of 10 years in order to predict the future sensitivity.

As far as this work is concerned, the most important effect in changing from Pass 7 to Pass 8 will be the increased effective area and observation time. We do not take into account the improvement in point spread function as we cannot model the PSF based only on estimates of its 68\% and 95\% containment angles (see e.g., slide 8 in \cite{Pass8}). For example, the 68\% containment angle may decrease by as much as 20\% at energies above 1 GeV and would be used to lower the background by down-weighting events far from the location of the dwarf. Therefore, including the improved PSF may result in a somewhat stronger gain in sensitivity than what we present.

For the LAT, the effective area $\Aeff(E,\theta)$ is a function of energy and off-axis angle (the angle of the source with respect to the spacecraft ``boresight''). The effective area curve $\Aeff(E)$, a function of energy only, is the livetime-weighted average of $\Aeff(E,\theta)$ and depends on the position of the source and on the spacecraft pointing history. We approximate the effective area curve for Pass 8 by multiplying the Pass 7 curve (for each dwarf position) by the ratio of on-axis effective areas $\Aeff(E,\theta=0)$ for {\tt P8\_SOURCE} vs. {\tt P7REP\_SOURCE}~\citep{2014AAS...22335203R}. Similar results are obtained by using the ratio of acceptances (solid angle-weighted average of $\Aeff(E,\theta)$)~\citep{2013arXiv1303.3514A,2014AAS...22325603G}. Note that we only include events with $E > 1\,\GeV$. We write ``Pass 8'' in quotes to indicate we are using an unofficial approximation to the Pass 8 instrument response functions.
 
 The new exposure $\epsilon(E)$ used in Eq.~\eqref{eqn:expectedcounts} is the current Pass 7 exposure multiplied by the (energy-dependent) ratio of on-axis effective areas and by the increase in observation time: $10\,\yr / 5.8\,\yr = 1.7$. The background spectrum used in the weighting Eq.~\eqref{eqn:bQestimate} is multiplied by the same ratio (i.e. the background, being mostly gamma-rays, will scale with effective area and time). The background sampling described in Sec.~\ref{sec:bgsampling} no longer describes the ``Pass 8'' background. Instead we use a compound Poisson distribution where the expected number of background events is given by $b_Q$ in Eq.~\eqref{eqn:bQestimate}.

\begin{figure*}
\includegraphics[scale=.9]{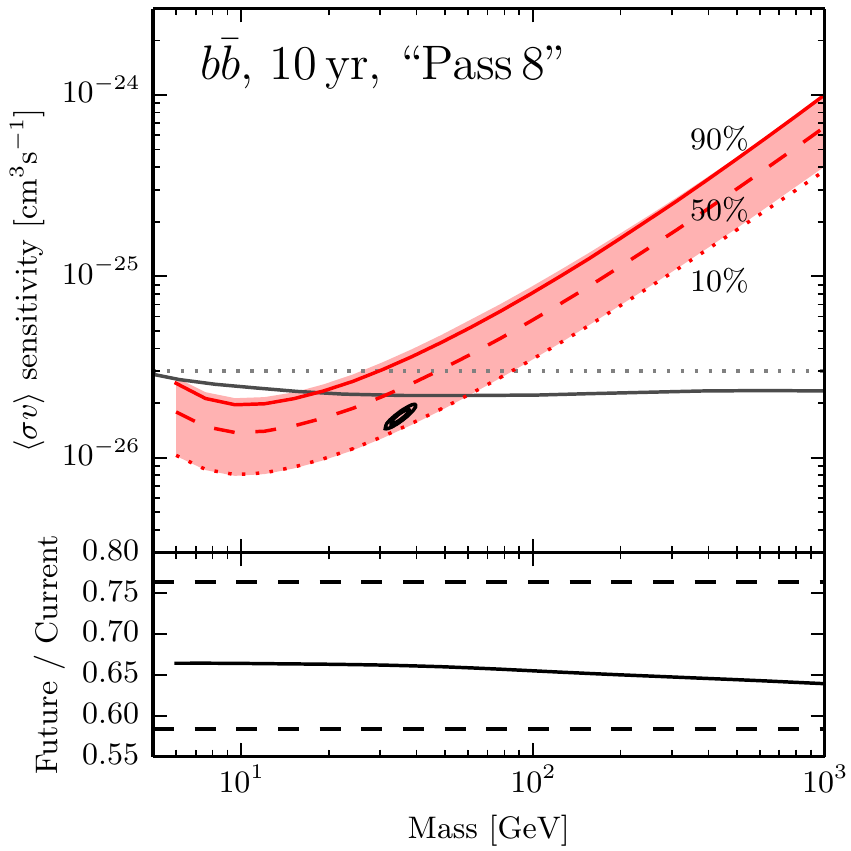}
\hspace{1cm}
\includegraphics[scale=.9]{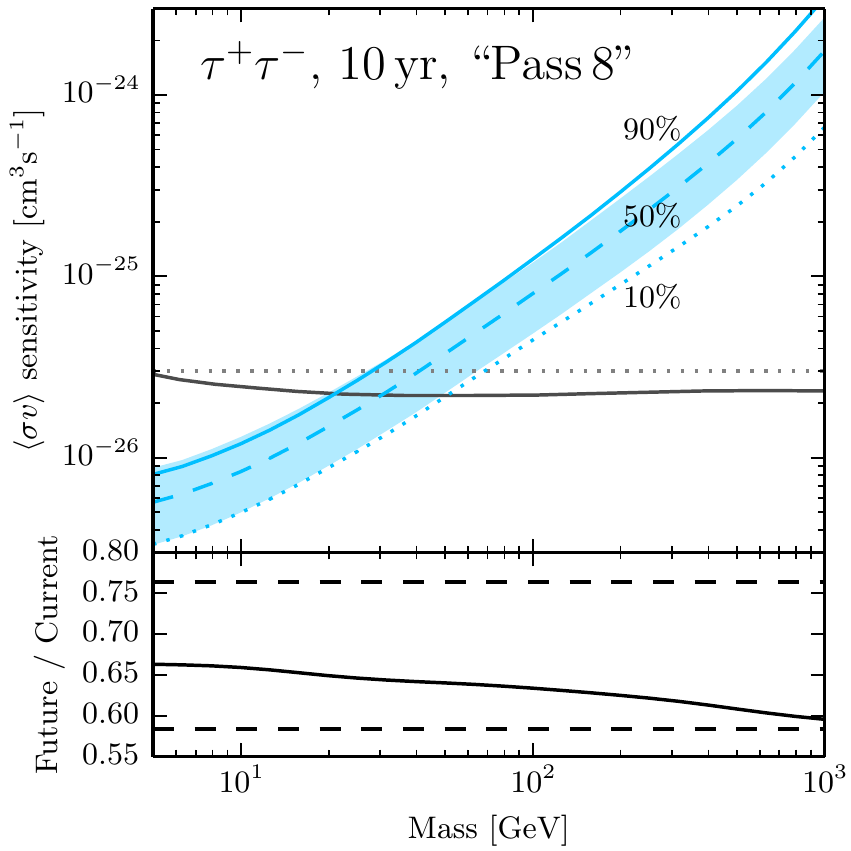}
\caption{Future sensitivity to dark matter annihilation in dwarf galaxies over a 10 year mission lifetime and using an approximation to the Pass 8 instrument response functions. The solid, dashed, and dotted lines show the cross sections for which there is a 90\%, 50\%, and 10\% chance of making a $3\sigma$ detection. The shaded band represents the uncertainty in the dark matter density profiles in the dwarfs. The band is centered on the 50\% curve and the systematic uncertainty shifts all three curves together. The lower panels show the cross section sensitivity of future observations divided by the sensitivity of current observations. The dashed lines show the ratio of current vs. Pass 8 observation times (lower line) and the square root of this ratio (upper line).
\label{fig:futuresensitivity}}
\end{figure*}

 For each realization of halo density profiles we find the sensitivity $\sigv_{90}$ with the current Pass 7 data and with 10 years of ``Pass 8'', assuming a compound Poisson background in both cases. The ratio of these two quantities gives the expected increase in sensitivity. We then find $\sigv_{90}$ for the current data using the correct empirical background. Scaling this latter sensitivity by the ``Pass 8''/Pass 7 ratio gives an estimate of the expected sensitivity for 10 years of Pass 8 observations. We are also interested in the cross section for which there is, say, a 50\% or 10\% chance of detection. To this end we noticed that the ratio between such a cross section and $\sigv_{90}$ does not strongly depend on the observation time, which effective areas are used, nor on which halo parameters are chosen (the relative variations in the cross section ratios are at the percent level when changing observation time, effective area, etc.). Therefore, we compute, for each mass and using a fiducial set of halo parameters, the cross section required for a 50\% and 10\% chance of detection for the current observations using the empirical sampled background. We use these to scale the ``Pass 8'' $\sigv_{90}$ values to estimate the cross sections for which there is a 50\% and 10\% chance of making a $3\sigma$ detection after 10 years with Pass 8 data.

Figure~\ref{fig:futuresensitivity} shows the results of this exercise for dark matter annihilation into $b$ quarks (left panel) and $\tau$ leptons (right panel). The solid, dashed, and dotted curves are the cross sections for which there is a 90\%, 50\%, and 10\% chance of making a $3\sigma$ detection with 10 years of ``Pass 8'' data. The lines may be interpreted simply: dark matter with a cross section above the 90\% curve is likely to be discovered; if the cross section is below the 10\% curve discovery in the dwarfs is unlikely. As in previous figures, the shaded band represents the systematic uncertainty in the dark matter density profiles in the dwarfs. That is, all three sensitivity curves may be shifted (together) up and down within the shaded region, depending on the actual density profiles. The band is shown centered on the 50\% curve. (It is a coincidence that the band's width is so similar to the distance between the 90\% and 10\% curves.) The horizontal lines show the relic abundance cross section (solid for the~\citet{2012PhRvD..86b3506S} value, dotted for the ``canonical'' $3\times 10^{-26}\, \cm^3{\rm s}^{-1}$). For the $b\bar{b}$ channel we also plot the $1\sigma$ and $2\sigma$ contours from~\citep{2014arXiv1402.6703D,HooperPC}, corresponding to a dark matter interpretation of Fermi observations of the Galactic Center. 

The lower panels of Fig.~\ref{fig:futuresensitivity} show the ratio between the cross section Fermi will be sensitive to in the future and the one it is sensitive to today (i.e. in the search performed in this work). The horizontal dashed lines show what might be expected from a naive scaling with time, either proportional to $t$ (lower line) or to $\sqrt{t}$ upper line. We see that there is a mass dependent improvement in sensitivity. For high mass dark matter the search is more signal dominated and the sensitivity scales more favorably with time.

Note that all of these projections are based on the current sample of 20 dwarf galaxies and our current understanding of the density profiles in these systems. The discovery of new, nearby dwarfs can only make the search more sensitive (see, for example,~\citep{2013arXiv1309.4780H}). A better understanding of the density profiles in the current sample of dwarfs will shrink the systematic bands.

\section{Conclusions}

We have developed and applied a new statistical framework for analyzing multiple datasets to search for dark matter annihilation. The method performs maximally powerful tests (in the frequentist sense) by weighting individual detected events based on their spatial and spectral properties. The weighting is influenced by knowledge of the instrument response, the particle physics properties affecting the expected signal, and an empirical determination of the background. The framework is general and likely can be applied to other studies searching for weak signals in noisy data and extracting particle physics constraints from multiple sets of data.

Using Fermi LAT gamma-ray observations, we have applied the method to search for dark matter annihilation in a joint analysis of 20 Milky Way dwarf galaxies, the complete population for which the dark matter distribution is constrained by stellar kinematics. No evidence for emission beyond background is found in any individual dwarf, nor in the combined sample. We therefore set upper limits on the annihilation cross section for various channels, carefully taking into account the systematic uncertainty in the dwarf density profiles. 
We find that for quarks and heavy leptons, the annihilation cross section upper limit is less than the relic abundance value ($2.2 \times 10^{-26}\,\cm^3\second^{-1}$) for dark matter masses less than a few tens of GeV, while for annihilation to a two-photon final state the cross section limit is around $10^{-27}\,\cm^3\second^{-1}$ in this mass range. 

These results show the importance of dwarf galaxies in searching for a dark matter signal from the sky. Current observations, combined with new methods of analysis, have reached the sensitivity to probe long-sought dark matter particle models. Such approaches are perhaps the most robust complements to the search for dark matter in the Galactic center. We therefore expect that future gamma-ray observations of dwarf galaxies will be crucial in discovering, confirming, or ruling out dark matter at the weak scale.

\acknowledgements
We acknowledge useful conversations with Gordon Blackadder, Jessi Cisewski, Ian Dell'Antonio, David van Dyk, Richard Gaitskell,  Deivid Ribeiro, Roberto Trotta, Larry Wasserman, Benjamin Zitzer, and Ying Zu. AGS was partially supported from a NASA-RI EPSCoR-RID grant NNX11AR21. SMK is supported by DOE DE-SC0010010, NSF PHYS-1417505, and NASA NNX13AO94G. MGW is supported by NSF grants AST-1313045 and AST-1412999. SMK thanks the Aspen Center for Physics and the Banff International Research Station for Mathematical Innovation and Discovery for hospitality where part of this work was completed.

%%%%%%%%%%%%%%%%%%% APPENDIX %%%%%%%%%%%%%%%%%%%%%%
\appendix
\section{Computation of PDFs}
\label{app:pdfs}

\subsection{Compound Poisson distributions}
We give a short derivation of the compound Poisson distribution function Eq.~\eqref{eqn:CPft}. The use of such distributions was introduced into astrophysics as ``$P(D)$ analysis'' \cite{1957PCPS...53..764S} (see also \cite{2009JCAP...07..007L} and references within).

Consider the sum over $N$ ``events'', where each event $i$ carries a weight $W_i$:
\begin{equation}
T = \sum\limits_{i=1}^N W_i.
\label{eqn:CPrandvar}
\end{equation}
Each random variable $W_i$ is drawn from the same probability distribution $f(w)$. The number of terms in the sum $N$ is a  random variable drawn from a discrete distribution: $q_j$ (for $j=0,1,\dots$) is the probability that $N=j$. We wish to find the PDF of the quantity $T$: $p(t)dt$ is the probability that $T$ has a value between $t$ and $t+dt$.

The PDF of the sum of $j$ independent random variables is the $j$-fold convolution of the PDFs of the individual variables. Therefore, the PDF of $T$ can be written as
\begin{equation}
p(t) = \sum\limits_{j=0}^\infty q_j f^{\ast j} (t).
\label{eqn:CPpdf}
\end{equation}
The quantity $f^{\ast j}$ is the PDF of the sum of exactly $j$ variables, each distributed according to $f$. In words, each term in the sum in Eq.~\eqref{eqn:CPpdf} represents the probability that the sum in Eq.~\eqref{eqn:CPrandvar} has $j$ terms and that the $j$ terms add up to $t$. Note that $f^{\ast 1}(t) = f(t)$ and that $f^{\ast 0}(t) = \delta(t)$, the Dirac delta function: i.e. if $N=0$ then the sum $T$ must be 0. Rigorously, the $j$-fold convolution $f^{\ast j}$ can be defined recursively by $f^{\ast j}(x) = \int f^{\ast (j-1)}(y) f(x-y)dy$ with $f^{\ast 0}(x) = \delta(x)$.  

We now take the Fourier transform $\mathcal{F}$ of both sides of Eq.~\eqref{eqn:CPpdf}:
\begin{equation}
\left[\mathcal{F}(p)\right](k) = \sum\limits_{j=0}^\infty q_j \left[\mathcal{F} \left(f^{\ast j}\right) \right] (k).
\end{equation}

Next, we use the fact that the Fourier transform of a convolution is the product of the Fourier transforms. Letting the symbols $\phi_T$ and $\phi_W$ denote the Fourier transforms of $p(t)$ and $f(w)$ (i.e. $\phi_T(k) \equiv \left[\mathcal{F}(p)\right](k)$ and $\phi_W(k) \equiv \left[\mathcal{F}(f)\right](k)$), this property is written as
\begin{equation}
\left[\mathcal{F} \left(f^{\ast j}\right) \right] (k) = \left[ \phi_W (k) \right]^j.
\end{equation}
Note that the $j=0$ term retains its meaning as the empty sum, $N=0$, since $\left[\phi_W(k)\right]^0 =1$ is the Fourier transform of the delta function. We can now write down the Fourier transform $\phi_T$ of the PDF of $T$:
\begin{equation}
\phi_T(k) = \sum\limits_{j=0}^\infty q_j \left[ \phi_W (k) \right]^j = G_N\left[ \phi_W(k) \right].
\label{eqn:CPftgeneral}
\end{equation}
The last equality introduces the probability-generating function $G_N$ of the discrete random variable $N$:
\begin{equation}
G_N(x) =  \sum\limits_{j=0}^\infty q_j x^j = \Ex\left[x^N \right].
\label{eqn:GN}
\end{equation}
If $N$ is a Poisson variable with mean $\mu$ the $q_j$ are given by
\begin{equation}
q_j = \mathrm{e}^{-\mu}\frac{\mu^j}{j!},
\label{eqn:poissonpmf}
\end{equation}
and the probability-generating function Eq.~\eqref{eqn:GN} is therefore
\begin{equation}
\begin{aligned}
G_N(x) &= \sum\limits_{j=0}^\infty \left( \mathrm{e}^{-\mu}\frac{\mu^j}{j!} \right) x^j,  \\
             &= \mathrm{e}^{-\mu} \sum\limits_{j=0}^\infty \frac{\left(\mu x \right)^j}{j!}, \\
             &= \mathrm{e}^{\mu \left(x-1\right)}.
\end{aligned}
\end{equation}
Inserting this result into Eq.~\eqref{eqn:CPftgeneral} gives the Fourier transform of the compound Poisson variable $T$:
\begin{equation}
\phi_T(k) = \exp \left[ \mu \left( \phi_W(k) -1 \right) \right].
\end{equation}

\subsection{Numerical evaluation}
The closed form of the Fourier transform of the compound Poisson distribution facilitates its numerical evaluation and manipulation. First, note that in our situation $\phi_T$ must still be convolved with the probability distribution governing the component of $T$ due to background Eq.~\eqref{eqn:Tdefsb}. This convolution is most easily done in Fourier space where $\phi_T$ is simply multiplied with the Fourier transform of the PDF of $T$ due to background.

In the current work Fourier transforms are carried out numerically by means of Fast Fourier Transforms (FFTs). An FFT is a discrete version of the Fourier transform that acts on a finite array. We must deal with two properties of FFTs that distinguish them from continuous Fourier transforms: discreteness and end effects.

\subsubsection{Discreteness}
The PDF of the single-event weight $f(w)$ is a continuous function. In order to apply the FFT (to find $\phi_W$) the PDF must be suitably discretized (or arithmetized) onto equally-spaced grid points. To do this, distributions are considered as being sequences of delta function spikes localized at the $M$ points $\{i \Delta \mid i=0,\dots,M-1\}$. That is, the weight is allowed to take the discrete values $0, \Delta, 2\Delta, \dots, (M-1)\Delta$. There is no unique way to assign an amplitude to the spike at $i\Delta$. Following \cite{Embrechts:2009ly} it suffices to consider three methods to arithmetize the function $f(w)$:
\begin{equation}
\begin{aligned}
f^{\rm mid}_i &\equiv \int\limits_{(i - \frac{1}{2})\Delta}^{(i + \frac{1}{2})\Delta} f(w) dw, \\
f^{\rm low}_i &\equiv \int\limits_{i \Delta}^{(i + 1)\Delta} f(w) dw, \\
f^{\rm high}_i &\equiv \int\limits_{(i-1)\Delta}^{i\Delta} f(w) dw. \\
\end{aligned}
\end{equation}
We choose $\Delta$ to be small enough so that results do not depend on the choice of arithmetization.

\subsubsection{Wrap-around effects}
Two arithemtized functions may be convolved using FFTs: the FFT of the convolution is the product of the FFTs of the individual functions. However, the convolution performed in this way is a circular convolution (see e.g. \cite[Ch.~13.1]{Press:2007:NRE:1403886}). If the distributions have tails which do not go to zero quickly enough this circular convolution induces a wrap-around error (also called an aliasing error)~\citep{Embrechts:2009ly}. 

There are two ways to mitigate this issue. \citet{grubel1999computation} suggest a tilting procedure. In this method the two functions to be convolved are initially multiplied by $\exp(-Mx)$, which dampens the tails substantially (the recommended $M$ is such that the functions are damped by a factor of $\exp(-20)$ for the largest values of $x$~\cite{Embrechts:2009ly}). The FFTs of the two functions are found, multiplied together, and followed by an inverse FFT. The tilting is undone by multiplying the resulting function by $\exp(Mx)$. It is straightforward to show that the tilting ``commutes'' with the convolution.

We have found that as long as we describe the PDFs with long enough arrays (i.e. compute the PDFs out to large enough values of test statistic) the aliasing does not affect the results and the tilting procedure is unnecessary. In our implementation, the PDFs are stored in vectors with lengths corresponding to a range from $T=0$ out to (at least) 10 standard deviations above the mean, where the mean and standard deviation are those for the final, convolved distribution. The PDFs become extremely small at such large $T$ and the wrap-around effect is negligible.

\section{Convolution of $J$-profiles with the point spread function}
\label{app:JconvPSF}
Computing the expected annihilation signal at a particular angular separation from the dwarf requires convolving the intrinsic emission ($J$-profile, Eq.~\ref{eqn:Jdef}) with Fermi's point spread function (see Eq.~\ref{eqn:expectedcounts}). Both the $J$-profile and PSF are localized over a small enough angle that the spherical sky can be approximated as flat. The convolution is most efficiently done in Fourier space: the $J$-profile and PSF are Fourier transformed (in 2 dimensions), multiplied, and the result is inverse transformed.

Because both the $J$-profile and Fermi's PSF are modeled as radially symmetric functions the problem is essentially one dimensional. The 2-d Fourier transform $F(\vec{k})$ of a radially symmetric function $f(\vec{x}) = f(r)$ depends only on the magnitude $k$ of the Fourier mode $\vec{k}$. In polar coordinates the $\theta$ integral can be done first:
\begin{align}
F(\vec{k}) &= \int f(\vec{x}) e^{-i \vec{k} \cdot \vec{x}} d^2 \vec{x} \nonumber \\
                  &= \int_0^\infty \int_0^{2\pi} f(r,\theta) e^{-i kr \cos\theta} r dr d\theta \nonumber \\
                  &= \int_0^\infty 2\pi r f(r)  J_0(kr) dr, \label{eqn:hankeltransform}
\end{align}
where $J_0$ is the 0-th order Bessel function of the first kind. The integral in Eq.~\eqref{eqn:hankeltransform} is (up to a factor of $2\pi$) the Hankel transform of the function $f(r)$. The same calculation shows that the inverse Fourier transformation is another Hankel transform:
\begin{equation}
f(r) = \frac{1}{2\pi} \int_0^\infty k F(k)  J_0(kr) dk, \label{eqn:invhankeltransform}
\end{equation}

To convolve the $J$-profile with the PSF we take the Hankel transform of the $J$-profile and of the PSF. The transforms are multiplied and then a another Hankel transform is applied to return the convolution $(J \ast \PSF)(r)$. This is repeated for each energy (the PSF being energy dependent).

The Hankel transforms are numerically performed using an implementation of Hamilton's FFTLog algorithm~\citep{2000MNRAS.312..257H,FFTlog}. The algorithm operates in log-space --- i.e. the function $f(r)$ is sampled at log-spaced values of $r$. This is beneficial since the $J$-profile has power law behavior in $r$. FFTLog treats the sampled values of $f(r)$ as defining a function that is periodic in log-space. Specifically, the function $g(r) = f(r)  r^{1-q}$ obeys the property $g(r e^L) = g(r)$ for some period $L$ and ``bias'' $q$. There is freedom to chose the edge $r=r_0$ and the length $L$ of the periodic interval --- i.e. so that $g(r)$ is defined between $r_0$ and $r_0 e^L$. The bias $q$ can be adjusted to try to make the function $g(r)$ go to 0 at large and small $r$, though in practice we found $q=0$ to be most suitable. We took $r_0=10^{-4}\deg$, $r_0 e^L = 5000\deg$, and the number of sample points (size of the FFT) to be $N=2^{14}$. To take into account the diverging $J$-profile at angles less than $r_0$ we computed the integral of the $J$-profile within $10^{-3}\deg$ (called $J_{\rm core}$). The $J$-profile is then set to a constant within $10^{-3} \deg$ before the convolution is performed. Afterwards we add a term $J_{\rm core} \PSF$ to the $J\ast\PSF$ function.

After multiplying the Hankel transforms of $J$ and PSF and taking the inverse, the resulting $J \ast \PSF$ is clipped to positive real values and smoothed using a moving average (with window length 21).

FFTLog is numerically efficient --- requiring just two FFTs to perform the Hankel transform in Eq.~\ref{eqn:hankeltransform} --- and simply implemented in Python. An alternative method for performing the convolution is described by \citet{2012ApJ...756....5L}. There, the radial part of the 2-d convolution integral can be performed analytically because of the specific functional form chosen to parameterize the LAT PSF. This leaves a single numerical integral to be performed at each energy. Using FFTLog has the benefit of performing all integrals using FFTs and works with generic radially-symmetric PSFs (even those not defined by a functional form). It may be useful in speeding up general searches for extended sources with Fermi.

%%%%%%%%%%%%%%%%%%%%%%%%%%%%%%%%%%%%%%%%%%%%%%
\vspace{2in}

% Bibliography
\bibliography{manuscript}

\begin{thebibliography}{119}
\expandafter\ifx\csname natexlab\endcsname\relax\def\natexlab#1{#1}\fi
\expandafter\ifx\csname bibnamefont\endcsname\relax
  \def\bibnamefont#1{#1}\fi
\expandafter\ifx\csname bibfnamefont\endcsname\relax
  \def\bibfnamefont#1{#1}\fi
\expandafter\ifx\csname citenamefont\endcsname\relax
  \def\citenamefont#1{#1}\fi
\expandafter\ifx\csname url\endcsname\relax
  \def\url#1{\texttt{#1}}\fi
\expandafter\ifx\csname urlprefix\endcsname\relax\def\urlprefix{URL }\fi
\providecommand{\bibinfo}[2]{#2}
\providecommand{\eprint}[2][]{\url{#2}}

\bibitem[{\citenamefont{{Gunn} et~al.}(1978)\citenamefont{{Gunn}, {Lee},
  {Lerche}, {Schramm}, and {Steigman}}}]{1978ApJ...223.1015G}
\bibinfo{author}{\bibfnamefont{J.~E.} \bibnamefont{{Gunn}}},
  \bibinfo{author}{\bibfnamefont{B.~W.} \bibnamefont{{Lee}}},
  \bibinfo{author}{\bibfnamefont{I.}~\bibnamefont{{Lerche}}},
  \bibinfo{author}{\bibfnamefont{D.~N.} \bibnamefont{{Schramm}}},
  \bibnamefont{and}
  \bibinfo{author}{\bibfnamefont{G.}~\bibnamefont{{Steigman}}},
  \bibinfo{journal}{\apj} \textbf{\bibinfo{volume}{223}}, \bibinfo{pages}{1015}
  (\bibinfo{year}{1978}).

\bibitem[{\citenamefont{{Steigman} et~al.}(1978)\citenamefont{{Steigman},
  {Sarazin}, {Quintana}, and {Faulkner}}}]{1978AJ.....83.1050S}
\bibinfo{author}{\bibfnamefont{G.}~\bibnamefont{{Steigman}}},
  \bibinfo{author}{\bibfnamefont{C.~L.} \bibnamefont{{Sarazin}}},
  \bibinfo{author}{\bibfnamefont{H.}~\bibnamefont{{Quintana}}},
  \bibnamefont{and}
  \bibinfo{author}{\bibfnamefont{J.}~\bibnamefont{{Faulkner}}},
  \bibinfo{journal}{\aj} \textbf{\bibinfo{volume}{83}}, \bibinfo{pages}{1050}
  (\bibinfo{year}{1978}).

\bibitem[{\citenamefont{{Steigman}}(1979)}]{1979ARNPS..29..313S}
\bibinfo{author}{\bibfnamefont{G.}~\bibnamefont{{Steigman}}},
  \bibinfo{journal}{Annual Review of Nuclear and Particle Science}
  \textbf{\bibinfo{volume}{29}}, \bibinfo{pages}{313} (\bibinfo{year}{1979}).

\bibitem[{\citenamefont{{Pagels} and {Primack}}(1982)}]{1982PhRvL..48..223P}
\bibinfo{author}{\bibfnamefont{H.}~\bibnamefont{{Pagels}}} \bibnamefont{and}
  \bibinfo{author}{\bibfnamefont{J.~R.} \bibnamefont{{Primack}}},
  \bibinfo{journal}{Physical Review Letters} \textbf{\bibinfo{volume}{48}},
  \bibinfo{pages}{223} (\bibinfo{year}{1982}).

\bibitem[{\citenamefont{{Alcock} et~al.}(2000)\citenamefont{{Alcock},
  {Allsman}, {Alves}, {Axelrod}, {Becker}, {Bennett}, {Cook}, and
  et~al.}}]{2000ApJ...542..281A}
\bibinfo{author}{\bibfnamefont{C.}~\bibnamefont{{Alcock}}},
  \bibinfo{author}{\bibfnamefont{R.~A.} \bibnamefont{{Allsman}}},
  \bibinfo{author}{\bibfnamefont{D.~R.} \bibnamefont{{Alves}}},
  \bibinfo{author}{\bibfnamefont{T.~S.} \bibnamefont{{Axelrod}}},
  \bibinfo{author}{\bibfnamefont{A.~C.} \bibnamefont{{Becker}}},
  \bibinfo{author}{\bibfnamefont{D.~P.} \bibnamefont{{Bennett}}},
  \bibinfo{author}{\bibfnamefont{K.~H.} \bibnamefont{{Cook}}},
  \bibnamefont{and} \bibinfo{author}{\bibnamefont{et~al.}},
  \bibinfo{journal}{\apj} \textbf{\bibinfo{volume}{542}}, \bibinfo{pages}{281}
  (\bibinfo{year}{2000}), \eprint{arXiv:astro-ph/0001272}.

\bibitem[{\citenamefont{Zel'dovich}(1965)}]{zel1965advances}
\bibinfo{author}{\bibfnamefont{Y.~B.} \bibnamefont{Zel'dovich}},
  \emph{\bibinfo{title}{Advances in Astronomy and Astrophysics}}
  (\bibinfo{publisher}{Academic Press}, \bibinfo{year}{1965}),
  vol.~\bibinfo{volume}{3}, p. \bibinfo{pages}{242}.

\bibitem[{\citenamefont{{Zeldovic} et~al.}(1965)\citenamefont{{Zeldovic},
  {Okun}, and {Pikelner}}}]{1965PhL....17..164Z}
\bibinfo{author}{\bibfnamefont{Y.~B.} \bibnamefont{{Zeldovic}}},
  \bibinfo{author}{\bibfnamefont{L.~B.} \bibnamefont{{Okun}}},
  \bibnamefont{and} \bibinfo{author}{\bibfnamefont{S.~B.}
  \bibnamefont{{Pikelner}}}, \bibinfo{journal}{Physics Letters}
  \textbf{\bibinfo{volume}{17}}, \bibinfo{pages}{164} (\bibinfo{year}{1965}).

\bibitem[{\citenamefont{{Chiu}}(1966)}]{1966PhRvL..17..712C}
\bibinfo{author}{\bibfnamefont{H.-Y.} \bibnamefont{{Chiu}}},
  \bibinfo{journal}{Physical Review Letters} \textbf{\bibinfo{volume}{17}},
  \bibinfo{pages}{712} (\bibinfo{year}{1966}).

\bibitem[{\citenamefont{{Lee} and {Weinberg}}(1977)}]{1977PhRvL..39..165L}
\bibinfo{author}{\bibfnamefont{B.~W.} \bibnamefont{{Lee}}} \bibnamefont{and}
  \bibinfo{author}{\bibfnamefont{S.}~\bibnamefont{{Weinberg}}},
  \bibinfo{journal}{Phys. Rev. Lett.} \textbf{\bibinfo{volume}{39}},
  \bibinfo{pages}{165} (\bibinfo{year}{1977}).

\bibitem[{\citenamefont{{Hut}}(1977)}]{1977PhLA...69...85H}
\bibinfo{author}{\bibfnamefont{P.}~\bibnamefont{{Hut}}},
  \bibinfo{journal}{Physics Letters A} \textbf{\bibinfo{volume}{69}},
  \bibinfo{pages}{85} (\bibinfo{year}{1977}).

\bibitem[{\citenamefont{{Wolfram}}(1979)}]{1979PhLB...82...65W}
\bibinfo{author}{\bibfnamefont{S.}~\bibnamefont{{Wolfram}}},
  \bibinfo{journal}{Physics Letters B} \textbf{\bibinfo{volume}{82}},
  \bibinfo{pages}{65} (\bibinfo{year}{1979}).

\bibitem[{\citenamefont{{Bernstein} et~al.}(1985)\citenamefont{{Bernstein},
  {Brown}, and {Feinberg}}}]{1985PhRvD..32.3261B}
\bibinfo{author}{\bibfnamefont{J.}~\bibnamefont{{Bernstein}}},
  \bibinfo{author}{\bibfnamefont{L.~S.} \bibnamefont{{Brown}}},
  \bibnamefont{and}
  \bibinfo{author}{\bibfnamefont{G.}~\bibnamefont{{Feinberg}}},
  \bibinfo{journal}{\prd} \textbf{\bibinfo{volume}{32}}, \bibinfo{pages}{3261}
  (\bibinfo{year}{1985}).

\bibitem[{\citenamefont{{Scherrer} and {Turner}}(1986)}]{1986PhRvD..33.1585S}
\bibinfo{author}{\bibfnamefont{R.~J.} \bibnamefont{{Scherrer}}}
  \bibnamefont{and} \bibinfo{author}{\bibfnamefont{M.~S.}
  \bibnamefont{{Turner}}}, \bibinfo{journal}{\prd}
  \textbf{\bibinfo{volume}{33}}, \bibinfo{pages}{1585} (\bibinfo{year}{1986}).

\bibitem[{\citenamefont{{Srednicki} et~al.}(1988)\citenamefont{{Srednicki},
  {Watkins}, and {Olive}}}]{1988NuPhB.310..693S}
\bibinfo{author}{\bibfnamefont{M.}~\bibnamefont{{Srednicki}}},
  \bibinfo{author}{\bibfnamefont{R.}~\bibnamefont{{Watkins}}},
  \bibnamefont{and} \bibinfo{author}{\bibfnamefont{K.~A.}
  \bibnamefont{{Olive}}}, \bibinfo{journal}{Nuclear Physics B}
  \textbf{\bibinfo{volume}{310}}, \bibinfo{pages}{693} (\bibinfo{year}{1988}).

\bibitem[{\citenamefont{{Gondolo} and {Gelmini}}(1991)}]{1991NuPhB.360..145G}
\bibinfo{author}{\bibfnamefont{P.}~\bibnamefont{{Gondolo}}} \bibnamefont{and}
  \bibinfo{author}{\bibfnamefont{G.}~\bibnamefont{{Gelmini}}},
  \bibinfo{journal}{Nuclear Physics B} \textbf{\bibinfo{volume}{360}},
  \bibinfo{pages}{145} (\bibinfo{year}{1991}).

\bibitem[{\citenamefont{{Georgi} et~al.}(1974)\citenamefont{{Georgi}, {Quinn},
  and {Weinberg}}}]{1974PhRvL..33..451G}
\bibinfo{author}{\bibfnamefont{H.}~\bibnamefont{{Georgi}}},
  \bibinfo{author}{\bibfnamefont{H.~R.} \bibnamefont{{Quinn}}},
  \bibnamefont{and}
  \bibinfo{author}{\bibfnamefont{S.}~\bibnamefont{{Weinberg}}},
  \bibinfo{journal}{Physical Review Letters} \textbf{\bibinfo{volume}{33}},
  \bibinfo{pages}{451} (\bibinfo{year}{1974}).

\bibitem[{\citenamefont{{Dimopoulos} et~al.}(1981)\citenamefont{{Dimopoulos},
  {Raby}, and {Wilczek}}}]{1981PhRvD..24.1681D}
\bibinfo{author}{\bibfnamefont{S.}~\bibnamefont{{Dimopoulos}}},
  \bibinfo{author}{\bibfnamefont{S.}~\bibnamefont{{Raby}}}, \bibnamefont{and}
  \bibinfo{author}{\bibfnamefont{F.}~\bibnamefont{{Wilczek}}},
  \bibinfo{journal}{\prd} \textbf{\bibinfo{volume}{24}}, \bibinfo{pages}{1681}
  (\bibinfo{year}{1981}).

\bibitem[{\citenamefont{{Steigman} et~al.}(2012)\citenamefont{{Steigman},
  {Dasgupta}, and {Beacom}}}]{2012PhRvD..86b3506S}
\bibinfo{author}{\bibfnamefont{G.}~\bibnamefont{{Steigman}}},
  \bibinfo{author}{\bibfnamefont{B.}~\bibnamefont{{Dasgupta}}},
  \bibnamefont{and} \bibinfo{author}{\bibfnamefont{J.~F.}
  \bibnamefont{{Beacom}}}, \bibinfo{journal}{\prd}
  \textbf{\bibinfo{volume}{86}}, \bibinfo{eid}{023506} (\bibinfo{year}{2012}),
  \eprint{1204.3622}.

\bibitem[{\citenamefont{{Griest} and
  {Kamionkowski}}(1990)}]{1990PhRvL..64..615G}
\bibinfo{author}{\bibfnamefont{K.}~\bibnamefont{{Griest}}} \bibnamefont{and}
  \bibinfo{author}{\bibfnamefont{M.}~\bibnamefont{{Kamionkowski}}},
  \bibinfo{journal}{Physical Review Letters} \textbf{\bibinfo{volume}{64}},
  \bibinfo{pages}{615} (\bibinfo{year}{1990}).

\bibitem[{\citenamefont{{Hui}}(2001)}]{2001PhRvL..86.3467H}
\bibinfo{author}{\bibfnamefont{L.}~\bibnamefont{{Hui}}},
  \bibinfo{journal}{Physical Review Letters} \textbf{\bibinfo{volume}{86}},
  \bibinfo{pages}{3467} (\bibinfo{year}{2001}), \eprint{astro-ph/0102349}.

\bibitem[{\citenamefont{{Kaplinghat} et~al.}(2000)\citenamefont{{Kaplinghat},
  {Knox}, and {Turner}}}]{2000PhRvL..85.3335K}
\bibinfo{author}{\bibfnamefont{M.}~\bibnamefont{{Kaplinghat}}},
  \bibinfo{author}{\bibfnamefont{L.}~\bibnamefont{{Knox}}}, \bibnamefont{and}
  \bibinfo{author}{\bibfnamefont{M.~S.} \bibnamefont{{Turner}}},
  \bibinfo{journal}{Physical Review Letters} \textbf{\bibinfo{volume}{85}},
  \bibinfo{pages}{3335} (\bibinfo{year}{2000}), \eprint{astro-ph/0005210}.

\bibitem[{\citenamefont{{Beacom} et~al.}(2007)\citenamefont{{Beacom}, {Bell},
  and {Mack}}}]{2007PhRvL..99w1301B}
\bibinfo{author}{\bibfnamefont{J.~F.} \bibnamefont{{Beacom}}},
  \bibinfo{author}{\bibfnamefont{N.~F.} \bibnamefont{{Bell}}},
  \bibnamefont{and} \bibinfo{author}{\bibfnamefont{G.~D.}
  \bibnamefont{{Mack}}}, \bibinfo{journal}{Physical Review Letters}
  \textbf{\bibinfo{volume}{99}}, \bibinfo{pages}{231301}
  (\bibinfo{year}{2007}), \eprint{arXiv:astro-ph/0608090}.

\bibitem[{\citenamefont{{Gondolo} and {Silk}}(1999)}]{1999PhRvL..83.1719G}
\bibinfo{author}{\bibfnamefont{P.}~\bibnamefont{{Gondolo}}} \bibnamefont{and}
  \bibinfo{author}{\bibfnamefont{J.}~\bibnamefont{{Silk}}},
  \bibinfo{journal}{Physical Review Letters} \textbf{\bibinfo{volume}{83}},
  \bibinfo{pages}{1719} (\bibinfo{year}{1999}), \eprint{astro-ph/9906391}.

\bibitem[{\citenamefont{{Ullio} et~al.}(2001)\citenamefont{{Ullio}, {Zhao}, and
  {Kamionkowski}}}]{2001PhRvD..64d3504U}
\bibinfo{author}{\bibfnamefont{P.}~\bibnamefont{{Ullio}}},
  \bibinfo{author}{\bibfnamefont{H.}~\bibnamefont{{Zhao}}}, \bibnamefont{and}
  \bibinfo{author}{\bibfnamefont{M.}~\bibnamefont{{Kamionkowski}}},
  \bibinfo{journal}{\prd} \textbf{\bibinfo{volume}{64}}, \bibinfo{eid}{043504}
  (\bibinfo{year}{2001}), \eprint{astro-ph/0101481}.

\bibitem[{\citenamefont{{Merritt} et~al.}(2002)\citenamefont{{Merritt},
  {Milosavljevi{\'c}}, {Verde}, and {Jimenez}}}]{2002PhRvL..88s1301M}
\bibinfo{author}{\bibfnamefont{D.}~\bibnamefont{{Merritt}}},
  \bibinfo{author}{\bibfnamefont{M.}~\bibnamefont{{Milosavljevi{\'c}}}},
  \bibinfo{author}{\bibfnamefont{L.}~\bibnamefont{{Verde}}}, \bibnamefont{and}
  \bibinfo{author}{\bibfnamefont{R.}~\bibnamefont{{Jimenez}}},
  \bibinfo{journal}{Physical Review Letters} \textbf{\bibinfo{volume}{88}},
  \bibinfo{eid}{191301} (\bibinfo{year}{2002}), \eprint{astro-ph/0201376}.

\bibitem[{\citenamefont{{Merritt}}(2004)}]{2004PhRvL..92t1304M}
\bibinfo{author}{\bibfnamefont{D.}~\bibnamefont{{Merritt}}},
  \bibinfo{journal}{Physical Review Letters} \textbf{\bibinfo{volume}{92}},
  \bibinfo{eid}{201304} (\bibinfo{year}{2004}), \eprint{astro-ph/0311594}.

\bibitem[{\citenamefont{{Fields} et~al.}(2014)\citenamefont{{Fields},
  {Shapiro}, and {Shelton}}}]{2014arXiv1406.4856F}
\bibinfo{author}{\bibfnamefont{B.~D.} \bibnamefont{{Fields}}},
  \bibinfo{author}{\bibfnamefont{S.~L.} \bibnamefont{{Shapiro}}},
  \bibnamefont{and}
  \bibinfo{author}{\bibfnamefont{J.}~\bibnamefont{{Shelton}}},
  \bibinfo{journal}{ArXiv e-prints}  (\bibinfo{year}{2014}),
  \eprint{1406.4856}.

\bibitem[{\citenamefont{{Lacroix}
  et~al.}(2014{\natexlab{a}})\citenamefont{{Lacroix}, {B{\o}hm}, and
  {Silk}}}]{2014PhRvD..89f3534L}
\bibinfo{author}{\bibfnamefont{T.}~\bibnamefont{{Lacroix}}},
  \bibinfo{author}{\bibfnamefont{C.}~\bibnamefont{{B{\o}hm}}},
  \bibnamefont{and} \bibinfo{author}{\bibfnamefont{J.}~\bibnamefont{{Silk}}},
  \bibinfo{journal}{\prd} \textbf{\bibinfo{volume}{89}}, \bibinfo{eid}{063534}
  (\bibinfo{year}{2014}{\natexlab{a}}), \eprint{1311.0139}.

\bibitem[{\citenamefont{{Hooper} and {Goodenough}}(2011)}]{2011PhLB..697..412H}
\bibinfo{author}{\bibfnamefont{D.}~\bibnamefont{{Hooper}}} \bibnamefont{and}
  \bibinfo{author}{\bibfnamefont{L.}~\bibnamefont{{Goodenough}}},
  \bibinfo{journal}{Physics Letters B} \textbf{\bibinfo{volume}{697}},
  \bibinfo{pages}{412} (\bibinfo{year}{2011}), \eprint{1010.2752}.

\bibitem[{\citenamefont{{Hooper} and {Linden}}(2011)}]{2011PhRvD..84l3005H}
\bibinfo{author}{\bibfnamefont{D.}~\bibnamefont{{Hooper}}} \bibnamefont{and}
  \bibinfo{author}{\bibfnamefont{T.}~\bibnamefont{{Linden}}},
  \bibinfo{journal}{\prd} \textbf{\bibinfo{volume}{84}}, \bibinfo{eid}{123005}
  (\bibinfo{year}{2011}), \eprint{1110.0006}.

\bibitem[{\citenamefont{{Abazajian} and
  {Kaplinghat}}(2012)}]{2012PhRvD..86h3511A}
\bibinfo{author}{\bibfnamefont{K.~N.} \bibnamefont{{Abazajian}}}
  \bibnamefont{and}
  \bibinfo{author}{\bibfnamefont{M.}~\bibnamefont{{Kaplinghat}}},
  \bibinfo{journal}{\prd} \textbf{\bibinfo{volume}{86}}, \bibinfo{eid}{083511}
  (\bibinfo{year}{2012}), \eprint{1207.6047}.

\bibitem[{\citenamefont{{Abazajian} and
  {Kaplinghat}}(2013)}]{2013PhRvD..87l9902A}
\bibinfo{author}{\bibfnamefont{K.~N.} \bibnamefont{{Abazajian}}}
  \bibnamefont{and}
  \bibinfo{author}{\bibfnamefont{M.}~\bibnamefont{{Kaplinghat}}},
  \bibinfo{journal}{\prd} \textbf{\bibinfo{volume}{87}}, \bibinfo{eid}{129902}
  (\bibinfo{year}{2013}).

\bibitem[{\citenamefont{{Gordon} and
  {Mac{\'{\i}}as}}(2013)}]{2013PhRvD..88h3521G}
\bibinfo{author}{\bibfnamefont{C.}~\bibnamefont{{Gordon}}} \bibnamefont{and}
  \bibinfo{author}{\bibfnamefont{O.}~\bibnamefont{{Mac{\'{\i}}as}}},
  \bibinfo{journal}{\prd} \textbf{\bibinfo{volume}{88}}, \bibinfo{eid}{083521}
  (\bibinfo{year}{2013}), \eprint{1306.5725}.

\bibitem[{\citenamefont{{Abazajian} et~al.}(2014)\citenamefont{{Abazajian},
  {Canac}, {Horiuchi}, and {Kaplinghat}}}]{2014arXiv1402.4090A}
\bibinfo{author}{\bibfnamefont{K.~N.} \bibnamefont{{Abazajian}}},
  \bibinfo{author}{\bibfnamefont{N.}~\bibnamefont{{Canac}}},
  \bibinfo{author}{\bibfnamefont{S.}~\bibnamefont{{Horiuchi}}},
  \bibnamefont{and}
  \bibinfo{author}{\bibfnamefont{M.}~\bibnamefont{{Kaplinghat}}},
  \bibinfo{journal}{ArXiv e-prints}  (\bibinfo{year}{2014}),
  \eprint{1402.4090}.

\bibitem[{\citenamefont{{Daylan} et~al.}(2014)\citenamefont{{Daylan},
  {Finkbeiner}, {Hooper}, {Linden}, {Portillo}, {Rodd}, and
  {Slatyer}}}]{2014arXiv1402.6703D}
\bibinfo{author}{\bibfnamefont{T.}~\bibnamefont{{Daylan}}},
  \bibinfo{author}{\bibfnamefont{D.~P.} \bibnamefont{{Finkbeiner}}},
  \bibinfo{author}{\bibfnamefont{D.}~\bibnamefont{{Hooper}}},
  \bibinfo{author}{\bibfnamefont{T.}~\bibnamefont{{Linden}}},
  \bibinfo{author}{\bibfnamefont{S.~K.~N.} \bibnamefont{{Portillo}}},
  \bibinfo{author}{\bibfnamefont{N.~L.} \bibnamefont{{Rodd}}},
  \bibnamefont{and} \bibinfo{author}{\bibfnamefont{T.~R.}
  \bibnamefont{{Slatyer}}}, \bibinfo{journal}{ArXiv e-prints}
  (\bibinfo{year}{2014}), \eprint{1402.6703}.

\bibitem[{\citenamefont{{Zhou} et~al.}(2014)\citenamefont{{Zhou}, {Liang},
  {Huang}, {Li}, {Fan}, {Feng}, and {Chang}}}]{2014arXiv1406.6948Z}
\bibinfo{author}{\bibfnamefont{B.}~\bibnamefont{{Zhou}}},
  \bibinfo{author}{\bibfnamefont{Y.-F.} \bibnamefont{{Liang}}},
  \bibinfo{author}{\bibfnamefont{X.}~\bibnamefont{{Huang}}},
  \bibinfo{author}{\bibfnamefont{X.}~\bibnamefont{{Li}}},
  \bibinfo{author}{\bibfnamefont{Y.-Z.} \bibnamefont{{Fan}}},
  \bibinfo{author}{\bibfnamefont{L.}~\bibnamefont{{Feng}}}, \bibnamefont{and}
  \bibinfo{author}{\bibfnamefont{J.}~\bibnamefont{{Chang}}},
  \bibinfo{journal}{ArXiv e-prints}  (\bibinfo{year}{2014}),
  \eprint{1406.6948}.

\bibitem[{\citenamefont{{Calore} et~al.}(2014)\citenamefont{{Calore}, {Cholis},
  and {Weniger}}}]{2014arXiv1409.0042C}
\bibinfo{author}{\bibfnamefont{F.}~\bibnamefont{{Calore}}},
  \bibinfo{author}{\bibfnamefont{I.}~\bibnamefont{{Cholis}}}, \bibnamefont{and}
  \bibinfo{author}{\bibfnamefont{C.}~\bibnamefont{{Weniger}}},
  \bibinfo{journal}{ArXiv e-prints}  (\bibinfo{year}{2014}),
  \eprint{1409.0042}.

\bibitem[{\citenamefont{{Lake}}(1990)}]{1990Natur.346...39L}
\bibinfo{author}{\bibfnamefont{G.}~\bibnamefont{{Lake}}},
  \bibinfo{journal}{\nat} \textbf{\bibinfo{volume}{346}}, \bibinfo{pages}{39}
  (\bibinfo{year}{1990}).

\bibitem[{\citenamefont{{Baltz} and {Wai}}(2004)}]{2004PhRvD..70b3512B}
\bibinfo{author}{\bibfnamefont{E.~A.} \bibnamefont{{Baltz}}} \bibnamefont{and}
  \bibinfo{author}{\bibfnamefont{L.}~\bibnamefont{{Wai}}},
  \bibinfo{journal}{\prd} \textbf{\bibinfo{volume}{70}}, \bibinfo{eid}{023512}
  (\bibinfo{year}{2004}), \eprint{astro-ph/0403528}.

\bibitem[{\citenamefont{{Bergstr{\"o}m} and
  {Hooper}}(2006)}]{2006PhRvD..73f3510B}
\bibinfo{author}{\bibfnamefont{L.}~\bibnamefont{{Bergstr{\"o}m}}}
  \bibnamefont{and} \bibinfo{author}{\bibfnamefont{D.}~\bibnamefont{{Hooper}}},
  \bibinfo{journal}{\prd} \textbf{\bibinfo{volume}{73}}, \bibinfo{eid}{063510}
  (\bibinfo{year}{2006}), \eprint{hep-ph/0512317}.

\bibitem[{\citenamefont{{Colafrancesco}
  et~al.}(2007)\citenamefont{{Colafrancesco}, {Profumo}, and
  {Ullio}}}]{2007PhRvD..75b3513C}
\bibinfo{author}{\bibfnamefont{S.}~\bibnamefont{{Colafrancesco}}},
  \bibinfo{author}{\bibfnamefont{S.}~\bibnamefont{{Profumo}}},
  \bibnamefont{and} \bibinfo{author}{\bibfnamefont{P.}~\bibnamefont{{Ullio}}},
  \bibinfo{journal}{\prd} \textbf{\bibinfo{volume}{75}}, \bibinfo{eid}{023513}
  (\bibinfo{year}{2007}), \eprint{astro-ph/0607073}.

\bibitem[{\citenamefont{{Profumo} and
  {Kamionkowski}}(2006)}]{2006JCAP...03..003P}
\bibinfo{author}{\bibfnamefont{S.}~\bibnamefont{{Profumo}}} \bibnamefont{and}
  \bibinfo{author}{\bibfnamefont{M.}~\bibnamefont{{Kamionkowski}}},
  \bibinfo{journal}{\jcap} \textbf{\bibinfo{volume}{3}}, \bibinfo{eid}{003}
  (\bibinfo{year}{2006}), \eprint{astro-ph/0601249}.

\bibitem[{\citenamefont{{Strigari} et~al.}(2007)\citenamefont{{Strigari},
  {Koushiappas}, {Bullock}, and {Kaplinghat}}}]{2007PhRvD..75h3526S}
\bibinfo{author}{\bibfnamefont{L.~E.} \bibnamefont{{Strigari}}},
  \bibinfo{author}{\bibfnamefont{S.~M.} \bibnamefont{{Koushiappas}}},
  \bibinfo{author}{\bibfnamefont{J.~S.} \bibnamefont{{Bullock}}},
  \bibnamefont{and}
  \bibinfo{author}{\bibfnamefont{M.}~\bibnamefont{{Kaplinghat}}},
  \bibinfo{journal}{\prd} \textbf{\bibinfo{volume}{75}},
  \bibinfo{pages}{083526} (\bibinfo{year}{2007}),
  \eprint{arXiv:astro-ph/0611925}.

\bibitem[{\citenamefont{{Scott} et~al.}(2010)\citenamefont{{Scott}, {Conrad},
  {Edsj{\"o}}, {Bergstr{\"o}m}, {Farnier}, and {Akrami}}}]{2010JCAP...01..031S}
\bibinfo{author}{\bibfnamefont{P.}~\bibnamefont{{Scott}}},
  \bibinfo{author}{\bibfnamefont{J.}~\bibnamefont{{Conrad}}},
  \bibinfo{author}{\bibfnamefont{J.}~\bibnamefont{{Edsj{\"o}}}},
  \bibinfo{author}{\bibfnamefont{L.}~\bibnamefont{{Bergstr{\"o}m}}},
  \bibinfo{author}{\bibfnamefont{C.}~\bibnamefont{{Farnier}}},
  \bibnamefont{and} \bibinfo{author}{\bibfnamefont{Y.}~\bibnamefont{{Akrami}}},
  \bibinfo{journal}{\jcap} \textbf{\bibinfo{volume}{1}}, \bibinfo{pages}{31}
  (\bibinfo{year}{2010}), \eprint{0909.3300}.

\bibitem[{\citenamefont{{Strigari} et~al.}(2008)\citenamefont{{Strigari},
  {Koushiappas}, {Bullock}, {Kaplinghat}, {Simon}, {Geha}, and
  {Willman}}}]{2008ApJ...678..614S}
\bibinfo{author}{\bibfnamefont{L.~E.} \bibnamefont{{Strigari}}},
  \bibinfo{author}{\bibfnamefont{S.~M.} \bibnamefont{{Koushiappas}}},
  \bibinfo{author}{\bibfnamefont{J.~S.} \bibnamefont{{Bullock}}},
  \bibinfo{author}{\bibfnamefont{M.}~\bibnamefont{{Kaplinghat}}},
  \bibinfo{author}{\bibfnamefont{J.~D.} \bibnamefont{{Simon}}},
  \bibinfo{author}{\bibfnamefont{M.}~\bibnamefont{{Geha}}}, \bibnamefont{and}
  \bibinfo{author}{\bibfnamefont{B.}~\bibnamefont{{Willman}}},
  \bibinfo{journal}{\apj} \textbf{\bibinfo{volume}{678}}, \bibinfo{pages}{614}
  (\bibinfo{year}{2008}), \eprint{0709.1510}.

\bibitem[{\citenamefont{{Wood} et~al.}(2008)\citenamefont{{Wood}, {Blaylock},
  {Bradbury}, {Buckley}, {Byrum}, {Chow}, {Cui}, {de la Calle Perez},
  {Falcone}, {Fegan} et~al.}}]{2008ApJ...678..594W}
\bibinfo{author}{\bibfnamefont{M.}~\bibnamefont{{Wood}}},
  \bibinfo{author}{\bibfnamefont{G.}~\bibnamefont{{Blaylock}}},
  \bibinfo{author}{\bibfnamefont{S.~M.} \bibnamefont{{Bradbury}}},
  \bibinfo{author}{\bibfnamefont{J.~H.} \bibnamefont{{Buckley}}},
  \bibinfo{author}{\bibfnamefont{K.~L.} \bibnamefont{{Byrum}}},
  \bibinfo{author}{\bibfnamefont{Y.~C.~K.} \bibnamefont{{Chow}}},
  \bibinfo{author}{\bibfnamefont{W.}~\bibnamefont{{Cui}}},
  \bibinfo{author}{\bibfnamefont{I.}~\bibnamefont{{de la Calle Perez}}},
  \bibinfo{author}{\bibfnamefont{A.~D.} \bibnamefont{{Falcone}}},
  \bibinfo{author}{\bibfnamefont{S.~J.} \bibnamefont{{Fegan}}},
  \bibnamefont{et~al.}, \bibinfo{journal}{\apj} \textbf{\bibinfo{volume}{678}},
  \bibinfo{pages}{594} (\bibinfo{year}{2008}), \eprint{0801.1708}.

\bibitem[{\citenamefont{{H.E.S.S.~Collaboration}
  et~al.}(2011)\citenamefont{{H.E.S.S.~Collaboration}, {Abramowski}, {Acero},
  {Aharonian}, {Akhperjanian}, {Anton}, {Barnacka}, {Barres de Almeida},
  {Bazer-Bachi}, {Becherini} et~al.}}]{2011APh....34..608H}
\bibinfo{author}{\bibnamefont{{H.E.S.S.~Collaboration}}},
  \bibinfo{author}{\bibfnamefont{A.}~\bibnamefont{{Abramowski}}},
  \bibinfo{author}{\bibfnamefont{F.}~\bibnamefont{{Acero}}},
  \bibinfo{author}{\bibfnamefont{F.}~\bibnamefont{{Aharonian}}},
  \bibinfo{author}{\bibfnamefont{A.~G.} \bibnamefont{{Akhperjanian}}},
  \bibinfo{author}{\bibfnamefont{G.}~\bibnamefont{{Anton}}},
  \bibinfo{author}{\bibfnamefont{A.}~\bibnamefont{{Barnacka}}},
  \bibinfo{author}{\bibfnamefont{U.}~\bibnamefont{{Barres de Almeida}}},
  \bibinfo{author}{\bibfnamefont{A.~R.} \bibnamefont{{Bazer-Bachi}}},
  \bibinfo{author}{\bibfnamefont{Y.}~\bibnamefont{{Becherini}}},
  \bibnamefont{et~al.}, \bibinfo{journal}{Astroparticle Physics}
  \textbf{\bibinfo{volume}{34}}, \bibinfo{pages}{608} (\bibinfo{year}{2011}),
  \eprint{1012.5602}.

\bibitem[{\citenamefont{{Geringer-Sameth} and
  {Koushiappas}}(2011)}]{2011PhRvL.107x1303G}
\bibinfo{author}{\bibfnamefont{A.}~\bibnamefont{{Geringer-Sameth}}}
  \bibnamefont{and} \bibinfo{author}{\bibfnamefont{S.~M.}
  \bibnamefont{{Koushiappas}}}, \bibinfo{journal}{Physical Review Letters}
  \textbf{\bibinfo{volume}{107}}, \bibinfo{eid}{241303} (\bibinfo{year}{2011}),
  \eprint{1108.2914}.

\bibitem[{\citenamefont{{Ackermann} et~al.}(2011)\citenamefont{{Ackermann},
  {Ajello}, {Albert}, {Atwood}, {Baldini}, {Ballet}, {Barbiellini}, {Bastieri},
  {Bechtol}, {Bellazzini} et~al.}}]{2011PhRvL.107x1302A}
\bibinfo{author}{\bibfnamefont{M.}~\bibnamefont{{Ackermann}}},
  \bibinfo{author}{\bibfnamefont{M.}~\bibnamefont{{Ajello}}},
  \bibinfo{author}{\bibfnamefont{A.}~\bibnamefont{{Albert}}},
  \bibinfo{author}{\bibfnamefont{W.~B.} \bibnamefont{{Atwood}}},
  \bibinfo{author}{\bibfnamefont{L.}~\bibnamefont{{Baldini}}},
  \bibinfo{author}{\bibfnamefont{J.}~\bibnamefont{{Ballet}}},
  \bibinfo{author}{\bibfnamefont{G.}~\bibnamefont{{Barbiellini}}},
  \bibinfo{author}{\bibfnamefont{D.}~\bibnamefont{{Bastieri}}},
  \bibinfo{author}{\bibfnamefont{K.}~\bibnamefont{{Bechtol}}},
  \bibinfo{author}{\bibfnamefont{R.}~\bibnamefont{{Bellazzini}}},
  \bibnamefont{et~al.}, \bibinfo{journal}{Physical Review Letters}
  \textbf{\bibinfo{volume}{107}}, \bibinfo{eid}{241302} (\bibinfo{year}{2011}),
  \eprint{1108.3546}.

\bibitem[{\citenamefont{{Mazziotta} et~al.}(2012)\citenamefont{{Mazziotta},
  {Loparco}, {de Palma}, and {Giglietto}}}]{2012APh....37...26M}
\bibinfo{author}{\bibfnamefont{M.~N.} \bibnamefont{{Mazziotta}}},
  \bibinfo{author}{\bibfnamefont{F.}~\bibnamefont{{Loparco}}},
  \bibinfo{author}{\bibfnamefont{F.}~\bibnamefont{{de Palma}}},
  \bibnamefont{and}
  \bibinfo{author}{\bibfnamefont{N.}~\bibnamefont{{Giglietto}}},
  \bibinfo{journal}{Astroparticle Physics} \textbf{\bibinfo{volume}{37}},
  \bibinfo{pages}{26} (\bibinfo{year}{2012}), \eprint{1203.6731}.

\bibitem[{\citenamefont{{Baushev} et~al.}(2012)\citenamefont{{Baushev},
  {Federici}, and {Pohl}}}]{2012PhRvD..86f3521B}
\bibinfo{author}{\bibfnamefont{A.~N.} \bibnamefont{{Baushev}}},
  \bibinfo{author}{\bibfnamefont{S.}~\bibnamefont{{Federici}}},
  \bibnamefont{and} \bibinfo{author}{\bibfnamefont{M.}~\bibnamefont{{Pohl}}},
  \bibinfo{journal}{\prd} \textbf{\bibinfo{volume}{86}}, \bibinfo{eid}{063521}
  (\bibinfo{year}{2012}), \eprint{1205.3620}.

\bibitem[{\citenamefont{{He} et~al.}(2013)\citenamefont{{He}, {Bechtol},
  {Hearin}, and {Hooper}}}]{2013arXiv1309.4780H}
\bibinfo{author}{\bibfnamefont{C.}~\bibnamefont{{He}}},
  \bibinfo{author}{\bibfnamefont{K.}~\bibnamefont{{Bechtol}}},
  \bibinfo{author}{\bibfnamefont{A.~P.} \bibnamefont{{Hearin}}},
  \bibnamefont{and} \bibinfo{author}{\bibfnamefont{D.}~\bibnamefont{{Hooper}}},
  \bibinfo{journal}{ArXiv e-prints}  (\bibinfo{year}{2013}),
  \eprint{1309.4780}.

\bibitem[{\citenamefont{{Ackermann} et~al.}(2013)\citenamefont{{Ackermann},
  {Ajello}, {Albert}, {Allafort}, {Baldini}, {Barbiellini}, {Bastieri},
  {Bechtol}, {Bellazzini}, {Bissaldi} et~al.}}]{2013PhRvD..88h2002A}
\bibinfo{author}{\bibfnamefont{M.}~\bibnamefont{{Ackermann}}},
  \bibinfo{author}{\bibfnamefont{M.}~\bibnamefont{{Ajello}}},
  \bibinfo{author}{\bibfnamefont{A.}~\bibnamefont{{Albert}}},
  \bibinfo{author}{\bibfnamefont{A.}~\bibnamefont{{Allafort}}},
  \bibinfo{author}{\bibfnamefont{L.}~\bibnamefont{{Baldini}}},
  \bibinfo{author}{\bibfnamefont{G.}~\bibnamefont{{Barbiellini}}},
  \bibinfo{author}{\bibfnamefont{D.}~\bibnamefont{{Bastieri}}},
  \bibinfo{author}{\bibfnamefont{K.}~\bibnamefont{{Bechtol}}},
  \bibinfo{author}{\bibfnamefont{R.}~\bibnamefont{{Bellazzini}}},
  \bibinfo{author}{\bibfnamefont{E.}~\bibnamefont{{Bissaldi}}},
  \bibnamefont{et~al.}, \bibinfo{journal}{\prd} \textbf{\bibinfo{volume}{88}},
  \bibinfo{eid}{082002} (\bibinfo{year}{2013}).

\bibitem[{\citenamefont{{Natarajan} et~al.}(2013)\citenamefont{{Natarajan},
  {Peterson}, {Voytek}, {Spekkens}, {Mason}, {Aguirre}, and
  {Willman}}}]{2013PhRvD..88h3535N}
\bibinfo{author}{\bibfnamefont{A.}~\bibnamefont{{Natarajan}}},
  \bibinfo{author}{\bibfnamefont{J.~B.} \bibnamefont{{Peterson}}},
  \bibinfo{author}{\bibfnamefont{T.~C.} \bibnamefont{{Voytek}}},
  \bibinfo{author}{\bibfnamefont{K.}~\bibnamefont{{Spekkens}}},
  \bibinfo{author}{\bibfnamefont{B.}~\bibnamefont{{Mason}}},
  \bibinfo{author}{\bibfnamefont{J.}~\bibnamefont{{Aguirre}}},
  \bibnamefont{and}
  \bibinfo{author}{\bibfnamefont{B.}~\bibnamefont{{Willman}}},
  \bibinfo{journal}{\prd} \textbf{\bibinfo{volume}{88}}, \bibinfo{eid}{083535}
  (\bibinfo{year}{2013}), \eprint{1308.4979}.

\bibitem[{\citenamefont{{Sming Tsai} et~al.}(2013)\citenamefont{{Sming Tsai},
  {Yuan}, and {Huang}}}]{2013JCAP...03..018S}
\bibinfo{author}{\bibfnamefont{Y.-L.} \bibnamefont{{Sming Tsai}}},
  \bibinfo{author}{\bibfnamefont{Q.}~\bibnamefont{{Yuan}}}, \bibnamefont{and}
  \bibinfo{author}{\bibfnamefont{X.}~\bibnamefont{{Huang}}},
  \bibinfo{journal}{\jcap} \textbf{\bibinfo{volume}{3}}, \bibinfo{eid}{018}
  (\bibinfo{year}{2013}), \eprint{1212.3990}.

\bibitem[{\citenamefont{{Spekkens} et~al.}(2013)\citenamefont{{Spekkens},
  {Mason}, {Aguirre}, and {Nhan}}}]{2013ApJ...773...61S}
\bibinfo{author}{\bibfnamefont{K.}~\bibnamefont{{Spekkens}}},
  \bibinfo{author}{\bibfnamefont{B.~S.} \bibnamefont{{Mason}}},
  \bibinfo{author}{\bibfnamefont{J.~E.} \bibnamefont{{Aguirre}}},
  \bibnamefont{and} \bibinfo{author}{\bibfnamefont{B.}~\bibnamefont{{Nhan}}},
  \bibinfo{journal}{\apj} \textbf{\bibinfo{volume}{773}}, \bibinfo{eid}{61}
  (\bibinfo{year}{2013}), \eprint{1301.5306}.

\bibitem[{\citenamefont{{Carlson} et~al.}(2014)\citenamefont{{Carlson},
  {Hooper}, and {Linden}}}]{2014arXiv1409.1572C}
\bibinfo{author}{\bibfnamefont{E.}~\bibnamefont{{Carlson}}},
  \bibinfo{author}{\bibfnamefont{D.}~\bibnamefont{{Hooper}}}, \bibnamefont{and}
  \bibinfo{author}{\bibfnamefont{T.}~\bibnamefont{{Linden}}},
  \bibinfo{journal}{ArXiv e-prints}  (\bibinfo{year}{2014}),
  \eprint{1409.1572}.

\bibitem[{\citenamefont{{Aleksi{\'c}} et~al.}(2014)\citenamefont{{Aleksi{\'c}},
  {Ansoldi}, {Antonelli}, {Antoranz}, {Babic}, {Bangale}, {Barres de Almeida},
  {Barrio}, {Becerra Gonz{\'a}lez}, {Bednarek} et~al.}}]{2014JCAP...02..008A}
\bibinfo{author}{\bibfnamefont{J.}~\bibnamefont{{Aleksi{\'c}}}},
  \bibinfo{author}{\bibfnamefont{S.}~\bibnamefont{{Ansoldi}}},
  \bibinfo{author}{\bibfnamefont{L.~A.} \bibnamefont{{Antonelli}}},
  \bibinfo{author}{\bibfnamefont{P.}~\bibnamefont{{Antoranz}}},
  \bibinfo{author}{\bibfnamefont{A.}~\bibnamefont{{Babic}}},
  \bibinfo{author}{\bibfnamefont{P.}~\bibnamefont{{Bangale}}},
  \bibinfo{author}{\bibfnamefont{U.}~\bibnamefont{{Barres de Almeida}}},
  \bibinfo{author}{\bibfnamefont{J.~A.} \bibnamefont{{Barrio}}},
  \bibinfo{author}{\bibfnamefont{J.}~\bibnamefont{{Becerra Gonz{\'a}lez}}},
  \bibinfo{author}{\bibfnamefont{W.}~\bibnamefont{{Bednarek}}},
  \bibnamefont{et~al.}, \bibinfo{journal}{\jcap} \textbf{\bibinfo{volume}{2}},
  \bibinfo{eid}{008} (\bibinfo{year}{2014}), \eprint{1312.1535}.

\bibitem[{\citenamefont{{Charbonnier} et~al.}(2011)\citenamefont{{Charbonnier},
  {Combet}, {Daniel}, {Funk}, {Hinton}, {Maurin}, {Power}, {Read}, {Sarkar},
  {Walker} et~al.}}]{2011MNRAS.418.1526C}
\bibinfo{author}{\bibfnamefont{A.}~\bibnamefont{{Charbonnier}}},
  \bibinfo{author}{\bibfnamefont{C.}~\bibnamefont{{Combet}}},
  \bibinfo{author}{\bibfnamefont{M.}~\bibnamefont{{Daniel}}},
  \bibinfo{author}{\bibfnamefont{S.}~\bibnamefont{{Funk}}},
  \bibinfo{author}{\bibfnamefont{J.~A.} \bibnamefont{{Hinton}}},
  \bibinfo{author}{\bibfnamefont{D.}~\bibnamefont{{Maurin}}},
  \bibinfo{author}{\bibfnamefont{C.}~\bibnamefont{{Power}}},
  \bibinfo{author}{\bibfnamefont{J.~I.} \bibnamefont{{Read}}},
  \bibinfo{author}{\bibfnamefont{S.}~\bibnamefont{{Sarkar}}},
  \bibinfo{author}{\bibfnamefont{M.~G.} \bibnamefont{{Walker}}},
  \bibnamefont{et~al.}, \bibinfo{journal}{\mnras}
  \textbf{\bibinfo{volume}{418}}, \bibinfo{pages}{1526} (\bibinfo{year}{2011}),
  \eprint{1104.0412}.

\bibitem[{\citenamefont{Geringer-Sameth
  et~al.}(2015)\citenamefont{Geringer-Sameth, Koushiappas, and
  Walker}}]{0004-637X-801-2-74}
\bibinfo{author}{\bibfnamefont{A.}~\bibnamefont{Geringer-Sameth}},
  \bibinfo{author}{\bibfnamefont{S.~M.} \bibnamefont{Koushiappas}},
  \bibnamefont{and} \bibinfo{author}{\bibfnamefont{M.}~\bibnamefont{Walker}},
  \bibinfo{journal}{The Astrophysical Journal} \textbf{\bibinfo{volume}{801}},
  \bibinfo{pages}{74} (\bibinfo{year}{2015}),
  \urlprefix\url{http://stacks.iop.org/0004-637X/801/i=2/a=74}.

\bibitem[{\citenamefont{{Aliu} et~al.}()\citenamefont{{Aliu}, {(VERITAS
  Collaboration)} et~al.}}]{VERITASjointdwarfs}
\bibinfo{author}{\bibfnamefont{E.}~\bibnamefont{{Aliu}}},
  \bibinfo{author}{\bibnamefont{{(VERITAS Collaboration)}}},
  \bibnamefont{et~al.}, \bibinfo{note}{in preparation}.

\bibitem[{\citenamefont{{Kolb} and {Turner}}(1990)}]{1990eaun.book.....K}
\bibinfo{author}{\bibfnamefont{E.~W.} \bibnamefont{{Kolb}}} \bibnamefont{and}
  \bibinfo{author}{\bibfnamefont{M.~S.} \bibnamefont{{Turner}}},
  \emph{\bibinfo{title}{{The Early Universe}}}
  (\bibinfo{publisher}{Addison-Wesley}, \bibinfo{address}{Reading, Mass.},
  \bibinfo{year}{1990}).

\bibitem[{\citenamefont{{Cirelli} et~al.}(2011)\citenamefont{{Cirelli},
  {Corcella}, {Hektor}, {H{\"u}tsi}, {Kadastik}, {Panci}, {Raidal}, {Sala}, and
  {Strumia}}}]{2011JCAP...03..051C}
\bibinfo{author}{\bibfnamefont{M.}~\bibnamefont{{Cirelli}}},
  \bibinfo{author}{\bibfnamefont{G.}~\bibnamefont{{Corcella}}},
  \bibinfo{author}{\bibfnamefont{A.}~\bibnamefont{{Hektor}}},
  \bibinfo{author}{\bibfnamefont{G.}~\bibnamefont{{H{\"u}tsi}}},
  \bibinfo{author}{\bibfnamefont{M.}~\bibnamefont{{Kadastik}}},
  \bibinfo{author}{\bibfnamefont{P.}~\bibnamefont{{Panci}}},
  \bibinfo{author}{\bibfnamefont{M.}~\bibnamefont{{Raidal}}},
  \bibinfo{author}{\bibfnamefont{F.}~\bibnamefont{{Sala}}}, \bibnamefont{and}
  \bibinfo{author}{\bibfnamefont{A.}~\bibnamefont{{Strumia}}},
  \bibinfo{journal}{\jcap} \textbf{\bibinfo{volume}{3}}, \bibinfo{eid}{051}
  (\bibinfo{year}{2011}), \eprint{1012.4515}.

\bibitem[{\citenamefont{{Sj{\"o}strand}
  et~al.}(2008)\citenamefont{{Sj{\"o}strand}, {Mrenna}, and
  {Skands}}}]{2008CoPhC.178..852S}
\bibinfo{author}{\bibfnamefont{T.}~\bibnamefont{{Sj{\"o}strand}}},
  \bibinfo{author}{\bibfnamefont{S.}~\bibnamefont{{Mrenna}}}, \bibnamefont{and}
  \bibinfo{author}{\bibfnamefont{P.}~\bibnamefont{{Skands}}},
  \bibinfo{journal}{Computer Physics Communications}
  \textbf{\bibinfo{volume}{178}}, \bibinfo{pages}{852} (\bibinfo{year}{2008}),
  \eprint{0710.3820}.

\bibitem[{\citenamefont{{Kachelrie{\ss}} and
  {Serpico}}(2007)}]{2007PhRvD..76f3516K}
\bibinfo{author}{\bibfnamefont{M.}~\bibnamefont{{Kachelrie{\ss}}}}
  \bibnamefont{and} \bibinfo{author}{\bibfnamefont{P.~D.}
  \bibnamefont{{Serpico}}}, \bibinfo{journal}{\prd}
  \textbf{\bibinfo{volume}{76}}, \bibinfo{eid}{063516} (\bibinfo{year}{2007}),
  \eprint{0707.0209}.

\bibitem[{\citenamefont{{Bell} et~al.}(2008)\citenamefont{{Bell}, {Dent},
  {Jacques}, and {Weiler}}}]{2008PhRvD..78h3540B}
\bibinfo{author}{\bibfnamefont{N.~F.} \bibnamefont{{Bell}}},
  \bibinfo{author}{\bibfnamefont{J.~B.} \bibnamefont{{Dent}}},
  \bibinfo{author}{\bibfnamefont{T.~D.} \bibnamefont{{Jacques}}},
  \bibnamefont{and} \bibinfo{author}{\bibfnamefont{T.~J.}
  \bibnamefont{{Weiler}}}, \bibinfo{journal}{\prd}
  \textbf{\bibinfo{volume}{78}}, \bibinfo{eid}{083540} (\bibinfo{year}{2008}),
  \eprint{0805.3423}.

\bibitem[{\citenamefont{{Kachelrie{\ss}}
  et~al.}(2009)\citenamefont{{Kachelrie{\ss}}, {Serpico}, and
  {Solberg}}}]{2009PhRvD..80l3533K}
\bibinfo{author}{\bibfnamefont{M.}~\bibnamefont{{Kachelrie{\ss}}}},
  \bibinfo{author}{\bibfnamefont{P.~D.} \bibnamefont{{Serpico}}},
  \bibnamefont{and} \bibinfo{author}{\bibfnamefont{M.~A.}
  \bibnamefont{{Solberg}}}, \bibinfo{journal}{\prd}
  \textbf{\bibinfo{volume}{80}}, \bibinfo{eid}{123533} (\bibinfo{year}{2009}),
  \eprint{0911.0001}.

\bibitem[{\citenamefont{{Ciafaloni} and {Urbano}}(2010)}]{2010PhRvD..82d3512C}
\bibinfo{author}{\bibfnamefont{P.}~\bibnamefont{{Ciafaloni}}} \bibnamefont{and}
  \bibinfo{author}{\bibfnamefont{A.}~\bibnamefont{{Urbano}}},
  \bibinfo{journal}{\prd} \textbf{\bibinfo{volume}{82}}, \bibinfo{eid}{043512}
  (\bibinfo{year}{2010}), \eprint{1001.3950}.

\bibitem[{\citenamefont{{Ciafaloni} et~al.}(2011)\citenamefont{{Ciafaloni},
  {Comelli}, {Riotto}, {Sala}, {Strumia}, and {Urbano}}}]{2011JCAP...03..019C}
\bibinfo{author}{\bibfnamefont{P.}~\bibnamefont{{Ciafaloni}}},
  \bibinfo{author}{\bibfnamefont{D.}~\bibnamefont{{Comelli}}},
  \bibinfo{author}{\bibfnamefont{A.}~\bibnamefont{{Riotto}}},
  \bibinfo{author}{\bibfnamefont{F.}~\bibnamefont{{Sala}}},
  \bibinfo{author}{\bibfnamefont{A.}~\bibnamefont{{Strumia}}},
  \bibnamefont{and} \bibinfo{author}{\bibfnamefont{A.}~\bibnamefont{{Urbano}}},
  \bibinfo{journal}{\jcap} \textbf{\bibinfo{volume}{3}}, \bibinfo{pages}{19}
  (\bibinfo{year}{2011}), \eprint{1009.0224}.

\bibitem[{\citenamefont{{Ackermann} et~al.}(2012)\citenamefont{{Ackermann},
  {Ajello}, {Albert}, {Allafort}, {Atwood}, {Axelsson}, {Baldini}, {Ballet},
  {Barbiellini}, {Bastieri} et~al.}}]{2012ApJS..203....4A}
\bibinfo{author}{\bibfnamefont{M.}~\bibnamefont{{Ackermann}}},
  \bibinfo{author}{\bibfnamefont{M.}~\bibnamefont{{Ajello}}},
  \bibinfo{author}{\bibfnamefont{A.}~\bibnamefont{{Albert}}},
  \bibinfo{author}{\bibfnamefont{A.}~\bibnamefont{{Allafort}}},
  \bibinfo{author}{\bibfnamefont{W.~B.} \bibnamefont{{Atwood}}},
  \bibinfo{author}{\bibfnamefont{M.}~\bibnamefont{{Axelsson}}},
  \bibinfo{author}{\bibfnamefont{L.}~\bibnamefont{{Baldini}}},
  \bibinfo{author}{\bibfnamefont{J.}~\bibnamefont{{Ballet}}},
  \bibinfo{author}{\bibfnamefont{G.}~\bibnamefont{{Barbiellini}}},
  \bibinfo{author}{\bibfnamefont{D.}~\bibnamefont{{Bastieri}}},
  \bibnamefont{et~al.}, \bibinfo{journal}{\apjs}
  \textbf{\bibinfo{volume}{203}}, \bibinfo{eid}{4} (\bibinfo{year}{2012}),
  \eprint{1206.1896}.

\bibitem[{\citenamefont{{Hamilton}}(2000)}]{2000MNRAS.312..257H}
\bibinfo{author}{\bibfnamefont{A.~J.~S.} \bibnamefont{{Hamilton}}},
  \bibinfo{journal}{\mnras} \textbf{\bibinfo{volume}{312}},
  \bibinfo{pages}{257} (\bibinfo{year}{2000}), \eprint{astro-ph/9905191}.

\bibitem[{\citenamefont{{Fermi-LAT Collaboration}}(2011)}]{FSSC}
\bibinfo{author}{\bibnamefont{{Fermi-LAT Collaboration}}}
  (\bibinfo{year}{2011}), \urlprefix\url{http://fermi.gsfc.nasa.gov/ssc/data/}.

\bibitem[{Cic()}]{Cicerone}
\urlprefix\url{http://fermi.gsfc.nasa.gov/ssc/data/analysis/LAT_caveats.html}.

\bibitem[{\citenamefont{{Simon} and {Geha}}(2007)}]{simon07}
\bibinfo{author}{\bibfnamefont{J.~D.} \bibnamefont{{Simon}}} \bibnamefont{and}
  \bibinfo{author}{\bibfnamefont{M.}~\bibnamefont{{Geha}}},
  \bibinfo{journal}{\apj} \textbf{\bibinfo{volume}{670}}, \bibinfo{pages}{313}
  (\bibinfo{year}{2007}), \eprint{0706.0516}.

\bibitem[{\citenamefont{{Walker, Mateo \& Olszewski}}(2009)}]{walker09a}
\bibinfo{author}{\bibnamefont{{Walker, Mateo \& Olszewski}}},
  \bibinfo{journal}{\aj} \textbf{\bibinfo{volume}{137}}, \bibinfo{pages}{3100}
  (\bibinfo{year}{2009}), \eprint{0811.0118}.

\bibitem[{\citenamefont{{Binney} and {Tremaine}}(2008)}]{2008gady.book.....B}
\bibinfo{author}{\bibfnamefont{J.}~\bibnamefont{{Binney}}} \bibnamefont{and}
  \bibinfo{author}{\bibfnamefont{S.}~\bibnamefont{{Tremaine}}},
  \emph{\bibinfo{title}{{Galactic Dynamics: Second Edition}}}
  (\bibinfo{publisher}{Princeton University Press}, \bibinfo{year}{2008}).

\bibitem[{\citenamefont{{Walker}}(2013)}]{2013pss5.book.1039W}
\bibinfo{author}{\bibfnamefont{M.}~\bibnamefont{{Walker}}},
  \emph{\bibinfo{title}{{Dark Matter in the Galactic Dwarf Spheroidal
  Satellites}}} (\bibinfo{year}{2013}), p. \bibinfo{pages}{1039}.

\bibitem[{\citenamefont{{Battaglia} et~al.}(2013)\citenamefont{{Battaglia},
  {Helmi}, and {Breddels}}}]{2013NewAR..57...52B}
\bibinfo{author}{\bibfnamefont{G.}~\bibnamefont{{Battaglia}}},
  \bibinfo{author}{\bibfnamefont{A.}~\bibnamefont{{Helmi}}}, \bibnamefont{and}
  \bibinfo{author}{\bibfnamefont{M.}~\bibnamefont{{Breddels}}},
  \bibinfo{journal}{\nar} \textbf{\bibinfo{volume}{57}}, \bibinfo{pages}{52}
  (\bibinfo{year}{2013}), \eprint{1305.5965}.

\bibitem[{\citenamefont{{Strigari}}(2013)}]{2013PhR...531....1S}
\bibinfo{author}{\bibfnamefont{L.~E.} \bibnamefont{{Strigari}}},
  \bibinfo{journal}{\physrep} \textbf{\bibinfo{volume}{531}},
  \bibinfo{pages}{1} (\bibinfo{year}{2013}), \eprint{1211.7090}.

\bibitem[{\citenamefont{{Feroz} and {Hobson}}(2008)}]{2008MNRAS.384..449F}
\bibinfo{author}{\bibfnamefont{F.}~\bibnamefont{{Feroz}}} \bibnamefont{and}
  \bibinfo{author}{\bibfnamefont{M.~P.} \bibnamefont{{Hobson}}},
  \bibinfo{journal}{\mnras} \textbf{\bibinfo{volume}{384}},
  \bibinfo{pages}{449} (\bibinfo{year}{2008}), \eprint{0704.3704}.

\bibitem[{\citenamefont{{Feroz} et~al.}(2009)\citenamefont{{Feroz}, {Hobson},
  and {Bridges}}}]{2009MNRAS.398.1601F}
\bibinfo{author}{\bibfnamefont{F.}~\bibnamefont{{Feroz}}},
  \bibinfo{author}{\bibfnamefont{M.~P.} \bibnamefont{{Hobson}}},
  \bibnamefont{and}
  \bibinfo{author}{\bibfnamefont{M.}~\bibnamefont{{Bridges}}},
  \bibinfo{journal}{\mnras} \textbf{\bibinfo{volume}{398}},
  \bibinfo{pages}{1601} (\bibinfo{year}{2009}), \eprint{0809.3437}.

\bibitem[{\citenamefont{{Bonnivard} et~al.}(2015)\citenamefont{{Bonnivard},
  {Combet}, {Maurin}, and {Walker}}}]{2015MNRAS.446.3002B}
\bibinfo{author}{\bibfnamefont{V.}~\bibnamefont{{Bonnivard}}},
  \bibinfo{author}{\bibfnamefont{C.}~\bibnamefont{{Combet}}},
  \bibinfo{author}{\bibfnamefont{D.}~\bibnamefont{{Maurin}}}, \bibnamefont{and}
  \bibinfo{author}{\bibfnamefont{M.~G.} \bibnamefont{{Walker}}},
  \bibinfo{journal}{\mnras} \textbf{\bibinfo{volume}{446}},
  \bibinfo{pages}{3002} (\bibinfo{year}{2015}), \eprint{1407.7822}.

\bibitem[{\citenamefont{{Willman} et~al.}(2011)\citenamefont{{Willman}, {Geha},
  {Strader}, {Strigari}, {Simon}, {Kirby}, {Ho}, and
  {Warres}}}]{2011AJ....142..128W}
\bibinfo{author}{\bibfnamefont{B.}~\bibnamefont{{Willman}}},
  \bibinfo{author}{\bibfnamefont{M.}~\bibnamefont{{Geha}}},
  \bibinfo{author}{\bibfnamefont{J.}~\bibnamefont{{Strader}}},
  \bibinfo{author}{\bibfnamefont{L.~E.} \bibnamefont{{Strigari}}},
  \bibinfo{author}{\bibfnamefont{J.~D.} \bibnamefont{{Simon}}},
  \bibinfo{author}{\bibfnamefont{E.}~\bibnamefont{{Kirby}}},
  \bibinfo{author}{\bibfnamefont{N.}~\bibnamefont{{Ho}}}, \bibnamefont{and}
  \bibinfo{author}{\bibfnamefont{A.}~\bibnamefont{{Warres}}},
  \bibinfo{journal}{\aj} \textbf{\bibinfo{volume}{142}}, \bibinfo{eid}{128}
  (\bibinfo{year}{2011}), \eprint{1007.3499}.

\bibitem[{\citenamefont{Stuart and Ord}(1987)}]{Kendall5th}
\bibinfo{author}{\bibfnamefont{A.}~\bibnamefont{Stuart}} \bibnamefont{and}
  \bibinfo{author}{\bibfnamefont{K.~J.} \bibnamefont{Ord}},
  \emph{\bibinfo{title}{{Kendall's advanced theory of statistics}}}
  (\bibinfo{publisher}{Oxford University Press}, \bibinfo{address}{New York},
  \bibinfo{year}{1987}), \bibinfo{edition}{5th} ed.

\bibitem[{\citenamefont{Neyman and Pearson}(1933)}]{NeymanPearson1933}
\bibinfo{author}{\bibfnamefont{J.}~\bibnamefont{Neyman}} \bibnamefont{and}
  \bibinfo{author}{\bibfnamefont{E.~S.} \bibnamefont{Pearson}},
  \bibinfo{journal}{Philosophical Transactions of the Royal Society of London.
  Series A, Containing Papers of a Mathematical or Physical Character}
  \textbf{\bibinfo{volume}{231}}, \bibinfo{pages}{pp. 289}
  (\bibinfo{year}{1933}), ISSN \bibinfo{issn}{02643952},
  \urlprefix\url{http://www.jstor.org/stable/91247}.

\bibitem[{\citenamefont{Adelson}(1966)}]{adelson1966}
\bibinfo{author}{\bibfnamefont{R.~M.} \bibnamefont{Adelson}},
  \bibinfo{journal}{OR} \textbf{\bibinfo{volume}{17}}, \bibinfo{pages}{pp. 73}
  (\bibinfo{year}{1966}), ISSN \bibinfo{issn}{14732858},
  \urlprefix\url{http://www.jstor.org/stable/3007241}.

\bibitem[{\citenamefont{Embrechts and Frei}(2009)}]{Embrechts:2009ly}
\bibinfo{author}{\bibfnamefont{P.}~\bibnamefont{Embrechts}} \bibnamefont{and}
  \bibinfo{author}{\bibfnamefont{M.}~\bibnamefont{Frei}},
  \bibinfo{journal}{Mathematical Methods of Operations Research}
  \textbf{\bibinfo{volume}{69}}, \bibinfo{pages}{497} (\bibinfo{year}{2009}),
  \urlprefix\url{http://dx.doi.org/10.1007/s00186-008-0249-2}.

\bibitem[{\citenamefont{{Nolan} et~al.}(2012)\citenamefont{{Nolan}, {Abdo},
  {Ackermann}, {Ajello}, {Allafort}, {Antolini}, {Atwood}, {Axelsson},
  {Baldini}, {Ballet} et~al.}}]{2012ApJS..199...31N}
\bibinfo{author}{\bibfnamefont{P.~L.} \bibnamefont{{Nolan}}},
  \bibinfo{author}{\bibfnamefont{A.~A.} \bibnamefont{{Abdo}}},
  \bibinfo{author}{\bibfnamefont{M.}~\bibnamefont{{Ackermann}}},
  \bibinfo{author}{\bibfnamefont{M.}~\bibnamefont{{Ajello}}},
  \bibinfo{author}{\bibfnamefont{A.}~\bibnamefont{{Allafort}}},
  \bibinfo{author}{\bibfnamefont{E.}~\bibnamefont{{Antolini}}},
  \bibinfo{author}{\bibfnamefont{W.~B.} \bibnamefont{{Atwood}}},
  \bibinfo{author}{\bibfnamefont{M.}~\bibnamefont{{Axelsson}}},
  \bibinfo{author}{\bibfnamefont{L.}~\bibnamefont{{Baldini}}},
  \bibinfo{author}{\bibfnamefont{J.}~\bibnamefont{{Ballet}}},
  \bibnamefont{et~al.}, \bibinfo{journal}{\apjs}
  \textbf{\bibinfo{volume}{199}}, \bibinfo{eid}{31} (\bibinfo{year}{2012}),
  \eprint{1108.1435}.

\bibitem[{\citenamefont{{Geringer-Sameth} and
  {Koushiappas}}(2012)}]{2012PhRvD..86b1302G}
\bibinfo{author}{\bibfnamefont{A.}~\bibnamefont{{Geringer-Sameth}}}
  \bibnamefont{and} \bibinfo{author}{\bibfnamefont{S.~M.}
  \bibnamefont{{Koushiappas}}}, \bibinfo{journal}{\prd}
  \textbf{\bibinfo{volume}{86}}, \bibinfo{eid}{021302} (\bibinfo{year}{2012}),
  \eprint{1206.0796}.

\bibitem[{\citenamefont{{Lee} et~al.}(2009)\citenamefont{{Lee}, {Ando}, and
  {Kamionkowski}}}]{2009JCAP...07..007L}
\bibinfo{author}{\bibfnamefont{S.~K.} \bibnamefont{{Lee}}},
  \bibinfo{author}{\bibfnamefont{S.}~\bibnamefont{{Ando}}}, \bibnamefont{and}
  \bibinfo{author}{\bibfnamefont{M.}~\bibnamefont{{Kamionkowski}}},
  \bibinfo{journal}{\jcap} \textbf{\bibinfo{volume}{7}}, \bibinfo{pages}{7}
  (\bibinfo{year}{2009}), \eprint{0810.1284}.

\bibitem[{\citenamefont{{Dodelson} et~al.}(2009)\citenamefont{{Dodelson},
  {Belikov}, {Hooper}, and {Serpico}}}]{2009PhRvD..80h3504D}
\bibinfo{author}{\bibfnamefont{S.}~\bibnamefont{{Dodelson}}},
  \bibinfo{author}{\bibfnamefont{A.~V.} \bibnamefont{{Belikov}}},
  \bibinfo{author}{\bibfnamefont{D.}~\bibnamefont{{Hooper}}}, \bibnamefont{and}
  \bibinfo{author}{\bibfnamefont{P.}~\bibnamefont{{Serpico}}},
  \bibinfo{journal}{\prd} \textbf{\bibinfo{volume}{80}},
  \bibinfo{pages}{083504} (\bibinfo{year}{2009}), \eprint{0903.2829}.

\bibitem[{\citenamefont{Baxter et~al.}(2010)\citenamefont{Baxter, Dodelson,
  Koushiappas, and Strigari}}]{2010PhRvD..82l3511B}
\bibinfo{author}{\bibfnamefont{E.~J.} \bibnamefont{Baxter}},
  \bibinfo{author}{\bibfnamefont{S.}~\bibnamefont{Dodelson}},
  \bibinfo{author}{\bibfnamefont{S.~M.} \bibnamefont{Koushiappas}},
  \bibnamefont{and} \bibinfo{author}{\bibfnamefont{L.~E.}
  \bibnamefont{Strigari}}, \bibinfo{journal}{\prd}
  \textbf{\bibinfo{volume}{82}}, \bibinfo{pages}{123511}
  (\bibinfo{year}{2010}), \eprint{1006.2399}.

\bibitem[{\citenamefont{{CMS Collaboration Chatrchyan}
  et~al.}(2012)\citenamefont{{CMS Collaboration Chatrchyan}, {Khachatryan},
  {Sirunyan}, {Tumasyan}, {Adam}, {Aguilo}, {Bergauer}, {Dragicevic},
  {Er{\"o}}, {Fabjan} et~al.}}]{2012PhLB..716...30C}
\bibinfo{author}{\bibfnamefont{S.}~\bibnamefont{{CMS Collaboration
  Chatrchyan}}},
  \bibinfo{author}{\bibfnamefont{V.}~\bibnamefont{{Khachatryan}}},
  \bibinfo{author}{\bibfnamefont{A.~M.} \bibnamefont{{Sirunyan}}},
  \bibinfo{author}{\bibfnamefont{A.}~\bibnamefont{{Tumasyan}}},
  \bibinfo{author}{\bibfnamefont{W.}~\bibnamefont{{Adam}}},
  \bibinfo{author}{\bibfnamefont{E.}~\bibnamefont{{Aguilo}}},
  \bibinfo{author}{\bibfnamefont{T.}~\bibnamefont{{Bergauer}}},
  \bibinfo{author}{\bibfnamefont{M.}~\bibnamefont{{Dragicevic}}},
  \bibinfo{author}{\bibfnamefont{J.}~\bibnamefont{{Er{\"o}}}},
  \bibinfo{author}{\bibfnamefont{C.}~\bibnamefont{{Fabjan}}},
  \bibnamefont{et~al.}, \bibinfo{journal}{Physics Letters B}
  \textbf{\bibinfo{volume}{716}}, \bibinfo{pages}{30} (\bibinfo{year}{2012}),
  \eprint{1207.7235}.

\bibitem[{\citenamefont{{Ackermann} et~al.}(2014)\citenamefont{{Ackermann},
  {Albert}, {Anderson}, {Baldini}, {Ballet}, {Barbiellini}, {Bastieri},
  {Bechtol}, {Bellazzini}, {Bissaldi} et~al.}}]{2014PhRvD..89d2001A}
\bibinfo{author}{\bibfnamefont{M.}~\bibnamefont{{Ackermann}}},
  \bibinfo{author}{\bibfnamefont{A.}~\bibnamefont{{Albert}}},
  \bibinfo{author}{\bibfnamefont{B.}~\bibnamefont{{Anderson}}},
  \bibinfo{author}{\bibfnamefont{L.}~\bibnamefont{{Baldini}}},
  \bibinfo{author}{\bibfnamefont{J.}~\bibnamefont{{Ballet}}},
  \bibinfo{author}{\bibfnamefont{G.}~\bibnamefont{{Barbiellini}}},
  \bibinfo{author}{\bibfnamefont{D.}~\bibnamefont{{Bastieri}}},
  \bibinfo{author}{\bibfnamefont{K.}~\bibnamefont{{Bechtol}}},
  \bibinfo{author}{\bibfnamefont{R.}~\bibnamefont{{Bellazzini}}},
  \bibinfo{author}{\bibfnamefont{E.}~\bibnamefont{{Bissaldi}}},
  \bibnamefont{et~al.}, \bibinfo{journal}{\prd} \textbf{\bibinfo{volume}{89}},
  \bibinfo{eid}{042001} (\bibinfo{year}{2014}).

\bibitem[{\citenamefont{{Lacroix}
  et~al.}(2014{\natexlab{b}})\citenamefont{{Lacroix}, {B{\AA}`hm}, and
  {Silk}}}]{2014PhRvD..90d3508L}
\bibinfo{author}{\bibfnamefont{T.}~\bibnamefont{{Lacroix}}},
  \bibinfo{author}{\bibfnamefont{C.}~\bibnamefont{{B{\AA}`hm}}},
  \bibnamefont{and} \bibinfo{author}{\bibfnamefont{J.}~\bibnamefont{{Silk}}},
  \bibinfo{journal}{\prd} \textbf{\bibinfo{volume}{90}}, \bibinfo{eid}{043508}
  (\bibinfo{year}{2014}{\natexlab{b}}), \eprint{1403.1987}.

\bibitem[{\citenamefont{{Boyarsky} et~al.}(2011)\citenamefont{{Boyarsky},
  {Malyshev}, and {Ruchayskiy}}}]{2011PhLB..705..165B}
\bibinfo{author}{\bibfnamefont{A.}~\bibnamefont{{Boyarsky}}},
  \bibinfo{author}{\bibfnamefont{D.}~\bibnamefont{{Malyshev}}},
  \bibnamefont{and}
  \bibinfo{author}{\bibfnamefont{O.}~\bibnamefont{{Ruchayskiy}}},
  \bibinfo{journal}{Physics Letters B} \textbf{\bibinfo{volume}{705}},
  \bibinfo{pages}{165} (\bibinfo{year}{2011}), \eprint{1012.5839}.

\bibitem[{\citenamefont{{Carlson} and {Profumo}}(2014)}]{2014PhRvD..90b3015C}
\bibinfo{author}{\bibfnamefont{E.}~\bibnamefont{{Carlson}}} \bibnamefont{and}
  \bibinfo{author}{\bibfnamefont{S.}~\bibnamefont{{Profumo}}},
  \bibinfo{journal}{\prd} \textbf{\bibinfo{volume}{90}}, \bibinfo{eid}{023015}
  (\bibinfo{year}{2014}), \eprint{1405.7685}.

\bibitem[{\citenamefont{{Mirabal}}(2013)}]{2013MNRAS.436.2461M}
\bibinfo{author}{\bibfnamefont{N.}~\bibnamefont{{Mirabal}}},
  \bibinfo{journal}{\mnras} \textbf{\bibinfo{volume}{436}},
  \bibinfo{pages}{2461} (\bibinfo{year}{2013}), \eprint{1309.3428}.

\bibitem[{\citenamefont{{Yuan} and {Zhang}}(2014)}]{2014JHEAp...3....1Y}
\bibinfo{author}{\bibfnamefont{Q.}~\bibnamefont{{Yuan}}} \bibnamefont{and}
  \bibinfo{author}{\bibfnamefont{B.}~\bibnamefont{{Zhang}}},
  \bibinfo{journal}{Journal of High Energy Astrophysics}
  \textbf{\bibinfo{volume}{3}}, \bibinfo{pages}{1} (\bibinfo{year}{2014}),
  \eprint{1404.2318}.

\bibitem[{\citenamefont{{Bringmann} et~al.}(2014)\citenamefont{{Bringmann},
  {Vollmann}, and {Weniger}}}]{2014arXiv1406.6027B}
\bibinfo{author}{\bibfnamefont{T.}~\bibnamefont{{Bringmann}}},
  \bibinfo{author}{\bibfnamefont{M.}~\bibnamefont{{Vollmann}}},
  \bibnamefont{and}
  \bibinfo{author}{\bibfnamefont{C.}~\bibnamefont{{Weniger}}},
  \bibinfo{journal}{ArXiv e-prints}  (\bibinfo{year}{2014}),
  \eprint{1406.6027}.

\bibitem[{\citenamefont{{Hooper}}()}]{HooperPC}
\bibinfo{author}{\bibfnamefont{D.}~\bibnamefont{{Hooper}}},
  \bibinfo{note}{private communication}.

\bibitem[{\citenamefont{{Conrad}}(2014)}]{2014arXiv1407.6617C}
\bibinfo{author}{\bibfnamefont{J.}~\bibnamefont{{Conrad}}},
  \bibinfo{journal}{ArXiv e-prints}  (\bibinfo{year}{2014}),
  \eprint{1407.6617}.

\bibitem[{\citenamefont{{Martinez}}(2013)}]{2013arXiv1309.2641M}
\bibinfo{author}{\bibfnamefont{G.~D.} \bibnamefont{{Martinez}}},
  \bibinfo{journal}{ArXiv e-prints}  (\bibinfo{year}{2013}),
  \eprint{1309.2641}.

\bibitem[{\citenamefont{Wasserman}(2004)}]{wasserman2004all}
\bibinfo{author}{\bibfnamefont{L.}~\bibnamefont{Wasserman}},
  \emph{\bibinfo{title}{All of Statistics: A Concise Course in Statistical
  Inference}}, Springer Texts in Statistics (\bibinfo{publisher}{Springer},
  \bibinfo{year}{2004}), ISBN \bibinfo{isbn}{9780387402727}.

\bibitem[{gai(2013)}]{gaiachallengewiki}
\emph{\bibinfo{title}{Gaia challenge wiki}} (\bibinfo{year}{2013}),
  \urlprefix\url{http://astrowiki.ph.surrey.ac.uk/dokuwiki/doku.php?id=tests:sphtri}.

\bibitem[{\citenamefont{{Walker} and
  {Pe{\~n}arrubia}}(2011)}]{2011ApJ...742...20W}
\bibinfo{author}{\bibfnamefont{M.~G.} \bibnamefont{{Walker}}} \bibnamefont{and}
  \bibinfo{author}{\bibfnamefont{J.}~\bibnamefont{{Pe{\~n}arrubia}}},
  \bibinfo{journal}{\apj} \textbf{\bibinfo{volume}{742}}, \bibinfo{eid}{20}
  (\bibinfo{year}{2011}), \eprint{1108.2404}.

\bibitem[{\citenamefont{{Dehnen}}(2009)}]{2009MNRAS.395.1079D}
\bibinfo{author}{\bibfnamefont{W.}~\bibnamefont{{Dehnen}}},
  \bibinfo{journal}{\mnras} \textbf{\bibinfo{volume}{395}},
  \bibinfo{pages}{1079} (\bibinfo{year}{2009}), \eprint{0902.2069}.

\bibitem[{\citenamefont{{Pullen} et~al.}(2007)\citenamefont{{Pullen}, {Chary},
  and {Kamionkowski}}}]{2007PhRvD..76f3006P}
\bibinfo{author}{\bibfnamefont{A.~R.} \bibnamefont{{Pullen}}},
  \bibinfo{author}{\bibfnamefont{R.-R.} \bibnamefont{{Chary}}},
  \bibnamefont{and}
  \bibinfo{author}{\bibfnamefont{M.}~\bibnamefont{{Kamionkowski}}},
  \bibinfo{journal}{\prd} \textbf{\bibinfo{volume}{76}}, \bibinfo{eid}{063006}
  (\bibinfo{year}{2007}), \eprint{astro-ph/0610295}.

\bibitem[{\citenamefont{{Bringmann} et~al.}(2011)\citenamefont{{Bringmann},
  {Calore}, {Vertongen}, and {Weniger}}}]{2011arXiv1106.1874B}
\bibinfo{author}{\bibfnamefont{T.}~\bibnamefont{{Bringmann}}},
  \bibinfo{author}{\bibfnamefont{F.}~\bibnamefont{{Calore}}},
  \bibinfo{author}{\bibfnamefont{G.}~\bibnamefont{{Vertongen}}},
  \bibnamefont{and}
  \bibinfo{author}{\bibfnamefont{C.}~\bibnamefont{{Weniger}}},
  \bibinfo{journal}{ArXiv e-prints}  (\bibinfo{year}{2011}),
  \eprint{1106.1874}.

\bibitem[{\citenamefont{{Bringmann} et~al.}(2012)\citenamefont{{Bringmann},
  {Huang}, {Ibarra}, {Vogl}, and {Weniger}}}]{2012JCAP...07..054B}
\bibinfo{author}{\bibfnamefont{T.}~\bibnamefont{{Bringmann}}},
  \bibinfo{author}{\bibfnamefont{X.}~\bibnamefont{{Huang}}},
  \bibinfo{author}{\bibfnamefont{A.}~\bibnamefont{{Ibarra}}},
  \bibinfo{author}{\bibfnamefont{S.}~\bibnamefont{{Vogl}}}, \bibnamefont{and}
  \bibinfo{author}{\bibfnamefont{C.}~\bibnamefont{{Weniger}}},
  \bibinfo{journal}{\jcap} \textbf{\bibinfo{volume}{7}}, \bibinfo{eid}{054}
  (\bibinfo{year}{2012}), \eprint{1203.1312}.

\bibitem[{\citenamefont{{Atwood} et~al.}(2013)\citenamefont{{Atwood}, {Albert},
  {Baldini}, {Tinivella}, {Bregeon}, {Pesce-Rollins}, {Sgr{\`o}}, {Bruel},
  {Charles}, {Drlica-Wagner} et~al.}}]{2013arXiv1303.3514A}
\bibinfo{author}{\bibfnamefont{W.}~\bibnamefont{{Atwood}}},
  \bibinfo{author}{\bibfnamefont{A.}~\bibnamefont{{Albert}}},
  \bibinfo{author}{\bibfnamefont{L.}~\bibnamefont{{Baldini}}},
  \bibinfo{author}{\bibfnamefont{M.}~\bibnamefont{{Tinivella}}},
  \bibinfo{author}{\bibfnamefont{J.}~\bibnamefont{{Bregeon}}},
  \bibinfo{author}{\bibfnamefont{M.}~\bibnamefont{{Pesce-Rollins}}},
  \bibinfo{author}{\bibfnamefont{C.}~\bibnamefont{{Sgr{\`o}}}},
  \bibinfo{author}{\bibfnamefont{P.}~\bibnamefont{{Bruel}}},
  \bibinfo{author}{\bibfnamefont{E.}~\bibnamefont{{Charles}}},
  \bibinfo{author}{\bibfnamefont{A.}~\bibnamefont{{Drlica-Wagner}}},
  \bibnamefont{et~al.}, \bibinfo{journal}{ArXiv e-prints}
  (\bibinfo{year}{2013}), \eprint{1303.3514}.

\bibitem[{Pas(2014)}]{Pass8}
 (\bibinfo{year}{2014}),
  \urlprefix\url{http://fermi.gsfc.nasa.gov/ssc/library/fug/140605/Fermi_LAT_Report.pdf}.

\bibitem[{\citenamefont{{Racusin} and {Fermi Large Area Telescope
  Collaboration}}(2014)}]{2014AAS...22335203R}
\bibinfo{author}{\bibfnamefont{J.~L.} \bibnamefont{{Racusin}}}
  \bibnamefont{and} \bibinfo{author}{\bibnamefont{{Fermi Large Area Telescope
  Collaboration}}}, in \emph{\bibinfo{booktitle}{American Astronomical Society
  Meeting Abstracts \#223}} (\bibinfo{year}{2014}), vol. \bibinfo{volume}{223}
  of \emph{\bibinfo{series}{American Astronomical Society Meeting Abstracts}},
  p. \bibinfo{pages}{\#352.03}.

\bibitem[{\citenamefont{{Grove} and {on behalf of the Fermi LAT
  Collaboration}}(2014)}]{2014AAS...22325603G}
\bibinfo{author}{\bibfnamefont{J.~E.} \bibnamefont{{Grove}}} \bibnamefont{and}
  \bibinfo{author}{\bibnamefont{{on behalf of the Fermi LAT Collaboration}}},
  in \emph{\bibinfo{booktitle}{American Astronomical Society Meeting Abstracts
  \#223}} (\bibinfo{year}{2014}), vol. \bibinfo{volume}{223} of
  \emph{\bibinfo{series}{American Astronomical Society Meeting Abstracts}}, p.
  \bibinfo{pages}{\#256.03}.

\bibitem[{\citenamefont{{Scheuer}}(1957)}]{1957PCPS...53..764S}
\bibinfo{author}{\bibfnamefont{P.~A.~G.} \bibnamefont{{Scheuer}}},
  \bibinfo{journal}{Proceedings of the Cambridge Philosophical Society}
  \textbf{\bibinfo{volume}{53}}, \bibinfo{pages}{764} (\bibinfo{year}{1957}).

\bibitem[{\citenamefont{Press et~al.}(2007)\citenamefont{Press, Teukolsky,
  Vetterling, and Flannery}}]{Press:2007:NRE:1403886}
\bibinfo{author}{\bibfnamefont{W.~H.} \bibnamefont{Press}},
  \bibinfo{author}{\bibfnamefont{S.~A.} \bibnamefont{Teukolsky}},
  \bibinfo{author}{\bibfnamefont{W.~T.} \bibnamefont{Vetterling}},
  \bibnamefont{and} \bibinfo{author}{\bibfnamefont{B.~P.}
  \bibnamefont{Flannery}}, \emph{\bibinfo{title}{Numerical Recipes 3rd Edition:
  The Art of Scientific Computing}} (\bibinfo{publisher}{Cambridge University
  Press}, \bibinfo{address}{New York, NY, USA}, \bibinfo{year}{2007}),
  \bibinfo{edition}{3rd} ed., ISBN \bibinfo{isbn}{0521880688, 9780521880688}.

\bibitem[{\citenamefont{Gr{\"u}bel and
  Hermesmeier}(1999)}]{grubel1999computation}
\bibinfo{author}{\bibfnamefont{R.}~\bibnamefont{Gr{\"u}bel}} \bibnamefont{and}
  \bibinfo{author}{\bibfnamefont{R.}~\bibnamefont{Hermesmeier}},
  \bibinfo{journal}{Astin Bulletin} \textbf{\bibinfo{volume}{29}},
  \bibinfo{pages}{197} (\bibinfo{year}{1999}).

\bibitem[{FFT()}]{FFTlog}
\urlprefix\url{http://casa.colorado.edu/~ajsh/FFTLog/}.

\bibitem[{\citenamefont{{Lande} et~al.}(2012)\citenamefont{{Lande},
  {Ackermann}, {Allafort}, {Ballet}, {Bechtol}, {Burnett}, {Cohen-Tanugi},
  {Drlica-Wagner}, {Funk}, {Giordano} et~al.}}]{2012ApJ...756....5L}
\bibinfo{author}{\bibfnamefont{J.}~\bibnamefont{{Lande}}},
  \bibinfo{author}{\bibfnamefont{M.}~\bibnamefont{{Ackermann}}},
  \bibinfo{author}{\bibfnamefont{A.}~\bibnamefont{{Allafort}}},
  \bibinfo{author}{\bibfnamefont{J.}~\bibnamefont{{Ballet}}},
  \bibinfo{author}{\bibfnamefont{K.}~\bibnamefont{{Bechtol}}},
  \bibinfo{author}{\bibfnamefont{T.~H.} \bibnamefont{{Burnett}}},
  \bibinfo{author}{\bibfnamefont{J.}~\bibnamefont{{Cohen-Tanugi}}},
  \bibinfo{author}{\bibfnamefont{A.}~\bibnamefont{{Drlica-Wagner}}},
  \bibinfo{author}{\bibfnamefont{S.}~\bibnamefont{{Funk}}},
  \bibinfo{author}{\bibfnamefont{F.}~\bibnamefont{{Giordano}}},
  \bibnamefont{et~al.}, \bibinfo{journal}{\apj} \textbf{\bibinfo{volume}{756}},
  \bibinfo{eid}{5} (\bibinfo{year}{2012}), \eprint{1207.0027}.

\end{thebibliography}

\end{document}